\documentclass[prx,a4paper,aps,twocolumn,superscriptaddress,nofootinbib,10pt]{revtex4-2}

\usepackage[utf8]{inputenc} 	
\usepackage[T1]{fontenc} 	
\usepackage[english]{babel}  
\usepackage{bbm, amsthm, amssymb, placeins, graphicx, wrapfig, subfigure, sidecap, float, verbatim, color, multirow, scalerel, makecell, mathrsfs, enumerate, amsmath, etoolbox, orcidlink, tikz, xparse, soul, dsfont, tabularx, booktabs, pifont}

\renewcommand{\thesection}{\arabic{section}}

\makeatletter
\renewcommand{\p@subsection}{}
\renewcommand{\p@subsubsection}{}

\makeatother

\usetikzlibrary{svg.path}

\hypersetup{
    unicode=true, breaklinks=false, pdfborder={0 0 1}, backref=false, colorlinks=true, linkcolor=blue, citecolor=blue
}

\allowdisplaybreaks

\numberwithin{equation}{section}

\usepackage[most]{tcolorbox} 

\newtcbtheorem[auto counter, number within=section]{Definition}{}{
    enhanced,
    sharp corners,
    attach boxed title to top left={
        xshift=-1mm,
        yshift=-5mm,
        yshifttext=-1mm
    },
    top=1.5em,
    colback=white,
    colframe=blue!75!black,
    fonttitle=\bfseries,
    boxed title style={
        sharp corners,
        size=small,
        colback=blue!75!black,
        colframe=blue!75!black,
    }
}{def}

\newcounter{definition}[section]
\renewcommand{\thedefinition}{\thesection.\arabic{definition}}

\NewDocumentEnvironment{myDefinition}{ o m o }{%
  \begin{Definition}[title=#2]%
    \IfValueT{#3}{\refstepcounter{definition}\label{#3}} 
    \IfValueT{#1}{\textbf{Definition \thedefinition. } #1}{\textbf{Definition \thedefinition. }}%
}{%
  \end{Definition}
}

\newenvironment{myDefinition*}[2]{ \begin{Definition}[adjusted title=#1]{}{#2}
  }{\end{Definition}}

\newtcbtheorem[auto counter, number within=section]{Proposition}{}{
    enhanced,
    sharp corners,
    attach boxed title to top left={
        xshift=-1mm,
        yshift=-5mm,
        yshifttext=-1mm
    },
    top=1.5em,
    colback=white,
    colframe=green!50!black,
    fonttitle=\bfseries,
    boxed title style={
        sharp corners,
        size=small,
        colback=green!50!black,
        colframe=green!50!black,
    }
}{def}

\newcounter{proposition}[section]
\renewcommand{\theproposition}{\thesection.\arabic{proposition}}

\NewDocumentEnvironment{myProposition}{ o m o }{%
  \begin{Proposition}[title=#2]%
    \IfValueT{#3}{\refstepcounter{proposition}\label{#3}} 
    \IfValueT{#1}{\textbf{Proposition \theproposition. } #1}{\textbf{Proposition \theproposition. }}%
}{%
  \end{Proposition}
}

\newenvironment{myProposition*}[2]{ \begin{Proposition}[adjusted title=#1]{}{#2}
  }{\end{Proposition}}

\newtcbtheorem[auto counter, number within=section]{Example}{}{
    enhanced,
    sharp corners,
    attach boxed title to top left={
        xshift=-1mm,
        yshift=-5mm,
        yshifttext=-1mm
    },
    top=1.5em,
    colback=white,
    colframe=red!50!black,
    fonttitle=\bfseries,
    boxed title style={
        sharp corners,
        size=small,
        colback=red!50!black,
        colframe=red!50!black,
    }
}{def}

\NewDocumentEnvironment{myRemark}{ o m o }{%
  \begin{Proposition}[title=#2]%
    \IfValueT{#3}{\refstepcounter{proposition}\label{#3}} 
    \IfValueT{#1}{\textbf{Remark \theproposition. } #1}{\textbf{Remark \theproposition. }}%
}{%
  \end{Proposition}
}

\newcounter{examplecount}[section]
\renewcommand{\theexamplecount}{\thesection.\arabic{examplecount}}

\NewDocumentEnvironment{myExample}{ o m o }{%
  \begin{Example}[title=#2]%
    \IfValueT{#3}{\refstepcounter{examplecount}\label{#3}} 
    \IfValueT{#1}{\textbf{Example \theexamplecount. } #1}{\textbf{Example \theexamplecount. }}%
}{%
  \end{Example}
}

\newenvironment{myExample*}[2]{ \begin{Example}[adjusted title=#1]{}{#2}
  }{\end{Example}}
  
\graphicspath{ {Images/} }

\newcommand{\inp}{\texttt{i}}
\newcommand{\out}{\texttt{o}}

\definecolor{darkorange}{RGB}{255,100,0}

\def\BraVert{\egroup\,\mid\,\bgroup}

\def\ketbra#1#2{\ket{#1\vphantom{#2}}\!\bra{#2\vphantom{#1}}}

\def\bra#1{\mathinner{\langle{#1}|}}

\def\ket#1{\mathinner{|{#1}\rangle}}

\def\braket#1{\mathinner{\langle{#1}\rangle}}

\makeatletter
\theoremstyle{plain}
\newtheorem{thm}{\protect\theoremname}
\theoremstyle{plain}
\newtheorem{lem}[thm]{\protect\lemmaname}
\theoremstyle{plain}

\theoremstyle{remark}
\newtheorem*{rem*}{\protect\remarkname}
\theoremstyle{plain}

\theoremstyle{plain}
\newtheorem{cor}[thm]{\protect\corollaryname}
\theoremstyle{definition}
\newtheorem{defn}[thm]{\protect\definitionname}
\theoremstyle{plain}
\newtheorem*{thm*}{\protect\theoremname}
\theoremstyle{plain}
\newtheorem*{lem*}{\protect\lemmaname}
\theoremstyle{definition}
\newtheorem{example}{\protect\examplename}
 
\providecommand{\propositionname}{Proposition}
\providecommand{\theoremname}{Theorem}
\providecommand{\lemmaname}{Lemma}
\providecommand{\remarkname}{Remark}
\providecommand{\conjecturename}{Conjecture}
\providecommand{\definitionname}{Definition}
\providecommand{\corollaryname}{Corollary}
\providecommand{\examplename}{Example}

\makeatletter
\renewcommand\@dotsep{500}   
\makeatother

\begin{document}

\title{Quantum Dynamics of Dynamics}

\author{Takano Taira\,\orcidlink{0000-0003-3083-9352}}
\email{takano.taira@riken.jp}
\affiliation{Analytical Quantum Complexity RIKEN Hakubi Research Team,
RIKEN Center for Quantum Computing (RQC), 2-1 Hirosawa, Wako, Saitama 351-0198, Japan}

\author{Hiroyasu Tajima\,\orcidlink{0000-0002-5983-9930}}
\email{hiroyasu.tajima@inf.kyushu-u.ac.jp}
\affiliation{Department of Informatics, Faculty of Information Science and Electrical Engineering,
Kyushu University, 744 Motooka, Nishi-ku, Fukuoka, 819-0395, Japan}

\author{Philip Taranto\,\orcidlink{0000-0002-4247-3901}}
\email{philip.taranto@manchester.ac.uk}
\affiliation{Department of Physics \& Astronomy, University of Manchester, Manchester M13 9PL, United Kingdom}

\date{\today}

\begin{abstract}
Quantum channels (in discrete time) and Gorini--Kossakowski--Sudarshan--Lindblad (GKSL) master equations (in continuous time) are the central tools of open quantum systems, but they rest on memoryless approximations (Born, Markov, secular) that fail in many physically relevant settings, most notably non-Markovian dynamics. There, tracking the state alone is insufficient: interventions at intermediate times disturb the environment, so multi-time correlations carry physical information that the state evolution cannot capture. The framework of higher-order quantum operations addresses this by treating controllable operations as \textit{inputs} to a process that maps operations to operations, so that the primitive object is a quantum \textit{dynamics} rather than a quantum \textit{state}. Here, we derive the most general continuous-time, CP-divisible evolution of such an object, the \textit{quantum comb master equation} (QCME), which governs the quantum ``dynamics of dynamics''. Its generator has a GKSL-like form supplemented by encoder--decoder correlation terms that couple different temporal legs of the comb. We also derive the QCME microscopically under a \textit{generalized Born--Markov coarse-graining approximation}, which identifies the physical conditions for its validity. Just as the GKSL equation determines the expectation value of any observable at any time, the QCME determines any two-time correlator, such as linear response functions and out-of-time-order correlators (OTOCs), and it extends naturally to more times. We illustrate this through the correlation of Schr\"odinger- and Heisenberg-picture evolutions, OTOCs as a diagnostic of quantum chaos, and open systems with initial system--environment correlations. We show that Schr\"odinger- and Heisenberg-picture evolutions can exhibit correlations that cannot be captured by independent GKSL equations; describe the evolution of quantum channels in a setting motivated by quantum process tomography; obtain time-local dynamical equations beyond the conventional GKSL framework for linear response functions and out-of-time-order correlators under environmental influences; and demonstrate, in a qubit model with initial system--environment correlations, that quantum comb evolution reproduces shifted-intervention statistics up to corrections of order $g^2\delta t^2$.

\end{abstract}

\maketitle

\setcounter{tocdepth}{1}
\tableofcontents

\section{Introduction}\label{sec:introduction}

The theory of open quantum systems is among the most powerful and widely applied frameworks in modern quantum physics~\cite{BreuerPetruccione,RivasHuelga,Vacchini}. By accounting for the inevitable coupling of a system to its environment, it extends closed quantum mechanics to the most general state transformations consistent with quantum theory. Absent initial system--environment correlations, these are described in discrete time by \textbf{completely positive, trace-preserving (CPTP)} maps $\mathcal{E}_{t,0}$, or \textit{quantum channels}, taking arbitrary initial states to states at a later time $t$. In continuous time, the natural counterpart is evolution that is \textit{CP-divisible}, $\mathcal{E}_{t,0} = \mathcal{E}_{t,s}\circ\mathcal{E}_{s,0}$ for all $t\geq s\geq 0$ with every constituent map CPTP. Such evolution is generated by the \textbf{Gorini--Kossakowski--Sudarshan--Lindblad (GKSL)} master equation~\cite{Gorini1976,Lindblad1976}, which has proven remarkably successful across quantum optics~\cite{Carmichael,WisemanMilburn}, condensed matter physics~\cite{CaldeiraLeggett,Weiss}, and quantum information processing~\cite{NielsonChuang}, to name but a few areas.

This paradigm rests on several assumptions, most prominently \textit{Markovianity} (memorylessness), that break down in a growing class of scenarios~\cite{Rivas2014,Breuer2016,deVega2017,quantumstochasticprocessesandquantumnon,Taranto2025}. In discrete time, initial system--environment correlations preclude a meaningful linear CPTP map on the system alone~\cite{Pechukas_1994,Alicki_1995,Pechukas_1995,RodriguezRosario_2010,modi_role_2010,modi_preparation_2011}. In continuous time, the microscopic derivation of the GKSL equation requires environmental correlations to decay quickly relative to the system's relaxation time, so that the environment effectively retains no memory of past interactions. Structured environments, strong system--bath coupling, and ultrafast probing push modern technologies beyond this regime, making memory effects unavoidable~\cite{Nelson2000,Sayrin2011,Chin2013,White2025}.

In such \textit{non-Markovian} processes, tracking the instantaneous state no longer suffices, because the relevant information is encoded in \textit{multi-time correlations}. Accessing these requires intervening at intermediate times, and any intervention on a system correlated with its environment disturbs the environment and feeds back into the subsequent evolution. This blurs the boundary between \textit{observation} and \textit{dynamics} on which the open-systems premise rests, in a way no single-time state description can capture~\cite{Pollock_2018_PRA,Taranto2020}. Nor do families of channels across pairs of times suffice, since pairwise descriptions lack complete information about correlations across multiple times~\cite{Pollock2018L,Milz2019_CP,Taranto2019_PRL,Taranto2019_PRA,Taranto2021,milz-spee-ent_2021,Taranto_2022,Taranto2026}. The framework of \textit{higher-order quantum operations} resolves this~\cite{chiribella2008quantum,Pollock_2018_PRA,Taranto2025}: By separating what is controllable (operations on the system) from what is not (the environment and its coupling to the system), it restores a clear delineation between observer and dynamics. It linearly maps any sequence of interventions to the correct multi-time statistics, capturing everything an experimenter could in principle observe~\cite{Pollock_2018_PRA,Milz2017,Milz2020,Milz2021}. As van Kampen remarked, \textit{``it is not proper to treat non-Markov processes merely as modifications or corrections of the Markov processes''}~\cite{vanKampen}.

A complete multi-time description is, however, computationally demanding, and often more than is needed~\cite{Milz_2018,White_2022}. In many physically relevant situations the quantities of interest are \textit{two-time} correlators such as $\langle A_t B_s\rangle$, which arise in response functions~\cite{Kubo_1957}, fluctuation relations~\cite{Esposito_2009,Campisi_2011}, error--disturbance tradeoffs~\cite{Ozawa_2003,Busch_2014,emori2023error}, and diagnostics of quantum chaos~\cite{Maldacena_2016,Swingle2018,Zonnios_2022,emori2025measuring}. Under the \textbf{quantum regression theorem (QRT)}, these can be obtained from the same reduced dynamical maps that govern the system's state~\cite{lax1963formal}. The QRT fails, however, when system--environment correlations retain memory of an earlier intervention, and CP-divisibility of the reduced dynamics alone does not guarantee that it holds~\cite{guarnieri2014quantum}. This raises a natural question: \textit{Can we formulate a GKSL-like equation for the evolution of a two-time process that retains these temporal correlations?}

Here we answer this in the affirmative. We promote the dynamical object from a quantum state to a \textit{quantum comb} (also called a process tensor) encoding a two-time process, such as a quantum channel, and derive the most general continuous-time, CP-divisible master equation for it: the \textbf{quantum comb master equation (QCME)}, which governs the quantum ``dynamics of dynamics''. Its generator has a GKSL-like structure supplemented by \textit{encoder--decoder correlation terms} that couple different temporal legs of the comb (Sec.~\ref{sec:quantumdynamicsofdynamics}). We also derive the QCME microscopically, under a generalized Born--Markov coarse-graining approximation, which specifies the physical conditions under which it holds (Sec.~\ref{sec:quantumdynamicsofdynamics-microscopic}). Just as the GKSL equation determines the expectation value of any observable at any time, the QCME determines any two-time correlator once the relevant preparations, interventions, and measurements are inserted. Although we focus on two-time processes, the construction extends straightforwardly to combs with more legs. We demonstrate its scope through three applications: correlations between Heisenberg-picture evolution of an observable and Schr\"odinger-picture evolution of a state (Example~\ref{(QDoD)eg:SchHeiCorr}); OTOCs, which probe information scrambling and diagnose quantum chaos (Sec.~\ref{sec:otoc}); and open-system dynamics with initial system--environment correlations (Sec.~\ref{sec:OQS}). Specifically, we find that correlations between Schr\"odinger- and Heisenberg-picture evolutions cannot generally be captured by independent GKSL equations; that quantum channels themselves can evolve in time, as illustrated by a setting motivated by quantum process tomography; that linear response functions and OTOCs admit time-local dynamical equations beyond the conventional GKSL framework under environmental influences; and that, in an initially correlated qubit model, the proposed comb evolution reproduces shifted-intervention statistics up to corrections of order $g^2\delta t^2$. Together, these show that the master equation derived here is not merely a formal curiosity but a broadly applicable and computationally tractable tool for the physics of open, noisy, and complex quantum systems.


\section{Background}
\label{sec:framework}

\subsection{Open Quantum Systems}
\label{subsec:openquantumsystems}

Consider an accessible system $S$ and an inaccessible environment $E$, with Hilbert spaces $\mathcal{H}_S$ and $\mathcal{H}_E$, respectively. The joint system--environment state evolves unitarily, so the reduced system state at time $t$ is
\begin{align}
    \rho^S(t)
    =
    \operatorname{Tr}_E\left(
    \mathcal{U}^{SE}_{t,0}[\rho^{SE}(0)]
    \right),
    \label{(QDoD)Eq:JointEvolution}
\end{align}
where $\mathcal{U}^{SE}_{t,0}[\rho]:=U_{t,0}\rho U_{t,0}^\dagger$ and $U_{t,0}$ acts on $\mathcal{H}_S\otimes\mathcal{H}_E$. At this stage, the initial joint state $\rho^{SE}(0)$ may contain system--environment correlations, with the reduced initial system state given by $\rho^S(0):=\operatorname{Tr}_E[\rho^{SE}(0)]$. In general, such correlations prevent the reduced evolution from being described by a linear map acting on $\rho^S(0)$ alone.

\begin{figure}
    \centering
    \includegraphics[width=\linewidth]{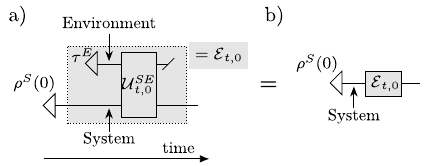}
    \caption{(a) Evolution of the initial system state $\rho^S(0)$ through the joint unitary map $\mathcal{U}^{SE}_{t,0}$ acting on $\mathcal{L}(\mathcal{H}_S\otimes\mathcal{H}_E)$. The system and environment wires correspond to $\mathcal{L}(\mathcal{H}_S)$ and $\mathcal{L}(\mathcal{H}_E)$, respectively. Triangles represent quantum states, the four-legged box represents the joint unitary map, and the slashed environment wire denotes the partial trace over the environment. (b) The corresponding reduced dynamical map acting on $\mathcal{L}(\mathcal{H}_S)$, obtained by fixing the initial environment state and tracing over the environment output. Panel (a) provides a unitary (Stinespring) dilation of this map.}
    \label{(QDoD) fig-intro}
\end{figure}

To avoid this difficulty, one commonly assumes that the system and environment are initially uncorrelated, with $\rho^{SE}(0)=\rho^S(0)\otimes\tau^E$, where $\tau^E$ is a fixed environmental state. Under this assumption, the joint evolution induces a linear map on the system:
\begin{align}
    \mathcal{E}_{t,0}[\rho^S(0)]
    :=
    \operatorname{Tr}_E\left(
    \mathcal{U}^{SE}_{t,0}
    [\rho^S(0)\otimes\tau^E]
    \right).
    \label{(QDoD)Eq:OQS1}
\end{align}
For fixed $\tau^E$ and $U_{t,0}$, this map is completely positive and trace-preserving (CPTP), and its definition is independent of the choice of $\rho^S(0)$. This provides the \textit{microscopic approach} to a dynamical map: we begin with a system--environment model and obtain a reduced description. The construction is illustrated in Fig.~\ref{(QDoD) fig-intro}(a), while panel (b) shows the resulting system map.

Alternatively, we can ask directly what conditions a deterministic state-to-state transformation must satisfy, without introducing a specific environment or unitary: it must be linear, completely positive, and trace-preserving, so that it maps physical states to physical states and preserves probabilities. This is the \textit{axiomatic approach}. Stinespring's dilation theorem connects the two descriptions: every finite-dimensional CPTP map admits a unitary realization of the form in Eq.~\eqref{(QDoD)Eq:OQS1}, with a suitable environment initially in a fixed state~\cite{stinespring1955positive}. Thus, the microscopic construction and the axiomatic requirements fully characterize the same class of state transformations, although the microscopic realization is not unique.

Having established the correspondence between the microscopic and axiomatic descriptions of a single dynamical map, we next ask how these two approaches describe continuous-time evolution. Microscopically, we start from a system--environment Hamiltonian and derive a master equation that evolves the reduced state of the system under suitable physical assumptions. Axiomatically, we impose conditions on a continuous-time family of dynamical maps and determine the allowed form of its generator. We outline these two approaches below.


\subsubsection{Microscopic Approach}

We begin with the microscopic approach, starting from closed system--environment dynamics and imposing the Born, Markov, and secular approximations to obtain the time-independent GKSL equation. Consider the Hamiltonian of the system and environment,
\begin{align}
    H=H_S+H_E+gH_{I},\qquad H_{I}=\sum_\alpha A_\alpha\otimes B_\alpha,
\end{align}
where $g$ is a time-independent coupling constant. We choose $A_\alpha$ and $B_\alpha$ to be Hermitian and assume $[H_E,\tau^E]=0$ for the initial environmental state $\tau^E$. We also take $\operatorname{Tr}_E(B_\alpha\tau^E)=0$, which can always be arranged by absorbing any nonzero mean of the interaction into $H_S$.

In the interaction picture, the joint state satisfies $\dot{\rho}^{SE}_{\text{int}}(t) =-ig[H_{\text{int}}(t),\rho^{SE}_{\text{int}}(t)]$, where
\begin{align}
    H_{\text{int}}(t) =e^{i(H_S+H_E)t}H_{I}e^{-i(H_S+H_E)t}.
\end{align}
Integrating this equation once and substituting it back into the differential equation gives an exact equation for $\dot\rho^{SE}_{\text{int}}(t)$ involving the joint state at earlier times. Tracing over the environment and applying the \textit{Born approximation}, $\rho^{SE}_{\text{int}}(s)\approx\rho^S_{\text{int}}(s)\otimes\tau^E$, we obtain
\begin{equation}
    \begin{split}
        &\frac{d\rho^S_{\text{int}}(t)}{dt}\\
        &= -g^2\int_0^t ds\,\operatorname{Tr}_E\Big( \big[H_{\text{int}}(t),\big[H_{\text{int}}(s),\rho^S_{\text{int}}(s)\otimes\tau^E\big]\big]\Big).
    \end{split}
\end{equation}
Here the first-order term vanishes because the initial state is a product state and $\operatorname{Tr}_E(B_\alpha\tau^E)=0$. The Born approximation neglects system--environment correlations in the integral and treats the environment state as unchanged by the interaction.

Changing the integration variable to $s\to t-s$ and relabeling, this becomes
\begin{align}
    &\frac{d\rho^S_{\text{int}}(t)}{dt}= \notag \\ 
    &-g^2\int_0^t ds\,\operatorname{Tr}_E\Big(\Big[H_{\text{int}}(t),\big[H_{\text{int}}(t-s),\rho^S_{\text{int}}(t-s)\otimes\tau^E\big]\Big]\Big).
\end{align}
Now suppose that the environment correlations decay on a timescale much shorter than that of the reduced system evolution. The \textit{Markov approximation} then replaces $\rho^S_{\text{int}}(t-s)$ by $\rho^S_{\text{int}}(t)$ inside the memory integral and extends its upper limit to infinity:
\begin{equation}
    \begin{split}
    &\frac{d\rho^S_{\text{int}}(t)}{dt} =\\
    &-g^2\int_0^\infty ds \operatorname{Tr}_E\Big(\Big[H_{\text{int}}(t),\big[H_{\text{int}}(t-s),\rho^S_{\text{int}}(t)\otimes\tau^E\big]\Big]\Big).\label{(QDoD)Eq:MicroGKSL1}
    \end{split}
\end{equation}
This is known as the \textit{time-local Born--Markov equation}.

Next, we resolve the system operators into their Bohr-frequency components:
\begin{align}
    A_\alpha(\omega)&:=\sum_{\epsilon'-\epsilon=\omega}\Pi_\epsilon A_\alpha\Pi_{\epsilon'},\label{(QDoD)Eq:MicroGKSL2-1}\\
    [H_S,A_\alpha(\omega)]&=-\omega A_\alpha(\omega),\label{(QDoD)Eq:MicroGKSL2-2}
\end{align}
where $\Pi_\epsilon$ projects onto the eigenspace of $H_S$ with eigenvalue $\epsilon$. Equation~\eqref{(QDoD)Eq:MicroGKSL2-2} implies
\begin{align}
    e^{iH_St}A_\alpha(\omega)e^{-iH_St}
    =
    e^{-i\omega t}A_\alpha(\omega).
\end{align}
The interaction Hamiltonian can therefore be written as
\begin{align}
    H_{\text{int}}(t)
    &=
    \sum_{\alpha,\omega}
    e^{-i\omega t}A_\alpha(\omega)\otimes B_\alpha(t),
    \\
    H_{\text{int}}(t-s)
    &=
    \sum_{\beta,\omega'}
    e^{i\omega'(t-s)}
    A_\beta^\dagger(\omega')\otimes B_\beta^\dagger(t-s),
\end{align}
where $B_\alpha(t)=e^{iH_Et}B_\alpha e^{-iH_Et}$. In the second line, we have used the Hermiticity of the interaction Hamiltonian to write its adjoint expansion. Substituting these expressions into Eq.~\eqref{(QDoD)Eq:MicroGKSL1}, each pair of frequency components contributes the phase
\begin{align}
    e^{-i\omega t}e^{i\omega'(t-s)}
    =
    e^{-i(\omega-\omega')t}e^{-i\omega's}.
\end{align}
The factor $e^{-i\omega's}$ enters the integral over the bath correlation functions, while $e^{-i(\omega-\omega')t}$ describes oscillations at the frequency difference $\omega-\omega'$. Under the \textit{secular approximation}, contributions with $\omega\ne\omega'$ average out on the timescale of the reduced dynamics. This secular averaging also follows from the weak-coupling limit, where $g\rightarrow0$ and $t\rightarrow\infty$ with $g^2t$ fixed. Under suitable assumptions on the environment correlations, Davies showed that the reduced dynamics converges to a Markovian semigroup in this limit~\cite{davies1974markovian}.

The remaining coefficients are determined by the bath correlation functions
\begin{align}
    C_{\alpha\beta}(s)
    &:=
    \operatorname{Tr}_E
    \big(B_\alpha(s)B_\beta\tau^E\big),
    \\
    \Gamma_{\alpha\beta}(\omega)
    &:=
    \int_0^\infty ds\,e^{i\omega s}C_{\alpha\beta}(s),
\end{align}
where $B_\alpha(s)=e^{iH_Es}B_\alpha e^{-iH_Es}$. Their Hermitian and anti-Hermitian parts define
\begin{align}
    \gamma_{\alpha\beta}(\omega)
    &:=
    \Gamma_{\alpha\beta}(\omega)
    +\Gamma_{\beta\alpha}^*(\omega),
    \\
    S_{\alpha\beta}(\omega)
    &:=
    \frac{
    \Gamma_{\alpha\beta}(\omega)
    -\Gamma_{\beta\alpha}^*(\omega)}{2i}.
\end{align}
The matrix $\gamma(\omega)$ is positive semidefinite, while $S(\omega)$ determines the Lamb-shift Hamiltonian
\begin{align}
    H_{\mathrm{LS}}
    =
    \sum_{\omega,\alpha,\beta}
    S_{\alpha\beta}(\omega)
    A_\alpha^\dagger(\omega)A_\beta(\omega).
\end{align}
Returning to the Schr\"odinger picture, we obtain
\begin{equation}
    \begin{split}
        \frac{d\rho^S(t)}{dt} =&-i[H_S+g^2H_{\mathrm{LS}},\rho^S(t)] \\
        &+g^2\sum_{\omega,\alpha,\beta}
    \gamma_{\alpha\beta}(\omega)
    \Big(
    A_\beta(\omega)\rho^S(t)A_\alpha^\dagger(\omega)\\
    &-\frac12
    \{A_\alpha^\dagger(\omega)A_\beta(\omega),\rho^S(t)\}
    \Big)=: \mathcal{L}^{\text{micro}}[\rho^S(t)],
    \label{Eq: Microscopic system GKSL}
    \end{split}
\end{equation}
where the superoperator $\mathcal{L}^{\text{micro}}$ is called the \textit{Liouvillian generator}.

Finally, we diagonalize the positive semidefinite coefficient matrix $\gamma(\omega)$ at each frequency as
\begin{align}
    \gamma_{\alpha\beta}(\omega)=\sum_k\kappa_k(\omega)u_{\alpha k}(\omega)u_{\beta k}^*(\omega),\quad \kappa_k(\omega)\geq0,
\end{align}
where $u_{\alpha k}(\omega)$ are the components of the $k$-th orthonormal eigenvector of $\gamma_{\alpha\beta}(\omega)$, with eigenvalue $\kappa_k(\omega)$. Defining the Lindblad operators
\begin{align}
    L_{\omega k}:=g\sqrt{\kappa_k(\omega)}\sum_\beta u_{\beta k}^*(\omega)A_\beta(\omega),\label{(QDoD)Eq:MicroGKSL5}
\end{align}
the generator becomes
\begin{equation}
    \begin{split}
        &\mathcal{L}^{\text{micro}}[\rho^S(t)]=-i[H_S+g^2H_{\mathrm{LS}},\rho^S(t)]\\
        &\quad+\sum_{\omega,k}\Big(
        L_{\omega k}\rho^S(t) L_{\omega k}^\dagger
        -\frac12\{L_{\omega k}^\dagger L_{\omega k},\rho^S(t)\}
        \Big).
    \end{split}
    \label{(QDoD)Eq:MicroGKSL4}
\end{equation}
The resulting master equation is the Gorini--Kossakowski--Sudarshan--Lindblad (GKSL) equation~\cite{Gorini1976,Lindblad1976}.

Using the Liouvillian generator, we define the state-to-state map
\begin{align}
    \rho^S(t)=\mathcal{E}_{t,0}[\rho^S(0)],
    \qquad
    \mathcal{E}_{t,0}=e^{t\mathcal{L}^{\text{micro}}}.\label{(QDoD)Eq:MicroGKSL4-1}
\end{align}
Since $\mathcal{L}^{\text{micro}}$ has GKSL form with positive semidefinite coefficient matrices $\gamma(\omega)$, exponentiating it yields a CPTP map at every time: $\mathcal{E}_{t,0}$ and $\mathcal{E}_{s,0}$ are CPTP for all $t,s\geq0$, and so is
\begin{align}
    \mathcal{E}_{t,s}=e^{(t-s)\mathcal{L}^{\text{micro}}},
    \qquad
    \mathcal{E}_{t,0}
    =
    \mathcal{E}_{t,s}\circ\mathcal{E}_{s,0}
    \quad \forall\, t\geq s\geq0.
    \label{(QDoD)Eq:MicroGKSL3}
\end{align}
The decomposition~\eqref{(QDoD)Eq:MicroGKSL3}, with a CPTP map $\mathcal{E}_{t,s}$ for every $t\geq s\geq0$, is called CP-divisibility. More specifically, since the evolution is time-homogeneous, $\mathcal{E}_{t,s}$ depends only on $t-s$. Writing $\mathcal{E}_t:=\mathcal{E}_{t,0}$, the maps satisfy $\mathcal{E}_{t+s}=\mathcal{E}_t\circ\mathcal{E}_s$ and $\mathcal{E}_0=\mathcal{I}$, and therefore form a continuous quantum dynamical semigroup. These properties provide the starting point for the axiomatic approach discussed next.

In this section, we have shown that the microscopic approach yields the GKSL equation through the Born, Markov, and secular approximations. This approach also extends to time-dependent Hamiltonians, provided the approximations remain valid, yielding a time-dependent Liouvillian generator and a state-to-state map expressed by a time-ordered exponential.


\subsubsection{Axiomatic Approach}\label{(QDoD)Sec:ReviewAxiom}

We now present the axiomatic approach to deriving the GKSL equation. Instead of starting from the Hamiltonian, the axiomatic approach imposes CP-divisibility directly on a continuous-time family of CPTP maps $\{\mathcal{E}_{t,0}\}_{t\geq0}$ of the form in Eq.~\eqref{(QDoD)Eq:OQS1}, without specifying the environment state or joint unitary, and without assuming time-homogeneity: we require only that
\begin{align}
    \mathcal{E}_{t+\epsilon,0}
    =
    \mathcal{E}_{t+\epsilon,t}\circ\mathcal{E}_{t,0}
\end{align}
for a CPTP map $\mathcal{E}_{t+\epsilon,t}$ at every $t,\epsilon\geq0$. The generator of this family is defined as
\begin{equation}
    \begin{split}
        \frac{d\rho^S(t)}{dt}
    &=
    \lim_{\epsilon\rightarrow0^+}
    \frac{\rho^S(t+\epsilon)-\rho^S(t)}{\epsilon}
    \\
    &=\lim_{\epsilon\rightarrow0^+}
    \frac{\mathcal{E}_{t+\epsilon,t}-\mathcal{I}}
    {\epsilon}[\rho^S(t)]=:\mathcal{L}^{\text{axiom}}(t)[\rho^S(t)],\label{(QDoD)Eq:OQSAx1}
    \end{split}
\end{equation}
where $\mathcal{I}$ is the identity map and $\mathcal{L}^{\text{axiom}}(t)$ is a time-dependent generator.

To determine the explicit form of the generator, we use the Kraus representation~\cite{kraus1971general}. The CPTP map $\mathcal{E}_{t+\epsilon,t}$ admits the representation
\begin{align}
    \mathcal{E}_{t+\epsilon,t}[\rho]
    &=
    \sum_\mu K_\mu(t,\epsilon)\rho K_\mu^\dagger(t,\epsilon),
    \\
    \sum_\mu K_\mu^\dagger(t,\epsilon)K_\mu(t,\epsilon)
    &=\mathds{1},
\end{align}
where the second equation expresses trace preservation. Since the Kraus representation is not unique, we choose the following convenient form for the infinitesimal evolution~\cite{preskill2015lecture}:
\begin{align}
    K_0(t,\epsilon)
    &=
    \mathds{1}
    +\epsilon\left(
    -iH(t)-\frac12\sum_{\mu\geq1}L_\mu^\dagger(t)L_\mu(t)
    \right)
    +\cdots,\\
    K_\mu(t,\epsilon)
    &=
    \sqrt{\epsilon}\,L_\mu(t)+\cdots,
    \qquad \mu\geq1,
\end{align}
with $H(t)=H^\dagger(t)$. The ellipses denote terms beyond order $\epsilon$ in the first equation and beyond order $\sqrt{\epsilon}$ in the second; these corrections need not involve integer powers of $\epsilon$.

Substituting these expressions into the Kraus representation and taking the infinitesimal limit gives
\begin{equation}
    \begin{split}
        &\frac{d\rho^S(t)}{dt}
    =\mathcal{L}^{\text{axiom}}(t)[\rho^S(t)]=-i[H(t),\rho^S(t)]\\
    &\quad+
    \sum_{\mu\geq1}
    \Big(
    L_\mu(t)\rho^S(t)L_\mu^\dagger(t)
    -\frac12
    \{L_\mu^\dagger(t)L_\mu(t),\rho^S(t)\}
    \Big).
    \end{split}\label{(QDoD)Eq:AxiomGKSL}
\end{equation}
This is the time-dependent GKSL equation. Its connection with the microscopic derivation is explicit upon comparison with Eq.~\eqref{(QDoD)Eq:MicroGKSL4}. Choosing
\begin{align}
    H(t)&=H_S+g^2H_{\mathrm{LS}},\\
    L_\mu(t)&=L_{\omega k},
    \qquad \mu=(\omega,k),
\end{align}
we recover $\mathcal{L}^{\text{axiom}}=\mathcal{L}^{\text{micro}}$.

Thus, both the axiomatic and microscopic approaches lead to the same GKSL structure under their respective assumptions. The advantage of the axiomatic approach is that it determines this structure solely from the properties of the dynamics, without specifying an environment state or joint unitary evolution. The microscopic approach, by contrast, provides an explicit expression for the Liouvillian generator in terms of the system Hamiltonian, decay rates, and Lindblad operators. These quantities also admit a physical interpretation: the operators $A_\alpha(\omega)$ used to construct the Lindblad operators~\eqref{(QDoD)Eq:MicroGKSL5} induce transitions that raise or lower the system energy, while the corresponding decay rates determine how rapidly these transitions occur. For interactions involving excitation exchange with the environment, these transitions describe the emission or absorption of excitations by the system. The two approaches therefore provide complementary descriptions of the same form of evolution.


\subsubsection{Relation Between the Two Approaches}\label{(QDoD)Sec:TwoQuestions}

The microscopic and axiomatic approaches address two complementary questions about quantum-state evolution. The axiomatic approach asks what form a dynamical equation can take when the evolution is CP-divisible and the limit~\eqref{(QDoD)Eq:OQSAx1} exists, leading to a time-local generator of GKSL form. The dynamical semigroup is the time-homogeneous special case, with a time-independent generator. The microscopic approach instead asks under what physical conditions such dynamics arises from a system--environment interaction. Under the Born, Markov, and secular approximations, it yields the same GKSL structure, with the Hamiltonian correction and dissipative coefficients determined by the interaction and the environment correlations.

We summarize these two questions and their answers as follows:
\begin{enumerate}
    \item \textbf{Axiomatic question:} What dynamical equation governs CP-divisible quantum evolution when the limit~\eqref{(QDoD)Eq:OQSAx1} exists? \textbf{Answer:} The time-dependent GKSL equation, whose time-independent special case generates a continuous dynamical semigroup.

    \item \textbf{Microscopic question:} Under what physical conditions does such evolution arise from system--environment dynamics? \textbf{Answer:} The Born--Markov--secular construction provides one realization, with the generator's coefficients fixed by the microscopic model.
\end{enumerate}
The significance of this connection is that the same mathematical structure can be understood both through general requirements on quantum evolution and through the physical interactions that produce it. This complementary description underpins the widespread use of the GKSL equation: its general structure can be studied independently of any particular model, while its coefficients can be connected to concrete physical processes.

In the next subsection, we present a simple example illustrating how this connection can appear to break down from an experimental perspective, when interventions are performed on the quantum system at closely spaced times, even though the underlying, unmonitored dynamics remains CP-divisible. In particular, CP-divisibility of the underlying dynamics does not, by itself, suffice to describe the resulting multi-time statistics. This motivates extending the two questions posed above to the more general setting of quantum combs.


\subsection{Limitations of the Dynamical-Map Description and Resolution Through Quantum Combs}\label{subsec:quantumsuperchannels}

This subsection motivates the two fundamental questions of this paper through a concrete example (Sec.~\ref{(QDoD)SubSubSec:Questions}), and then introduces the quantum comb framework needed to address them (Sec.~\ref{(QDoD)SubSubSec:QuantumCombFramework}).

\subsubsection{Fundamental Questions}\label{(QDoD)SubSubSec:Questions}

The GKSL equation provides an approximate description of an open quantum system under suitable Born, Markov, and secular approximations~\cite{BreuerPetruccione}. The resulting dynamics is CP-divisible. However, this does not fully capture all possible operational aspects of an open quantum system. For example, when an experimentalist intervenes in or measures the system, it is not known \textit{a priori} whether the subsequent dynamics remains describable by a time-dependent or time-independent GKSL equation~\cite{Milz2019_CP}. In particular, CP-divisibility of the underlying dynamics does not establish whether correlations retained in the environment can influence the system's response to a later intervention. This is because the dynamical-map description above tracks only the evolution of the density operator in time and cannot encode interventions that occur during that evolution. If we truly wish for a general open-systems theory that captures all possible observable phenomena, we need to extend the framework to allow for interventions. This motivates the \textit{quantum comb}.

\begin{example}[Open quantum system evolution with interventions]\label{(QDoD) Eg: evolution of state} Assume that the system and environment are initially uncorrelated and that the system dynamics are described by the CP-divisible family of CPTP maps $\{\mathcal{E}_{t,0}\}_{t\geq0}$ introduced in Eq.~\eqref{(QDoD)Eq:MicroGKSL3}. We will compare interventions performed over a fixed duration $\Delta t$, chosen such that $t_E \ll \Delta t \ll t_S$, where $t_E$ is the environment correlation time and $t_S$ is the system relaxation time, as in the Born approximation.

Let $A^S$ and $B^S$ be observables acting on the system, and introduce the following maps:
\begin{align}
    \mathcal{A}^S[\bullet] &= \operatorname{Tr}_S(A^S\bullet), \quad
    \mathcal{B}^S_\epsilon[\bullet] = e^{-i\epsilon B^S}\bullet e^{i\epsilon B^S},
\end{align}
where $\epsilon$ is an infinitesimal real parameter. The map $\mathcal{A}^S$ is a linear functional that yields the expectation value of the observable $A^S$, while $\mathcal{B}^S_\epsilon$ describes a weak unitary perturbation generated by $B^S$. Let us now consider two experiments.

\textit{Experiment by Alice}: An experimentalist Alice performs a weak unitary intervention $\mathcal{B}^S_\epsilon$ at time $0$, before the system interacts with the environment. The post-intervention state then evolves in time according to the CPTP map $\mathcal{E}_{t,0}$. At time $t$, Alice performs a second measurement of the observable $A^S$.

The complete experiment yields the linear response function of the observable $A^S$ to a perturbation generated by $B^S$:
\begin{equation}
    \begin{split}
        F_{\text{Alice}}(t):=&i\partial_\epsilon\left.\operatorname{Tr}_S\left(A^S\mathcal{E}_{t,0}\circ\mathcal{B}^S_\epsilon[\rho^S(0)]\right)\right|_{\epsilon=0}\\
    =&\braket{\left[A^{\mathrm{red}}(t),B^S\right]},
    \end{split}
\end{equation}
where the expectation value is with respect to $\rho^S(0)$. This is illustrated in Fig.~\ref{(QDoD) fig1-1}(a). Here, $A^{\text{red}}(t)$ denotes the reduced Heisenberg-picture observable, whose time evolution is governed by the completely positive unital (CPU) map~\cite{Lindblad1976,settimo2026divisibility}
\begin{align}
    A^{\mathrm{red}}(t):=\mathcal{E}^*_{t,0}[A^S],
\end{align}
where $\mathcal{E}_{t,0}$ is the CPTP map describing the reduced system dynamics, and its adjoint $\mathcal{E}^*_{t,0}$ is defined through the trace duality relation
\begin{align}
    \operatorname{Tr}_S\left(X^S\mathcal{E}_{t,0}[\rho^S]\right)
    =\operatorname{Tr}_S\left(\mathcal{E}^*_{t,0}[X^S]\rho^S\right).
\end{align}

Taking the time derivative of $F_{\text{Alice}}(t)$ and assuming that the reduced dynamics is governed by the GKSL equation, we obtain
\begin{equation}
    \begin{split}
        \frac{\partial F_{\text{Alice}}(t)}{\partial t}
        &= \operatorname{Tr}_S\Big(A^S\mathcal{L}[\sigma^S(t)]\Big)\\
        &= \operatorname{Tr}_S\Big(\mathcal{L}^*[A^S]\,\sigma^S(t)\Big),
        \label{(QDoD)Eq:AliceExp1}
    \end{split}
\end{equation}
where $\sigma^S(t):=\mathcal{E}_{t,0}\big([B^S,\rho^S(0)]\big)$ denotes the evolution, under the same linear map $\mathcal{E}_{t,0}$, of the (traceless) commutator $[B^S,\rho^S(0)]$, and $\mathcal{L}$ is the Liouvillian generator in the Schr\"odinger picture~\cite{Gorini1976}, with adjoint $\mathcal{L}^*$ generating the evolution in the Heisenberg picture~\cite{Lindblad1976}. Since $\mathcal{L}$ is linear, it governs the evolution of $\sigma^S(t)$ just as it would govern the evolution of a genuine state. Thus, to determine the dynamics of $F_{\text{Alice}}(t)$, Alice needs only to solve the GKSL equation in either the Schr\"odinger or the Heisenberg picture.

\begin{figure}
    \centering
    \includegraphics[width=\linewidth]{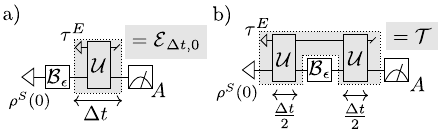}
    \caption{(a) Pictorial representation of experiment 1. The shaded box represents the dilation, as in Fig.~\ref{(QDoD) fig-intro}, of the CPTP map $\mathcal{E}_{\Delta t,0}$, with unitary superoperator $\mathcal{U}^{SE}_{\Delta t,0}$. (b) Pictorial representation of experiment 2, where the shaded region corresponds to the four-legged quantum comb $\mathcal{T}$. The two unitaries are $\mathcal{U}^{SE}_{\Delta t/2,0}$ and $\mathcal{U}^{SE}_{\Delta t,\Delta t/2}$.}
    \label{(QDoD) fig1-1}
\end{figure}

In contrast, consider an experiment by Bob in which the second intervention occurs partway through the comparison window, rather than at its end.

\textit{Experiment by Bob}: An experimentalist Bob repeats a similar experiment to Alice's, but performs the weak unitary intervention $\mathcal{B}^S_\epsilon$ at time $\Delta t/2$ instead of $t=0$, after the system has had a chance to evolve with the environment and possibly develop correlations with it. Because the system and environment are correlated at $\Delta t/2$, the evolution beyond this point cannot be given by a CPTP map acting on the system alone.

Immediately after this intervention, Bob performs the second measurement at $\Delta t$. Since the system and environment remain correlated at $\Delta t/2$, as noted above, this second step likewise cannot be written as a CPTP map, and the Born approximation---which neglects system--environment correlations---cannot be invoked at this intermediate time. The linear response function of Bob's experiment is given by
\begin{equation}
    \begin{split}
        F_{\text{Bob}}(\Delta t):=&i \partial_\epsilon \operatorname{Tr}_{SE} \Big[(A^S\otimes \mathds{1}^E) \,\mathcal{U}^{SE}_{\Delta t, \tfrac{\Delta t}{2}}\circ (\mathcal{B}^S_\epsilon \otimes \mathcal{I}_E)\\
    &\circ  \mathcal{U}^{SE}_{ \tfrac{\Delta t}{2}, 0}[\rho^S(0)\otimes \tau^E]\Big]\Big|_{\epsilon=0}.\label{(QDoD) Eq:eg1 step1}
    \end{split}
\end{equation}
This is illustrated schematically in Fig.~\ref{(QDoD) fig1-1}(b).

Neither experimentalist knows \textit{a priori} whether the system exhibits memoryless dynamics. For Alice, the GKSL description suffices~\eqref{(QDoD)Eq:AliceExp1}, whereas for Bob, it does not. This naturally raises the question of whether concepts such as CP-divisibility, the Born, Markov, and secular approximations, and the GKSL master equation can be extended to incorporate interventions into the dynamics.

For this purpose, consider the expression~\eqref{(QDoD) Eq:eg1 step1}. Ignoring the $\epsilon$-derivative for a moment, note that it is a linear function of the weak unitary map $\mathcal{B}^S_\epsilon$ and the system's initial state $\rho^S(0)$. We may therefore introduce a map $\mathcal{T}$, defined by
\begin{align}
    \mathcal{T}[\mathcal{B}^S_\epsilon&,\,\rho^S(0)] := \notag \\
    &\operatorname{Tr}_E\Big(\mathcal{U}^{SE}_{\Delta t, \frac{\Delta t}{2}}\circ (\mathcal{B}^S_\epsilon \otimes \mathcal{I}_E)\circ  \mathcal{U}^{SE}_{ \frac{\Delta t}{2}, 0}[\rho^S(0)\otimes\tau^E]\Big) .\label{(QDoD) Eq:eg1 step2}
\end{align}
Bob's linear response function is then
\begin{align}
    F_{\text{Bob}}(\Delta t)=i \partial_\epsilon \operatorname{Tr}_S \Big(A^S \mathcal{T} [\mathcal{B}^S_\epsilon,  \rho^S(0)] \Big)\Big|_{\epsilon=0}.
\end{align}
The linear map $\mathcal{T}$ is called a \textit{quantum comb}~\cite{chiribella2008quantum}, also known as a \textit{process tensor}~\cite{Pollock_2018_PRA}. It takes any map $\mathcal{B}^S_\epsilon$ and any state $\rho^S(0)$ and returns the system state at time $\Delta t$, encoding the fact that the system was intervened on by $\mathcal{B}^S_\epsilon$ at time $\Delta t/2$. Where no ambiguity arises we write $\mathcal{T}$ alone to refer to this map as an object in its own right, reserving the bracket notation $\mathcal{T}[\mathcal{B}^S_\epsilon,\rho^S(0)]$ for its evaluation on specific inputs.

Let us assume that the quantum comb $\mathcal{T}$ satisfies a dynamical equation of the form
\begin{align}
    \frac{\partial \mathcal{T}}{\partial t}
    = \mathcal{L}_{\text{comb}}[\mathcal{T}],\label{(QDoD)Eq:BobExp2}
\end{align}
where $\mathcal{L}_{\text{comb}}$ acts on the comb $\mathcal{T}$ itself, and so, by construction, cannot depend on any particular choice of inputs $\mathcal{B}^S_\epsilon$ and $\rho^S(0)$. For the sake of argument, we set aside two subtleties for the moment. First, the quantum comb $\mathcal{T}$ generally depends on multiple time variables, so a precise definition of the time derivative is required; we address this in Sec.~\ref{(QDoD)Sec:QDoD1}. Second, one must specify whether $\mathcal{L}_{\text{comb}}$ acts on $\mathcal{T}$ directly, as written above, or only after one or both of its legs have been evaluated on $\mathcal{B}^S_\epsilon$ and/or $\rho^S(0)$---that is, whether $\mathcal{L}_{\text{comb}}$ and evaluation on inputs commute. This distinction depends on how the initial condition is imposed on the comb, as discussed in Sec.~\ref{(QDoD)Sec:QDoD3}.

Using the dynamical equation for the quantum comb proposed above, and noting that $\mathcal{B}^S_\epsilon$ and $\rho^S(0)$ are fixed, $t$-independent inputs, we obtain
\begin{align}
    \frac{\partial F_{\text{Bob}}(t)}{\partial t} = i \partial_\epsilon \operatorname{Tr}_S \Big(A^S\, \mathcal{L}_{\text{comb}}[\mathcal{T}]\,[\mathcal{B}^S_\epsilon,\rho^S(0)]\Big)\Big|_{\epsilon=0}.\label{(QDoD)Eq:BobExp1}
\end{align}

This equation is analogous to Alice's dynamical equation~\eqref{(QDoD)Eq:AliceExp1}. Thus, to determine the dynamics of $F_{\text{Bob}}(t)$, Bob needs to solve Eq.~\eqref{(QDoD)Eq:BobExp2}. In fact, we will show that the dynamics observed by both Alice and Bob can be described by a dynamical equation of the form~\eqref{(QDoD)Eq:BobExp2}: the standard dynamical-map description governed by the GKSL equation emerges as a special case of our generalized master equation for quantum combs, as shown in Example~\ref{(QDoD)eg:SchHeiCorr}.
\end{example}

The main result of this work is a family of dynamical equations of the form~\eqref{(QDoD)Eq:BobExp2} for general quantum combs, extending their applicability beyond the experimental setup considered above. We call the resulting central result the \textit{quantum comb master equation} (QCME), given in Eq.~\eqref{(QDoD)Eq:CombME}. In Sec.~\ref{(QDoD)Sec:QDoD2}, we generalize CP-divisibility to quantum combs. With this generalization, we can revisit the two fundamental questions of open quantum systems introduced in Sec.~\ref{(QDoD)Sec:TwoQuestions} and extend them to quantum combs:
\begin{enumerate}
    \item \textbf{Axiomatic question:} What dynamical equation governs CP-divisible quantum-comb evolution? \textbf{Answer:} The QCME, a time-dependent, GKSL-like master equation for quantum combs, derived in Sec.~\ref{sec:quantumdynamicsofdynamics}.

    \item \textbf{Microscopic question:} Under what physical conditions can such CP-divisible quantum-comb evolution be realized? \textbf{Answer:} A generalized Born--Markov coarse-graining approximation, extending the Born--Markov--secular construction, derived in Sec.~\ref{sec:quantumdynamicsofdynamics-microscopic}.
\end{enumerate}
In this work, we address both questions for four-legged quantum combs, establishing the corresponding dynamical equations in Sec.~\ref{sec:quantumdynamicsofdynamics} and their microscopic foundations in Sec.~\ref{sec:quantumdynamicsofdynamics-microscopic}. The broader applicability of our framework is illustrated in Example~\ref{(QDoD)eg:SchHeiCorr}, Sec.~\ref{sec:otoc}, and Sec.~\ref{sec:OQS}.


\subsubsection{Quantum Comb Framework}\label{(QDoD)SubSubSec:QuantumCombFramework}
Let us now define the quantum comb. For this paper, we restrict our attention to the one-slot quantum comb; for more general constructions, see the review~\cite{Taranto2025}.

First, note that the quantum operator introduced above is a special case of a more general object constructed from two isometries, called the encoder and decoder. Recall that a unitary operator is a linear map $U:\mathcal{H}\rightarrow\mathcal{H}$ satisfying $U^\dagger U=UU^\dagger=\mathds{1}$. An isometry generalizes this notion to a linear embedding into a (generally larger) Hilbert space: a linear map $V:\mathcal{H}_A\rightarrow\mathcal{H}_B$ is called an isometry if and only if $V^\dagger V=\mathds{1}_A$, while in general $VV^\dagger\neq\mathds{1}_B$.

Let $V_\mathcal{E}:\mathcal{H}_{s^\inp}\rightarrow\mathcal{H}_{s^\out}\otimes\mathcal{H}_E$ and $V_\mathcal{D}:\mathcal{H}_{t^\inp}\otimes\mathcal{H}_E\rightarrow\mathcal{H}_{t^\out}$ be two such isometries, where $\mathcal{H}_E$ denotes the environmental Hilbert space, and the remaining four labels specify the Hilbert spaces associated with the open legs of the two isometries. Throughout this paper, we label each open leg by a time variable associated with its Hilbert space; we refer to this convention as \textit{Hilbert-space indexing}. We use the superscripts $\inp$ and $\out$ to distinguish legs that receive input quantum states from those that output quantum states, respectively. Pictorially, the input legs $t^\inp$ and $s^\inp$ always face left, whereas the output legs $t^\out$ and $s^\out$ always face right. This convention does not impose any temporal ordering among the four legs.

These isometries induce the encoder and decoder superoperators
\begin{align}
    \mathcal{E}_{s^\out,s^\inp}[\bullet]&:=V_\mathcal{E}\,\bullet\, V_\mathcal{E}^\dagger: \mathcal{L}(\mathcal{H}_{s^\inp}) \rightarrow \mathcal{L}(\mathcal{H}_{s^\out})\otimes \mathcal{L}(\mathcal{H}_E),\\
    \mathcal{D}_{t^\out,t^\inp}[\bullet]&:=V_\mathcal{D}\,\bullet\, V_\mathcal{D}^\dagger: \mathcal{L}(\mathcal{H}_{t^\inp})\otimes \mathcal{L}(\mathcal{H}_E) \rightarrow \mathcal{L}(\mathcal{H}_{t^\out}),
\end{align}
which act on the corresponding operator spaces. For generality, we do not impose a specific temporal ordering among the four legs at this stage. However, when describing a particular physical setup, we fix the temporal ordering of the legs according to the chronology of the underlying physical process; different temporal orderings then correspond to different physical setups, as illustrated by the examples and applications throughout this paper. In particular, in Example~\ref{(QDoD) Eg: evolution of channel} below, we consider the evolution of a quantum channel as a simple example of a temporally ordered quantum comb.

\begin{figure}
    \centering
    \includegraphics[width=\linewidth]{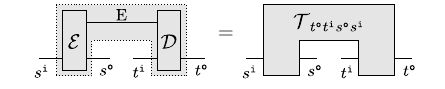}
    \caption{A general form of a four-legged quantum comb constructed from two isometries $\mathcal{E}$ and $\mathcal{D}$, contracted via the environmental space.}
    \label{(QDoD) Fig2.2-1}
\end{figure}

We construct a general four-legged quantum operator $\mathcal{T}$ from the above encoder and decoder as
\begin{align}
    \mathcal{T}_{t^\out t^\inp s^\out s^\inp} [\bullet]= \mathcal{D}_{t^\out , t^\inp}\circ(\bullet\otimes \mathcal{I}_{E})\circ  \mathcal{E}_{s^\out , s^\inp},\label{(QDoD)Eq:GeneralComb}
\end{align}
where the bullet $\bullet$ denotes the input to the open slot of the quantum comb, and $\mathcal{I}_E$ is the identity map on the environmental space $\mathcal{L}(\mathcal{H}_E)$. Its pictorial representation is given in Fig.~\ref{(QDoD) Fig2.2-1}.

For this operator to be a valid quantum comb, it must satisfy the generalized CPTP conditions. To formulate these clearly, we introduce the Choi representation. Recall that the Choi representation of a quantum channel $\mathcal{C}: \mathcal{L}(\mathcal{H}_\inp) \rightarrow \mathcal{L}(\mathcal{H}_\out)$ is defined as
\begin{align}
    \mathsf{C}
    &:=\operatorname{Choi}[\mathcal{C}]
    :=(\mathcal{C}\otimes\mathcal{I})[\Phi^+]\nonumber\\
    &=\sum_{i,j}\mathcal{C}[\ketbra{i}{j}]
    \otimes\ketbra{i}{j},
\end{align}
where $\Phi^+:=\sum_{i,j}\ketbra{i}{j}\otimes\ketbra{i}{j}$ is the unnormalized maximally entangled state. Throughout this paper, we use sans-serif letters to denote Choi representations of maps (which we denote by calligraphic letters).

Following Ref.~\cite{gour2019comparison}, the Choi representation of the four-legged quantum supermap $\mathcal{T}$ is given by
\begin{align}
    \mathsf{T}
    =\sum_{i,j,\alpha,\beta}
    \operatorname{Choi}(\mathcal{T}[\mathcal{M}_{(ij)(\alpha\beta)}])\otimes \operatorname{Choi}(\mathcal{M}_{(ij)(\alpha\beta)}),
\end{align}
where $\{\mathcal{M}_{(ij)(\alpha\beta)}\}$ forms a basis for the space of linear maps from $\mathcal{L}(\mathcal{H}_{s^\out})$ to $\mathcal{L}(\mathcal{H}_{t^\inp})$, defined by
\begin{align}
    \mathcal{M}_{(ij)(\alpha\beta)}[\bullet]
    :=\bra{i}\bullet\ket{j}\ketbra{\alpha}{\beta}.
\end{align}

\begin{defn}[Quantum comb~\cite{chiribella2008quantum,Taranto2025}]\label{(QDoD) Def: quantum comb}
    Let $\mathsf{T}_{t^\out t^\inp s^\out s^\inp}$ denote the Choi representation of a four-legged quantum map $\mathcal{T}_{t^\out t^\inp s^\out s^\inp}$, with temporal ordering $t^\out> t^\inp > s^\out >s^\inp$, as shown in Fig.~\ref{(QDoD) Fig2.2-1}. Then $\mathcal{T}$ is a quantum comb if and only if it satisfies the following two conditions:
    \begin{enumerate}
        \item (Generalized complete positivity) 
        \begin{align}
            \mathsf{T}_{t^\out t^\inp s^\out s^\inp}\geq0.\label{(QDoD)Eq:generalizedCP}
        \end{align}
        \item (Generalized trace preservation)
        \begin{equation}
            \begin{split}
                \operatorname{Tr}_{t^\out}(\mathsf{T}_{t^\out t^\inp s^\out s^\inp}) &= \mathds{1}_{t^\inp}\otimes \mathsf{T}_{s^\out s^\inp},\\
                \operatorname{Tr}_{s^\out}(\mathsf{T}_{s^\out s^\inp}) &=  \mathds{1}_{s^\inp},
            \end{split}\label{(QDoD)Eq:generalizedTP}
        \end{equation}
        where $\mathsf{T}_{s^\out s^\inp} \geq 0$.
    \end{enumerate}
\end{defn}
These conditions admit a simple reading: tracing out the latest leg, $t^\out$, must return the identity on the leg immediately preceding it in time, $t^\inp$, tensored with the remaining two-legged object $\mathsf{T}_{s^\out s^\inp}$; tracing out the next-latest leg, $s^\out$, must then return the identity on the earliest leg, $s^\inp$. In other words, the generalized trace-preservation condition \textit{trickles down} the temporal hierarchy: starting from the latest leg and working backward, tracing out each leg must leave the identity on the leg directly preceding it, together with a valid lower-order comb (or channel) on the remaining, earlier legs.

Other temporal orderings, such as $t^\inp>t^\out>s^\out>s^\inp$, $t^\out>t^\inp>s^\inp>s^\out$, and $t^\inp>t^\out>s^\inp>s^\out$, can also be considered with the same generalized CPTP conditions~\eqref{(QDoD)Eq:generalizedCP} and \eqref{(QDoD)Eq:generalizedTP}. However, the generalized TP condition~\eqref{(QDoD)Eq:generalizedTP} no longer necessarily follows a simple backward-in-time interpretation. For example, under the ordering $t^\out>t^\inp>s^\inp>s^\out$, the leg $s^\out$ corresponds to the earliest time, although it is traced out in the second generalized TP condition $\operatorname{Tr}_{s^\out}(\mathsf{T}_{s^\out s^\inp}) =  \mathds{1}_{s^\inp}$. We discuss these temporal orderings and derive their corresponding conditions from explicit dilations in Appendix~\ref{(QDoD) Apx: generalized TP}.

The generalized CP condition has an operational meaning analogous to CP for ordinary quantum channels. Let $\mathcal{T}$ be a one-slot quantum comb and consider a joint quantum operation
\begin{align}
    \mathcal{J}:\mathcal{L}(\mathcal{H}_{s^\out}\otimes\mathcal{H}_{A_\inp})\rightarrow\mathcal{L}(\mathcal{H}_{t^\inp}\otimes\mathcal{H}_{A_\out}).
\end{align}
Suppose $\mathcal{T}$ acts only on the legs associated with $\mathcal{H}_{s^\out}$ and $\mathcal{H}_{t^\inp}$, leaving the auxilary legs unchanged. The resulting joint operation is
\begin{align}
    \mathcal{J}'&:=(\mathcal{T}\otimes\mathcal{I}_A)[\mathcal{J}],\\
    \mathcal{J}'&:\mathcal{L}(\mathcal{H}_{s^\inp}\otimes\mathcal{H}_{A_\inp})\rightarrow\mathcal{L}(\mathcal{H}_{t^\out}\otimes\mathcal{H}_{A_\out}),
\end{align}
where $\mathcal{I}_A$ denotes the identity supermap on the auxilary legs, and $\mathcal{H}_{s^\inp}$ and $\mathcal{H}_{t^\out}$ are the external input and output spaces of the comb. Generalized CP ensures that $\mathcal{J}'$ is completely positive. Together with the generalized trace-preservation conditions, it ensures that $\mathcal{J}'$ is CPTP whenever $\mathcal{J}$ is CPTP, and \textbf{CP-trace-non-increasing (CPTNI)} whenever $\mathcal{J}$ is CPTNI. These statements hold for arbitrary auxillary spaces and joint operations, and this property ensures that the quantum comb can be given by encoder and decoder isometries~\cite{chiribella2008transforming}.

To understand the time evolution of a generic four-legged quantum comb, recall how we previously considered the time evolution desribed by a quantum channel.

\begin{example}\label{(QDoD) Fundamental example 2}(Schr{\" o}dinger and Heisenberg evolutions) Let $\mathcal{E}_{t^\out , 0^\inp}: \mathcal{L}(\mathcal{H}_{0^\inp})\rightarrow \mathcal{L}(\mathcal{H}_{t^\out})$ be a CPTP map. Acting on an initial state $\rho^S(0)\in\mathcal{L}(\mathcal{H}_{0^\inp})$, it gives a time-evolved state $\rho^S(t)\in\mathcal{L}(\mathcal{H}_{t^\out})$. On the other hand, its adjoint $\mathcal{E}^*_{0^\inp,t^\out}: \mathcal{L}(\mathcal{H}_{t^\out}) \rightarrow\mathcal{L}(\mathcal{H}_{0^\inp})$ acts on an initial observable $A^S\in \mathcal{L}(\mathcal{H}_{t^\out})$ and gives a time-evolved observable $A^S(t) \in \mathcal{L}(\mathcal{H}_{0^\inp})$. Pictorially, this is a box with two legs (see Fig.~\ref{(QDoD) Fig2.2-2}): one corresponds to the initial condition, and the other to the time evolution. Quantum channel evolution therefore comes in two types: \textit{Schr{\" o}dinger and Heisenberg evolutions}.

\begin{figure}
    \centering
    \includegraphics[width=\linewidth]{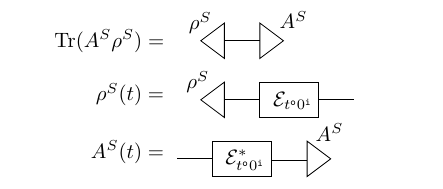}
    \caption{Pictorial representation of the Hilbert--Schmidt inner product between an observable $A^S$ and a quantum state $\rho^S$, each represented by triangles. Schr\"odinger-type evolution is given by the application of the CPTP map $\mathcal{E}_{t^\out,0^\inp}$ to the quantum state. Heisenberg-type evolution is given by its adjoint, defined with respect to the Hilbert--Schmidt inner product.}
    \label{(QDoD) Fig2.2-2}
\end{figure}

Furthermore, if the quantum channel satisfies CP-divisibility, the GKSL-type evolution can be given by repeated application of infinitesimal CPTP maps, as discussed in the previous subsection. We therefore observe two important properties of the quantum channel, expressed as a two-legged box:
\begin{enumerate}
    \item One leg corresponds to the initial condition (the \textit{initial-condition leg}), and the remaining leg corresponds to the evolution (the \textit{evolution leg}).
    \item The evolution leg can be connected to the initial-condition leg of another quantum channel to give another two-legged quantum channel.
\end{enumerate}
\end{example}

Drawing an analogy with the example above, for a general four-legged quantum comb one can consider the following properties of the evolution:
\begin{enumerate}
    \item Two legs correspond to the initial condition, and two legs correspond to the evolution.
    \item Connecting the evolution legs with the initial-condition legs of another comb gives another four-legged quantum comb.
\end{enumerate}
For a quantum channel, we had two types of evolution, namely the Schr\"odinger and Heisenberg evolutions, dictated by which leg was chosen as the initial condition and which for the evolution. For a quantum comb, one can deduce from the above two rules that there are six possible types of evolution. Note that this freedom already appears without imposing any strict temporal ordering among the four legs. Once a strict temporal ordering is imposed, there are twenty types of evolution, summarized in Appendix~\ref{(QDoD)Appendix:EvolutionFamily}.

Furthermore, for a quantum channel, the initial condition is either a state or an observable. For the four-legged quantum comb, many two-time objects can serve as an initial condition, such as a quantum channel in Example~\ref{(QDoD) Eg: evolution of channel}, a quantum instrument in Sec.~\ref{sec:otoc}, and an initially correlated open quantum dynamics in Sec.~\ref{sec:OQS}.

Before proceeding to our formalism, we close this section with one final example that genuinely departs from the evolution of states and observables: the evolution of a quantum channel itself.


\subsection{A Simple Example: Evolution of a Quantum Channel}\label{sec:motivatingexamples}
    In the previous section, we saw that the quantum comb arises naturally from a multi-time quantum operator. Furthermore, we have observed that by fixing which legs of the quantum comb correspond to the initial condition, we obtain several different evolution types. As a simple non-trivial example, we take a quantum channel as the initial condition and examine how imposing a CP-divisibility-inspired composition law on the quantum comb can generate its time evolution. Note that channel evolution is one of many possible evolution types; we explore others in the application sections. 
    
    \begin{example}(Evolution of channel)\label{(QDoD) Eg: evolution of channel}
        To motivate the evolution of quantum channels, let us consider an experimental setting based on quantum process tomography (QPT). In QPT, one prepares a set of input quantum states, lets them evolve through a physical sample (e.g., a superconducting qubit coupled to a noisy environment or a driven quantum system interacting with a thermal reservoir), and measures the resulting output states. By repeating this procedure with a tomographically complete set of input states and measurement settings, one can reconstruct the CPTP map. This setting is also relevant to experimental drift in quantum information processing~\cite{wan2019quantum,burnett2019decoherence,bylander2011noise,proctor2020detecting}, where low-frequency noise can cause the implemented quantum channel to change slowly over time, even when the same operation is repeatedly performed. The following example suggests how our formalism may provide an alternative approach to characterizing such time-dependent quantum-channel dynamics.
    
        Consider two experiments performed on identically prepared copies of the same physical sample, which are assumed to undergo identical system--environment dynamics (e.g., nonequilibrium dynamics). Alice performs QPT over the time interval $[s^\out,s^\inp]$ and reconstructs the quantum channel $\mathcal{E}_{s^\inp,s^\out}$. Bob performs the same procedure over a different time interval $[t^\inp,t^\out]$ and reconstructs the channel $\mathcal{E}_{t^\out,t^\inp}$. Fig.~\ref{(QDoD)Fig:Tomo} illustrates the two experimental protocols. In general, these two channels need not coincide, $\mathcal{E}_{s^\inp,s^\out}\neq\mathcal{E}_{t^\out,t^\inp}$, since the underlying system and its environment may evolve between the two time intervals. This suggests that a quantum channel can itself evolve over time. This naturally raises the question of whether a dynamical equation governs the evolution from $\mathcal{E}_{s^\inp,s^\out}$ to $\mathcal{E}_{t^\out,t^\inp}$.\footnote{We note the related work of Pollock and Modi~\cite{Pollock_2018_Quantum}, who showed that quantum process tomography performed at a sequence of times can reconstruct a generalized Nakajima--Zwanzig memory-kernel master equation in the continuous-time limit. Although their approach differs from ours, it shows how quantum process tomography can reconstruct dynamics beyond the conventional GKSL framework.}
        \begin{figure}
            \centering
            \includegraphics[width=\linewidth]{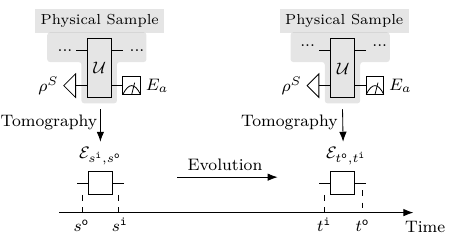}
            \caption{Schematic illustration of quantum process tomography performed over different time intervals. The input system states $\rho^S$ and the positive operator-valued measure (POVM) elements $E_a$ represent tomographically complete sets of state preparations and measurements, respectively, used to reconstruct the corresponding CPTP maps $\mathcal{E}_{s^\inp,s^\out}$ and $\mathcal{E}_{t^\out,t^\inp}$.}
            \label{(QDoD)Fig:Tomo}
        \end{figure}
 
        To construct a quantum comb that maps one quantum channel to another, we consider the quantum comb defined in Eq.~\eqref{(QDoD) Def: quantum comb} and impose the strict temporal ordering $t^\out>t^\inp>s^\inp>s^\out$. To distinguish this temporally ordered comb from the general one, we denote it by $\overleftarrow{\mathcal{T}}_{t^\out,t^\inp,s^\inp,s^\out}$, where the arrow indicates the imposed temporal ordering. Acting on the input quantum channel $\mathcal{E}_{s^\inp,s^\out}$, this comb produces the output quantum channel $\mathcal{E}_{t^\out,t^\inp}$, as
        \begin{align}
            \overleftarrow{\mathcal{T}}_{t^\out,t^\inp,s^\inp,s^\out}\circ \mathcal{E}_{s^\inp,s^\out}
            = \mathcal{E}_{t^\out,t^\inp}.\label{(QDoD)Eq:EgChannelEvo1}
        \end{align}
        The generalized CPTP conditions for quantum combs~\cite{chiribella2008quantum}, discussed in Appendix~\ref{(QDoD) Apx: generalized TP}, ensure that the output is a valid quantum channel. This is shown in Fig.~\ref{(QDoD) fig:Channel evolution example 2-1}(a).
        \begin{figure}[h]
            \centering
            \includegraphics[width=\linewidth]{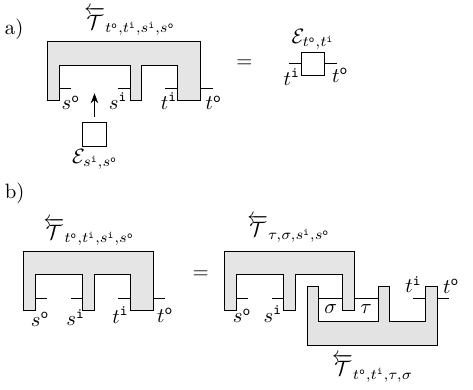}
            \caption{(a) A strictly temporally ordered quantum comb $\overleftarrow{\mathcal{T}}_{t^\out,t^\inp,s^\inp,s^\out}$ acting on the input quantum channel $\mathcal{E}_{s^\inp,s^\out}$ to produce the output channel $\mathcal{E}_{t^\out,t^\inp}$. (b) Illustration of the CP-divisibility condition for the strictly temporally ordered quantum comb.}
            \label{(QDoD) fig:Channel evolution example 2-1}
        \end{figure}
        
        Analogous to the CP-divisibility of the dynamical map, one can consider the quantum channel that can be decomposed as 
        \begin{align}\label{(QDoD) Eq: CP-divisibility of example2}
            \overleftarrow{\mathcal{T}}_{t^\out , t^\inp ,s^\inp ,s^\out}=\overleftarrow{\mathcal{T}}_{t^\out , t^\inp ,\tau,\sigma}\circ \overleftarrow{\mathcal{T}}_{\tau,\sigma , s^\inp , s^\out},
        \end{align}
        for any $t^\out \geq t^\inp \geq \tau \geq \sigma \geq s^\inp \geq s^\out$, where the superchannel $\overleftarrow{\mathcal{T}}_{t^\out , t^\inp ,\tau,\sigma}$ satisfies the generalized CPTP conditions. This is illustrated in Fig.~\ref{(QDoD) fig:Channel evolution example 2-1}(b). This decomposition is similar to the operational CP-divisibility introduced by Miltz et al.~\cite{Milz2019_CP}, where it is a decomposition property of the superchannel rather than the quantum channel. Furthermore, the above contraction is the evolution type labeled $1C$ in appendix~\ref{(QDoD)Appendix:EvolutionFamily}.

        Next, we comment on the explicit physical realization of the four-legged superchannel shown in Fig.~\ref{(QDoD) fig:Channel evolution example 2-1}. By the generalized dilation theorem for quantum combs~\cite {chiribella2008quantum}, we can construct a possible dilation of this evolution. Let us first consider the following conditional density matrix of the environment
        \begin{align}
            \rho^{E(\mathcal{E}_{s^\inp ,s^\out})} := \text{Tr}_S \left(\mathcal{U}^{\text{SE}}_{\tau ,s^\out} \circ \mathcal{E}_{s^\inp ,s^\out}\otimes \mathcal{I}_E [\rho^{SE}]\right),
        \end{align}
        where $\mathcal{I}_E$ is an identity operation on the environmental state. An arbitrary time variable $ \tau$ indicates the time at which the trace over the system is taken. Therefore, this is dynamically irrelevant as long as $t^\inp \geq \tau \geq s^\inp$. Using this conditional density matrix, one can obtain one possible dilation of the quantum comb as 
        \begin{equation}
            \begin{split}
                &\overleftarrow{\mathcal{T}}_{t^\out,t^\inp, s^\inp,s^\out}\circ \mathcal{E}_{s^\inp ,s^\out} [\rho_\text{S}(t^\inp )] \\
                &= \text{Tr}_\text{E} \left(\mathcal{U}^{\text{SE}}_{t^\out,t^\inp}[\rho_{\text{S}}(t^\inp)\otimes \rho^{(\mathcal{E}_{s^\inp ,s^\out})}]\right).\label{(QDoD) Eq:Channel evolution comb}
            \end{split}
        \end{equation}
        Fig.~\ref{(QDoD) fig:eg2-2} illustrates this.
        \begin{figure}[h]
            \centering
            \includegraphics[width=\linewidth]{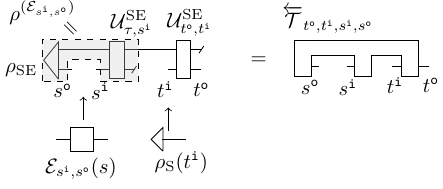}
            \caption{Dilation of the four-legged quantum comb $ \overleftarrow{\mathcal{T}}_{t^\out,t^\inp, s^\inp,s^\out}$. shaded region corresponds to the conditional density matrix $\rho^{(\mathcal{E}_{s^\inp ,s^\out})} $. An arbitrary quantum channel $\mathcal{E}_{s^\out, s^\inp}(s)$ and a state $\rho_{\text{S}}(t^\inp)$ can be inserted into the quantum comb to give a system state defined in $\mathcal{L}(\mathcal{H}_{t^\out})$.}
            \label{(QDoD) fig:eg2-2}
        \end{figure}
        
    \end{example}

    Finally, let us return to the two fundamental questions introduced in Sec.~\ref{(QDoD)SubSubSec:Questions}. As illustrated in Fig.~\ref{(QDoD) fig:eg2-2}, the quantum comb admits a system--environment dilation. This raises the \textit{microscopic question}: What additional physical assumptions on the underlying system--environment dynamics are sufficient to ensure the CP-divisibility condition~\eqref{(QDoD) Eq: CP-divisibility of example2}? On the other hand, taking this composition law as an axiom leads to the \textit{axiomatic question}: What is the most general dynamical generator compatible with this condition? In the following sections, Sec.~\ref{sec:quantumdynamicsofdynamics} and Sec.~\ref{sec:quantumdynamicsofdynamics-microscopic}, we address these two questions within a more general framework for quantum combs.


\section{Quantum Dynamics of Dynamics: Axiomatic Approach}\label{sec:quantumdynamicsofdynamics}

In this section, we extend the notion of the evolution map to quantum combs. First, recall the examples from the previous section. We observed that a quantum state represented pictorically as a one-legged object time-evolves under a two-legged quantum channel. Similarly, a quantum channel represented as two-legged quantum objects evolves under a four-legged quantum comb. Similarly, for an $n$--legged quantum comb, the evolution of $m\leq n$ of its legs can be realized by repeatedly applying a $2m$--legged quantum comb. 

In this work, we will explicitly focus on the four-time-dependent four-legged quantum comb denoted as 
\begin{align}
    \mathcal{T}_{t^\out , t^\inp , s^\out,s^\inp} (t^\out , t^\inp , s^\out,s^\inp),
\end{align}
where its pictorial representation is shown in Fig.~\ref{(QDoD) Fig2.2-1}. The subscripts are the Hilbert-space indices for each open leg, with $\inp$ and $\out$ indicating the input and output legs. 

\textbf{Temporal ordering.}
Temporal ordering specifies the temporal inequalities among the physical times, which can change depending on the situation. We have already encountered the ordering $t^\out>t^\inp>s^\out>s^\inp$ in Example~\ref{(QDoD) Eg: evolution of state} and $t^\out>t^\inp>s^\inp>s^\out$ in Example~\ref{(QDoD) Eg: evolution of channel}. Throughout this paper, unless explicitly stated, we fix the legs from right to left as $\alpha = (t^\out,t^\inp,s^\out,s^\inp)$. Accordingly, we write the quantum comb as
\begin{equation}
     \mathcal{T}_\alpha = \mathcal{T}_{t^\out, t^\inp ,s^\out, s^\inp}.
\end{equation} 
If the temporal ordering is fixed, we indicate this by an arrow, for example $\overleftarrow{\mathcal{T}}_{t^\out,t^\inp, s^\inp , s^\out}$, in Example~\ref{(QDoD) Eg: evolution of channel}

\begin{figure}
    \centering
    \includegraphics[width=\linewidth]{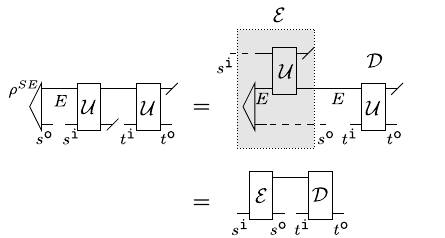}
    \caption{Diagram showing the quantum comb of temporal ordering $t^\out>t^\inp>s^\inp>s^\out$ consdiered in Fig.~\ref{(QDoD) fig:eg2-2}. An equivalent representation in terms of delayed-input construction is given, which gives the quantum comb in terms of two isometries of the encoder and decoders.}
    \label{(QDoD) Fig3-0}
\end{figure}

Recall that the four legs of a quantum comb can be arranged in different temporal orderings, giving rise to six distinct evolution types, as discussed in Appendix~\ref{(QDoD)Appendix:EvolutionFamily}. Rather than deriving a separate dynamical equation for each evolution type, we use the delayed-input construction~\cite{quintino2019probabilistic}---together with auxiliary spaces where necessary---to express all of them in the common one-slot comb form shown in Fig.~\ref{(QDoD) Fig2.2-1}. For example, the comb in Fig.~\ref{(QDoD) fig:eg2-2} has temporal ordering $t^\out>t^\inp>s^\inp>s^\out$, but can be rewritten as shown in Fig.~\ref{(QDoD) Fig3-0}. In its original form (shown in Fig.~\ref{(QDoD) fig:eg2-2}), the legs $s^\out$ and $s^\inp$ refer to the system spaces at different times, meaning $\mathcal{L}(\mathcal{H}_{s^\out})\cong\mathcal{L}(\mathcal{H}_{s^\inp})\cong\mathcal{L}(\mathcal{H}_S)$. In the delayed-input construction, we instead introduce a separate copy of the system space for the $s^\inp$ input. The space used for the dilation is therefore enlarged from $\mathcal{L}(\mathcal{H}_S)\otimes\mathcal{L}(\mathcal{H}_E)$ to $\mathcal{L}(\mathcal{H}_{s^\out})\otimes\mathcal{L}(\mathcal{H}_{s^\inp})\otimes \mathcal{L}(\mathcal{H}_E)$. The preparation, routing, and auxiliary contractions recover the original comb while preserving its physical temporal ordering. 

As illustrated in Fig.~\ref{(QDoD) Fig3-0}, when $s^\out<s^\inp$, causality requires that the isometric channel from $\mathcal{L}(\mathcal{H}_{s^\inp})$ to $\mathcal{L}(\mathcal{H}_{s^\out})\otimes\mathcal{L}(\mathcal{H}_E)$ does not allow signalling from $s^\inp$ to $s^\out$. In this paper, we consider both signaling and non-signaling isometric channels, depending on the temporal ordering. When necessary, we write $\mathcal{E}_{s^\out\nleftarrow s^\inp}$ to explicitly indicate that signaling from $s^\inp$ to $s^\out$ is forbidden.

One may wonder whether it is necessary to distinguish all 20 possible evolution types, since they can all be represented as quantum combs of the form shown in Fig.~\ref{(QDoD) Fig2.2-1}, with signaling or non-signaling isometries such as $\mathcal{E}_{s^\out\nleftarrow s^\inp}$. However, this common representation does not make the different evolution types equivalent: each specifies a different physical arrangement of the evolving and initial-condition legs and the corresponding causality constraints. The two examples encountered so far illustrate this distinction: Example~\ref{(QDoD) Eg: evolution of state} describes the dynamics of the linear response function~\eqref{(QDoD) Eq:eg1 step1}, while Example~\ref{(QDoD) Eg: evolution of channel} describes the evolution of a quantum channel~\eqref{(QDoD)Eq:EgChannelEvo1}. Therefore, combs with different temporal orders are physically distinct due to causality. 

\subsection{CP-Divisible Quantum Combs}\label{(QDoD)Sec:CP-DivisibleQuantumComb}
Let us first recall what we learned about the evolution induced by quantum combs from the preceding examples. As discussed in Example~\ref{(QDoD) Fundamental example 2}, each evolution type is specified by selecting two legs as the evolution legs with the remaining two serving as the initial-condition legs. Furthermore, from the CP-divisibility condition considered in Example~\ref{(QDoD) Eg: evolution of channel} and illustrated in Fig.~\ref{(QDoD) fig:Channel evolution example 2-1}(b), we learned that the evolution of quantum combs can be described by contracting the evolving legs of one comb with the initial-condition legs of another.

Let us now formalize what we observed in the preceding examples as a mathematical contraction between quantum combs.  If a comb $\mathcal{T}_{\alpha_0}$ evolves through the action of another comb $\mathcal{T}_{\alpha_1}$, the time-evolving legs of $\mathcal{T}_{\alpha_0}$ are contracted with the initial-condition legs of $\mathcal{T}_{\alpha_1}$. Assigning the same label to each pair of connected legs and distinct labels to all remaining legs, we have
\begin{equation}
    \alpha_0\cap\alpha_1=\{\tau,\sigma\}.
\end{equation}
Here $\tau$ and $\sigma$ label the two shared legs. Therefore, the contraction of two quantum combs is given by 
\begin{align}
        \mathcal T_{\alpha_0}\circ\mathcal T_{\alpha_1}=\mathcal T_{\alpha_0\Delta\alpha_1},
    \end{align}
where $\alpha_0\Delta\alpha_1:=(\alpha_0\setminus\alpha_1)\cup(\alpha_1\setminus\alpha_0)$ or equivalently, $\alpha_0\Delta\alpha_1=(\alpha_0\cup\alpha_1)\setminus\{\tau,\sigma\}$. Using this contraction rule, we now define CP-divisibility for quantum combs in terms of their decomposition into two quantum combs at discrete times.

\begin{defn}[CP-divisibile quantum combs]\label{(QDoD) Def: evolution comb}
Let $\mathcal{T}_{\alpha}$ be a four-time-dependent, four-legged quantum comb with $\alpha=(t^\out,t^\inp,s^\out,s^\inp)$, where two legs are designated as evolution legs and the remaining two as initial-condition legs.

The quantum comb $\mathcal{T}_{\alpha}$ is said to be \textit{CP-divisible} if, for any two label sets $\alpha_0$ and $\alpha_1$ satisfying
\begin{enumerate}
    \item $\alpha_0$ (or $\alpha_1$) contains the labels of the evolution legs of $\alpha$, while $\alpha_1$ (or $\alpha_0$) contains those of its initial-condition legs, and
    \item $|\alpha_0\cap\alpha_1|=2$,
\end{enumerate}
there exist quantum combs $\mathcal{T}_{\alpha_0}$ and $\mathcal{T}_{\alpha_1}$ such that
\begin{align}\label{(QDOD) Eq: evolution comb}
    \mathcal{T}_{\alpha}
    =\mathcal{T}_{\alpha_1}\circ\mathcal{T}_{\alpha_0}
    =\mathcal{T}_{\alpha_0\Delta\alpha_1}.
\end{align}
\end{defn}
\noindent In Appendix~\ref{(QDoD)Appendix:EvolutionFamily}, we enumerate all possible decompositions of four-legged quantum combs satisfying the above constraints. 

Note that the above definition extends to a more general $2n$-time-dependent, $2n$-legged quantum comb by assigning $n$ legs as evolution legs and the remaining $n$ as initial-condition legs, with the intersection condition generalized to $|\alpha_0\cap\alpha_1|=n$.

Before introducing the master equation that generates the CP-divisible quantum comb, we discuss a possible symmetry of its time dependence.
\begin{defn}[Time homogeneity]\label{(QDoD)Def:TimeHomogeneity}
A four-time-dependent, four-legged quantum comb $\mathcal{T}_{t^\out,t^\inp,s^\out,s^\inp}(t^\out,t^\inp,s^\out,s^\inp)$ is said to be \textit{time homogeneous} with respect to the pairs $(t^\out,t^\inp)$ and $(s^\out,s^\inp)$ if its explicit time dependence is only through the corresponding time differences, i.e.,
\begin{equation}
    \mathcal{T}_{t^\out,t^\inp,s^\out,s^\inp}(t^\out,t^\inp,s^\out,s^\inp)
    =
    \mathcal{T}_{t^\out,t^\inp,s^\out,s^\inp}(t^\out-t^\inp,s^\out-s^\inp).
\end{equation}
\end{defn}
Time homogeneity can also be defined with respect to other pairings of the four time variables. In this work, however, we restrict our attention to the pairing specified above.

When time homogeneity is imposed, we distinguish the explicit time arguments of the quantum comb from the Hilbert-space labels of its legs. This is because, e.g., if a four-legged quantum comb is time-homogeneous with respect to the pairs $(t^\out,t^\inp)$ and $(s^\out,s^\inp)$, then although it has four time variables associated with its legs through Hilbert-space indexing, the comb itself explicitly depends only on the two time differences $t^\out-t^\inp$ and $s^\out-s^\inp$.


\subsection{Quantum Dynamics of Dynamics}\label{(QDoD)Sec:QDoD}
Recall that in deriving the conventional master equation for quantum states (reviewed in Sec.~\ref{(QDoD)Sec:ReviewAxiom}) we considered a CP-divisible dynamical map $\mathcal{E}_{t,0}$ acting on an initial state $\rho^S(0)$. This map can be regarded as a two-legged object, with one leg serving as the evolution leg and the other as the initial-condition leg. In the Heisenberg picture, the roles of these two legs are interchanged, as discussed in Example~\ref{(QDoD) Fundamental example 2}. In either picture, the dynamical equation that time-evolves the initial condition is derived by considering the infinitesimal evolution of a single time variable associated with the evolution leg. For a quantum comb, on the other hand, we have seen that two of its four legs are designated as evolution legs, with the other two serving as the initial condition. Consequently, its infinitesimal evolution involves two time variables rather than one. To describe such evolution, we first introduce a path in the two-dimensional time plane along which these variables evolve.

\subsubsection{Evolution Path on the Time Plane}\label{(QDoD)Sec:QDoD1}
To make this notion precise, consider a CP-divisible quantum comb $\mathcal{T}_{\alpha}$. Denote the time variables associated with its two evolving legs by $(\gamma_1,\gamma_2)$. Their simultaneous variation is parametrized by a path in the corresponding time plane,
\begin{align}
    \vec{\gamma}(\lambda)=\bigl(\gamma_1(\lambda),\gamma_2(\lambda)\bigr),\label{Eq: def for path A}
\end{align}
where $\lambda$ is a \textit{path parameter}.

Denote the initial-condition legs by $(\delta_1,\delta_2)$, ordered so that the evolving-leg $\gamma_i$ of one comb contracts with the initial-condition-leg $\delta_i$ of the next comb, for $i=1,2$, following the contraction rule introduced in Sec.~\ref{(QDoD)Sec:CP-DivisibleQuantumComb}. The contraction pattern is therefore
\begin{align}
    \gamma_1\leftrightarrow\delta_1,
    \qquad
    \gamma_2\leftrightarrow\delta_2.
\end{align} To see this explicitly, let us revisit Example~\ref{(QDoD) Eg: evolution of channel}, which illustrates channel evolution. 

\begin{example}[Revisiting quantum channel evolution]
First, we are given a quantum comb $\mathcal{T}_{\alpha}$ where $\alpha = (t^\out,t^\inp,s^\out,s^\inp)$ with strict temporal ordering 
\begin{align}
    t^\out > t^\inp > s^\inp > s^\out\label{(QDoD)Eq:ExampleTime-1}
\end{align}
with $(t^\out, t^\inp)$ and $(s^\out , s^\inp)$ pairs corresponding to the time-evolving and initial-condition legs respectively. From Fig.~\ref{(QDoD) fig:Channel evolution example 2-1}(b), we see that the contracting legs between two combs are 
\begin{align}
    t^\out \leftrightarrow s^\inp,\quad t^\inp \leftrightarrow s^\out\label{(QDoD)Eq:ExampleTime-2},
\end{align}
and therefore we have 
\begin{align}
    ( \gamma_1 =t^\out , \gamma_2 =t^\inp)  , \quad (\delta_1= s^\inp ,\delta_2=s^\out  ).
\end{align}

Let us consider the following two consecutive flows of the two--time variables as the path parameter varies from $\lambda=0 $ to $\lambda=1$:
\begin{equation}
    \begin{split}
    (\delta_1 (0), \delta_2 (0))&=(s^\inp (0) , s^\out (0) ) \\
    &\xrightarrow{\text{(1)}} (s^\inp (1) , s^\out (1) ) = (t^\out (0), t^\inp(0)) \\
    &\xrightarrow{\text{(2)}} (t^\out (1) , t^\inp(1)) = (\gamma_1 (1) , \gamma_2(1))
    \end{split}
\end{equation}
Note that the equality $(s^\inp(1),s^\out(1))=(t^\out(0),t^\inp(0))$ follows from the contraction rule~\eqref{(QDoD)Eq:ExampleTime-2}. This is illustrated in Fig.~\ref{(QDoD)Fig:ExampleTime}. The shaded region is where the temporal ordering~\eqref{(QDoD)Eq:ExampleTime-1} of this comb is preserved. The path must remain within this region to respect the causal constraints of the chosen evolution type. Different temporal orderings of the legs correspond to different allowed regions in the time plane.
\begin{figure}
    \centering
    \includegraphics[width=\linewidth]{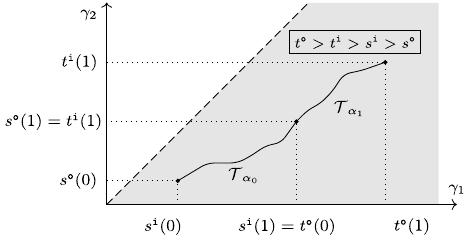}
    \caption{Time plane with the vertical and horizontal axes corresponding to the evolution times $\gamma_1$ and $\gamma_2$, respectively. The shaded region preserves the temporal ordering of the chosen evolution type. Evolution into the unshaded region is excluded by the causal constraints of that type.}
    \label{(QDoD)Fig:ExampleTime}
\end{figure}

Two consecutive flows of the time pair in parts (1) and (2) are given by the following quantum combs 
\begin{align}
    \text{(1): }& \mathcal{T}_{\alpha_0}=\mathcal{T}_{t^\out(0),t^\inp(0),s^\out(0),s^\inp(0)}\\
    \text{(2): }& \mathcal{T}_{\alpha_1}=\mathcal{T}_{t^\out(1),t^\inp(1),s^\out(1),s^\inp(1)}
\end{align}
Let us denote $\alpha_\lambda := (t^\out(\lambda),t^\inp(\lambda),s^\out(\lambda),s^\inp(\lambda)) $. Furthermore, let us employ the following relabelings 
\begin{align}
    \alpha_0=(t^\out(0),t^\inp(0),s^\out(0),s^\inp(0)) =& (\sigma ,\tau ,s^\out ,s^\inp) , \\
    \alpha_1=(t^\out(1),t^\inp(1),s^\out(1),s^\inp(1)) =& (t^\out, t^\inp ,\tau ,\sigma)
\end{align}
Then, using Def.~\eqref{(QDOD) Eq: evolution comb}, we obtain the decomposition of the quantum comb as 
\begin{align}
    \mathcal{T}_{t^\out ,t^\inp ,s^\out ,s^\inp}= \mathcal{T}_{t^\out ,t^\inp ,\tau ,\sigma}\circ \mathcal{T}_{\sigma ,\tau ,s^\out ,s^\inp}\label{(QDoD)Eq:ContractionExample-1}
\end{align}
After temporal ordering, the above contraction coincides with Eq.~(\ref{(QDoD) Eq: CP-divisibility of example2}). 
\end{example}

\subsubsection{CP-Divisibility of Quantum Combs}\label{(QDoD)Sec:QDoD2}
To derive the dynamical equation that generates the CP-divisible quantum comb, we need to characterize how a quantum comb evolves over an infinitesimal time interval. In the conventional axiomatic derivation of the master equation for quantum states reviewed in Sec.~\ref{(QDoD)Sec:ReviewAxiom}, one assumes CP-divisibility and considers its infinitesimal form to obtain the corresponding differential equation. In Def.~\ref{(QDoD) Def: evolution comb}, we introduced CP-divisibility for quantum combs at discrete times. In this subsubsection, we similarly consider its infinitesimal form to derive the dynamical equation for quantum combs.

Recall the contraction Eq.~\eqref{(QDoD)Eq:ContractionExample-1} from the above example. To express an infinitesimal evolution, let us first relabel the indices as
\begin{align}
    t^\out\rightarrow t^\out+\epsilon,\quad
    t^\inp\rightarrow t^\inp+0,\quad
    \sigma\rightarrow t^\out,\quad
    \tau\rightarrow t^\inp.
    \label{(QDoD)Eq:ContractionExample-2}
\end{align}
The contraction then becomes
\begin{align}
    \mathcal{T}_{t^\out+\epsilon,t^\inp+0,s^\out,s^\inp}
    =
    \mathcal{T}_{t^\out+\epsilon,t^\inp+0,t^\inp,t^\out}
    \circ
    \mathcal{T}_{t^\out,t^\inp,s^\out,s^\inp}.
    \label{(QDoD)Eq:ContractionExample-3}
\end{align}
The evolving legs of the comb $\mathcal{T}_{t^\out,t^\inp,s^\out,s^\inp}$ are $(\gamma_1,\gamma_2)=(t^\out,t^\inp)$. Thus, the left-hand side $\mathcal{T}_{t^\out+\epsilon,t^\inp+0,s^\out,s^\inp}$ represents an infinitesimal advance of $\gamma_1$, with $\gamma_2$ held fixed. Similarly, advancing the second evolving-leg $\gamma_2$ gives
\begin{align}
    \mathcal{T}_{t^\out+0,t^\inp+\epsilon,s^\out,s^\inp}
    =
    \mathcal{T}_{t^\out+0,t^\inp+\epsilon,t^\inp,t^\out}
    \circ
    \mathcal{T}_{t^\out,t^\inp,s^\out,s^\inp}.
    \label{(QDoD)Eq:ContractionExample-4}
\end{align}
Note that $\epsilon>0$ in this example. In other examples, however, $\epsilon$ may be negative, depending on the temporal ordering and the associated causality constraints.

We now generalize the above infinitesimal decomposition so that it does not depend on a particular choice of evolution legs. To this end, let us introduce the following ordered sets for Eq.~\eqref{(QDoD)Eq:ContractionExample-3}:
\begin{align}
    &\alpha=(t^\out,t^\inp,s^\out,s^\inp),\quad \alpha_{+\epsilon}:=(t^\out+\epsilon,t^\inp+0,s^\out,s^\inp),\nonumber\\
    &\beta_{+\epsilon}:=(t^\out+\epsilon,t^\inp+0,t^\inp,t^\out).
\end{align}
The contraction can then be written as
\begin{align}
    \mathcal{T}_{\alpha_{+\epsilon}}
    =
    \mathcal{T}_{\beta_{+\epsilon}}\circ\mathcal{T}_{\alpha}.
\end{align}
Since the evolution legs are $(\gamma_1,\gamma_2)=(t^\out,t^\inp)$, we can express $\alpha_{+\epsilon}$ as
\begin{align}
    \alpha_{+\epsilon}
    :=
    \left.\alpha\right|_{\substack{
        \gamma_1\rightarrow\gamma_1+\epsilon\\
        \gamma_2\rightarrow\gamma_2+0}}.\label{(QDoD)Eq:DefForAlpha}
\end{align}
The corresponding initial-condition legs are $(\delta_1,\delta_2)=(s^\inp,s^\out)$, ordered according to the contraction pairing $\gamma_1\leftrightarrow\delta_1$ and $\gamma_2\leftrightarrow\delta_2$. The labels of the comb $ \mathcal{T}_{\beta_{+\epsilon}}$ are obtained by replacing the initial-condition legs labels with the evolving legs labels:
\begin{align}
    \beta_{+\epsilon}
    :=
    \left.\alpha_{+\epsilon}\right|_{\substack{
        \delta_1\rightarrow\gamma_1\\
        \delta_2\rightarrow\gamma_2}}.\label{(QDoD)Eq:DefForBeta}
\end{align}
The corresponding labels for the infinitesimal evolution of $\gamma_2$ are defined analogously. Importantly, the prescriptions~\eqref{(QDoD)Eq:DefForAlpha} and~\eqref{(QDoD)Eq:DefForBeta} apply independently of which legs are designated as evolution and initial conditions. We therefore use them to formulate the infinitesimal form of CP-divisibility for general quantum combs in what follows.

\begin{defn}[Local CP-divisibility]\label{(QDoD)Def:CP-divisibleComb}
    Let $\mathcal{T}_{\alpha}$ be a CP-divisible comb,
    with $\alpha=(t^\out,t^\inp,s^\out,s^\inp)$.
    Choose two evolving legs with time variables
    $(\gamma_1,\gamma_2)$ and the corresponding
    initial-condition legs $(\delta_1,\delta_2)$.

    Keeping the other time variables fixed, define
    \begin{align}
        \alpha_{+\epsilon_{1}}
        &:=
        \left.\alpha\right|_{\substack{
            \gamma_1\rightarrow\gamma_1+\epsilon_{1}\\
            \gamma_2\rightarrow\gamma_2+0}},\\
        \alpha^{+\epsilon_{2}}
        &:=
        \left.\alpha\right|_{\substack{
            \gamma_1\rightarrow\gamma_1+0\\
            \gamma_2\rightarrow\gamma_2+\epsilon_{2}}},
    \end{align}
    We refer to $\mathcal{T}_{\alpha_{+\epsilon_1}}$ and $\mathcal{T}_{\alpha^{+\epsilon_2}}$ as the \textit{infinitesimally evolved combs of $\mathcal{T}_{\alpha}$} along the evolution legs $\gamma_1$ and $\gamma_2$, respectively. To formulate their CP-divisible decompositions, we introduce the following sets
    \begin{align}
        \beta_{+\epsilon_{1}}
        &:=
        \left.\alpha_{+\epsilon_{1}}\right|_{\substack{
            \delta_1\rightarrow\gamma_1\\
            \delta_2\rightarrow\gamma_2}},\\
        \beta^{+\epsilon_{2}}
        &:=
        \left.\alpha^{+\epsilon_{2}}\right|_{\substack{
            \delta_1\rightarrow\gamma_1\\
            \delta_2\rightarrow\gamma_2}}.
    \end{align}
    Then, the infinitesimal evolution of $\mathcal{T}_{\alpha}$ is said to be \textit{locally CP-divisible} if, for all sufficiently small $\epsilon_1$ and $\epsilon_2$ (whose signs may be chosen independently subject to the temporal-ordering and causality constraints), there exist quantum combs $\mathcal{T}_{\beta_{+\epsilon_{1}}}$ and $\mathcal{T}_{\beta^{+\epsilon_{2}}}$ that are completely positive, satisfy the generalized TP conditions of the chosen realization (see Def.~\ref{(QDoD) Def: quantum comb}), and obey
    \begin{align}
        \mathcal{T}_{\alpha_{+\epsilon_{1}}}&=\mathcal{T}_{\beta_{+\epsilon_{1}}}\circ\mathcal{T}_{\alpha},\label{(QDoD)Eq:DefintionForCP-divisbility1}\\
        \mathcal{T}_{\alpha^{+\epsilon_{2}}}&=\mathcal{T}_{\beta^{+\epsilon_{2}}}\circ\mathcal{T}_{\alpha}.
        \label{(QDoD)Eq:DefintionForCP-divisbility2}
    \end{align}
\end{defn}

With Def.~\ref{(QDoD)Def:CP-divisibleComb} at hand, we are now in the position to define the time-derivative of quantum combs and consequently their master equation. 

\subsubsection{Quantum Comb Master Equation}\label{(QDoD)Sec:QDoD3}
To obtain the master equation for quantum combs, we next consider their time derivatives. However, since the evolution involves two time variables, the question arises as to which derivative one should consider. Recall that in Sec.~\ref{(QDoD)Sec:QDoD1} (see also Fig.~\ref{(QDoD)Fig:ExampleTime}), we represented the evolution of a quantum comb as a path in the two-dimensional time plane. It is therefore natural to consider the infinitesimal motion along such a path. 

To this end, we first recall the definition of the path derivative:
\begin{align}\label{(QDoD) Eq: path-derivative}
    \frac{\partial \mathcal{T}_\alpha (\lambda)}{\partial\lambda} := \frac{d \gamma_1}{d \lambda} \frac{\partial\mathcal{T}_\alpha (\lambda)}{\partial \gamma_1}+\frac{d \gamma_2}{d \lambda} \frac{\partial\mathcal{T}_\alpha (\lambda)}{\partial \gamma_2},
\end{align}
where the derivatives with respect to the path-dependent variables $\{\gamma_1 , \gamma_2\}$ are defined as follows. 
\begin{align}
    \frac{\partial\mathcal{T}_\alpha (\lambda)}{\partial \gamma_1} &:= \lim_{\epsilon\rightarrow 0^+} \frac{\mathcal{T}_{\alpha_{+\epsilon}} (\gamma_1 + \epsilon , \gamma_2+0) -\mathcal{T}_{\alpha} (\gamma_1  , \gamma_2)  }{\epsilon}\\
    \frac{\partial\mathcal{T}_\alpha (\lambda)}{\partial \gamma_2} &:= \lim_{\epsilon\rightarrow 0^+} \frac{\mathcal{T}_{\alpha^{+\epsilon}} (\gamma_1 + 0 , \gamma_2+\epsilon) -\mathcal{T}_{\alpha} (\gamma_1  , \gamma_2)  }{\epsilon}
\end{align}
Using the local CP-divisibility~(\ref{(QDoD)Eq:DefintionForCP-divisbility1}) and (\ref{(QDoD)Eq:DefintionForCP-divisbility2}) of the infinitesimal evolution of the quantum comb $\mathcal{T}_{\alpha}$, the path-dependent time variable derivatives can be written as 
\begin{align}
    \frac{\partial\mathcal{T}_\alpha (\lambda)}{\partial \gamma_1} &= \lim_{\epsilon\rightarrow 0^+} \frac{\mathcal{T}_{\beta_{+\epsilon}} (\gamma_1 + \epsilon , \gamma_2+0) -\mathcal{I} }{\epsilon} \circ \mathcal{T}_{\alpha} (\gamma_1  , \gamma_2) \\
    \frac{\partial\mathcal{T}_\alpha (\lambda)}{\partial \gamma_2} &= \lim_{\epsilon\rightarrow 0^+} \frac{\mathcal{T}_{\beta^{+\epsilon}} (\gamma_1 + 0 , \gamma_2+\epsilon) -\mathcal{I}  }{\epsilon}\circ \mathcal{T}_{\alpha} (\gamma_1  , \gamma_2)
\end{align}
From the right-hand side of the above equations, we can define the generators of the dynamics. 

\begin{defn}[Dynamical decoding and encoding generators]\label{Def: dynamical decoding/encoding generators}
    Let $\mathcal{T}_\alpha(\lambda)$ be a CP-divisible quantum comb whose infinitesimal evolution is locally CP-divisible. The pair of path-dependent operators $\mathcal{L}_D(\lambda)$ and $\mathcal{L}_E(\lambda)$ acting on $\mathcal{T}_\alpha(\lambda)$ are called the \textit{dynamical decoding and encoding generators}, respectively, if the following limits exist:
    \begin{align}
        \mathcal{L}_{D}(\lambda) &=\lim_{\epsilon\rightarrow 0^+} \frac{\mathcal{T}_{\beta_{+\epsilon}}(\gamma_1+\epsilon,
        \gamma_2+0)- \mathcal{I}}{\epsilon},\label{(QDoD) Eq: defintion of decoding generators}\\
        \mathcal{L}_E(\lambda)  &= \lim_{\epsilon\rightarrow 0^+} \frac{\mathcal{T}_{\beta^{+\epsilon}} (\gamma_1+0,
        \gamma_2+\epsilon)-\mathcal{I}}{\epsilon}.\label{(QDoD) Eq: defintion of encoding generators}
    \end{align}
    Here, $\beta_{+\epsilon}$ and $\beta^{+\epsilon}$ are the label sets defined in Def.~\ref{(QDoD)Def:CP-divisibleComb}.
\end{defn}

For the semigroup case, the decoding and encoding generators take the same form as in Eqs.~\eqref{(QDoD) Eq: defintion of decoding generators} and~\eqref{(QDoD) Eq: defintion of encoding generators} with no explicit time-dependence. However, particular care must be taken with their explicit time dependence. To illustrate this, consider the evolution type with contraction pairings $\{t^\inp \leftrightarrow t^\out, s^\out \leftrightarrow s^\inp\}$, where the evolution legs are $\gamma_1=t^\inp$ and $\gamma_2=s^\out$.\footnote{Note that this choice corresponds to the OTOC evolution considered in Sec.~\ref{sec:otoc}.} Following the definitions introduced above, we obtain $\alpha_{+\epsilon}=(t^\out,\gamma_1+\epsilon,\gamma_2+0,s^\inp)$. Since the initial-condition legs are $\delta_1=t^\out$ and $\delta_2=s^\inp$, we have $\beta_{+\epsilon}=(\gamma_1,\gamma_1+\epsilon,\gamma_2+0,\gamma_2)$. Thus, the quantum comb defining the decoding generator has the explicit time dependence
\begin{align}
    \mathcal{T}_{\beta_{+\epsilon}}
    =\mathcal{T}_{\beta_{+\epsilon}}
    (t^\inp,t^\inp+\epsilon,s^\out+0,s^\out).
\end{align}
Now assume time homogeneity with respect to the pairs $(t^\out,t^\inp)$ and $(s^\out,s^\inp)$. The explicit time dependence then reduces to
\begin{align}
    \mathcal{T}_{\beta_{+\epsilon}}
    &=\mathcal{T}_{\beta_{+\epsilon}}
    (t^\inp-(t^\inp+\epsilon),s^\out+0-s^\out)=\mathcal{T}_{\beta_{+\epsilon}}(-\epsilon,0).
\end{align}
Therefore, the infinitesimal comb defining the decoding generator depends explicitly only on $\epsilon$, so the resulting generator has no explicit dependence on the path parameter. A similar argument holds for the encoding generator. Thus, for semigroup quantum combs satisfying the corresponding time-homogeneity conditions, the decoding and encoding generators are defined by Eqs.~\eqref{(QDoD) Eq: defintion of decoding generators} and~\eqref{(QDoD) Eq: defintion of encoding generators}, respectively, without any explicit dependence on the path parameter.

Using the above definition of dynamical decoding and encoding generators, we obtain the master equation for the quantum comb.
\begin{thm}(Quantum Comb-Generating Master Equation)\label{(QDoD) Def: Comb master equation}
   Let $\mathcal{T}_\alpha(\lambda)$ be a CP-divisible quantum comb whose infinitesimal evolution is locally CP-divisible. Let $\vec{\gamma} = (\gamma_1(\lambda),\gamma_2(\lambda))$ be the corresponding evolving legs. Then the \textit{quantum comb-generating master equation} (QCME) is given by 
    \begin{align}
        \frac{\partial \mathcal{T}_\alpha (\lambda)}{\partial \lambda} &=  \left[\frac{d \gamma_1}{d \lambda} \mathcal{L}_{D}(\lambda) +\frac{d \gamma_2}{d \lambda} \mathcal{L}_{E}(\lambda)  \right] \circ \mathcal{T}_\alpha (\lambda)\label{(QDoD)Eq:CombME}.\\
        &=: \mathcal{L}(\gamma_1 ,\gamma_2) \circ \mathcal{T}_\alpha (\lambda)  \label{(QDoD) Eq: Quantum comb generator}
    \end{align}
    where $\mathcal{L}_{D}(\lambda) $ and $\mathcal{L}_{E}(\lambda)$ are decoding/encoding generators defined in Def.~\ref{Def: dynamical decoding/encoding generators}. Here, $\mathcal{L}$ is the Liouvillian-like generator of the QCME, which we dub the \emph{quantum comb generator}. For brevity, we will also write $\mathcal{L}(\lambda)$.
\end{thm}
\begin{proof}
    The result follows directly from the preceding discussion.
\end{proof}

Note that a semigroup quantum comb satisfies the same QCME, with no explicit path dependence in its decoding and encoding generators. Although the individual generators are path-independent, the coefficients $d\gamma_1/d\lambda$ and $d\gamma_2/d\lambda$ still depend on the chosen path. Consequently, the full quantum-comb generator remains path-dependent.

\subsubsection{Choi Matrix Representation}\label{(QDoD)Sec:QDoD4}
In general, directly analyzing the dynamical equation from the previous subsubsection is challenging because it is a linear map acting on the quantum comb (which is itself a linear map). To derive an explicit form of the QCME, this subsubsection provides a detailed, step-by-step derivation of its matrix representation using the Choi representation~\cite{Choi1975CompletelyMatrices,jamiolkowski1972linear}. 

The outline of the derivation is as follows. We first introduce a Kraus representation of the quantum comb in Corollary~\ref{(QDoD)Coro:Kraus Representation}, followed by the coefficients entering the master equation in Def.~\ref{(QDoD)Def: path-dependent POB}. These provide the necessary ingredients for the main results of this subsubsection---Theorem~\ref{Lemma: GKSL form} and Corollary~\ref{Alternative form of GKSL}---which give explicit forms of the QCME in the Choi representation. Finally, in Example~\ref{(QDoD)eg:SchHeiCorr}, we illustrate how our QCME gives rise to nontrivial correlations between the Heisenberg-picture evolution of an observable and the Schr\"odinger-picture evolution of a state, which could never be seen by considering standard state/observable evolutions alone.
\begin{cor}\label{(QDoD)Coro:Kraus Representation}
     Let $\mathcal{T}_\alpha$ be a four-legged quantum comb acting on an $n$-legged quantum comb $\mathcal{M}_\beta$, where $\alpha$ and $\beta$ denote the sets of Hilbert-space labels associated with their respective open legs. Assume that the two combs are composed by contracting exactly two corresponding legs, such that $|\alpha\cap\beta|=2$. We denote the contracted legs by $\gamma:=\alpha\cap\beta$, and the remaining legs of $\mathcal{T}_\alpha$ and $\mathcal{M}_\beta$ by $\gamma_+:=\alpha\setminus\beta$ and $\gamma_-:=\beta\setminus\alpha$, respectively. Thus, $\gamma_+$ labels the two remaining legs of $\mathcal{T}_\alpha$, whereas $\gamma_-$ labels the $n-2$ legs of $\mathcal{M}_\beta$ that are left untouched by the composition.

    Denote the corresponding Choi representations by $\mathsf{T}_\alpha$ and $\mathsf{M}_\beta$. The Choi representation of their composition admits the following partial Kraus representation:
    \begin{equation}
        \begin{split}
            &\textup{Choi}\left(\mathcal{T}_\alpha\circ\mathcal{M}_\beta\right)\\
            &=\sum_{\substack{pq\\p'q'}}\chi_{pqp'q'}^\alpha(K_{pq}^{\gamma}\otimes\mathds{1}^{\gamma_-})\mathsf{M}_\beta({K_{p'q'}^{\gamma}}^\dagger\otimes\mathds{1}^{\gamma_-}).
        \end{split}
    \end{equation}
    
    Here, $\{K_{pq}\}$ is a polarization-operator basis of $\mathcal{L}(\mathcal{H}_\gamma)$, with $p=0,1,\dots,\dim(\mathcal{H}_\gamma)-1$ and $q=-p,\dots,p$. The operators $K_{pq}$ act on the contracted legs, while $\mathds{1}^{\gamma_-}$ acts on the untouched legs of $\mathcal{M}_\beta$. The explicit construction of the polarization-operator basis is given in Appendix~\ref{(QDoD)Appendix:Proofs-1}, although other orthonormal operator bases may also be chosen.
    
    The dependence on $\mathcal{T}_\alpha$ is entirely encoded in the coefficient matrix $\chi^\alpha$, whose entries are determined by its Choi representation $\mathsf{T}_\alpha$ as
    \begin{align}\label{(QDoD) Eq: Definition of chi}
        \chi_{pqp'q'}^\alpha:=\sum_{\substack{i_{\gamma_+}i_\gamma\\j_{\gamma_+}j_\gamma}}\mathsf{T}_{i_{\gamma_+}i_\gamma j_{\gamma_+}j_\gamma}(K_{pq})_{i_{\gamma_+}i_\gamma}^{*}(K_{p'q'})_{j_{\gamma_+}j_\gamma}.
    \end{align}
    Here, $i_\gamma$ and $i_{\gamma_+}$ denote basis indices of $\mathcal{H}_\gamma$ and $\mathcal{H}_{\gamma_+}$, respectively, and the matrix elements are defined by
    \begin{align}
        &(K_{pq})_{i_{\gamma_+}i_\gamma}=\bra{i_{\gamma_+}}K_{pq}\ket{i_\gamma},\\
        &\mathsf{T}_{i_{\gamma_+}i_\gamma j_{\gamma_+}j_\gamma}=\bra{i_{\gamma_+}i_\gamma}\mathsf{T}_\alpha\ket{j_{\gamma_+}j_\gamma},
    \end{align}
    where $\gamma_+\cup\gamma=\alpha$. Since $\mathsf{T}_\alpha$ is positive semidefinite, so is the coefficient matrix $\chi^\alpha$.
    
    For brevity, we introduce the composite indices $\mathfrak{p}:=(pq)$ and $\mathfrak{p}':=(p'q')$.
\end{cor}
\begin{proof}
    The proof is given in Appendix~\ref{(QDoD)Appendix:Proofs-1}.
\end{proof}

Deriving the Kraus representation does not require CP-divisibility or any additional assumptions on the dynamical evolution. Thus, this corollary holds for a general four-legged quantum comb.

Next, let us introduce the coefficients and operators that serve as the building blocks of the QCME.
\begin{defn}\label{(QDoD)Def: path-dependent POB}
    Let $\chi_{\mathfrak{p}\mathfrak{p}'}$ be the polarization operator coefficient of the quantum comb $\mathsf{T}_{\alpha} (\lambda)$ defined in Eq.~(\ref{(QDoD) Eq: Definition of chi}). Then, we define the following functions 
    \begin{align}
        d_{\mathfrak{p}\mathfrak{p}'}(\lambda)& :=\lim_{\epsilon\rightarrow 0^+} \frac{\chi_{\mathfrak{p}\mathfrak{p}'}(\gamma_1 (\lambda) + \epsilon , \gamma_2 (\lambda)  + 0 )}{\epsilon},\label{(QDoD) Eq: POB function d}\\
        e_{\mathfrak{p}\mathfrak{p}'}(\lambda) &:=  \lim_{\epsilon\rightarrow 0^+} \frac{\chi_{\mathfrak{p}\mathfrak{p}'}(\gamma_1(\lambda)  + 0 , \gamma_2(\lambda)  + \epsilon)}{\epsilon}. \label{(QDoD) Eq: POB function e}
    \end{align}
    Using these functions, we define the following path-dependent polarization operator basis 
    \begin{align}
        D_{\alpha} (\lambda) &:= \sum_{\mathfrak{p}}\frac{d_{\mathfrak{p}\mathfrak{0}}(\lambda)}{d_S}(K_{\mathfrak{p}}^{\beta_{+\epsilon}\cap \alpha} \otimes \mathds{1}_{\alpha\setminus\beta_{+\epsilon}}),\label{(QDoD) Eq: path-dependent POB D}\\
        E_{\alpha}(\lambda) &:= \sum_{\mathfrak{p}}\frac{e_{\mathfrak{p}\mathfrak{0}}(\lambda)}{d_S}(K_{\mathfrak{p}}^{\beta^{+\epsilon}\cap \alpha} \otimes \mathds{1}_{\alpha\setminus\beta^{+\epsilon}}),\label{(QDoD) Eq: path-dependent POB E}
    \end{align}
    where $\mathfrak{0}= (p=0 , q=0)$, $d_S$ is a dimension of the open legs and $\{K_{\mathfrak{p}}\}$ are polarization operator basis. The indices are $\alpha = (t^\out,t^\inp,s^\out,s^\inp)$ with $\beta_{+\epsilon}$ and $\beta^{+\epsilon}$ defined in Def.~\ref{(QDoD)Def:CP-divisibleComb}.
\end{defn}

For a semigroup quantum comb, the same functions~(\ref{(QDoD) Eq: POB function d})--(\ref{(QDoD) Eq: POB function e}) and basis~(\ref{(QDoD) Eq: path-dependent POB D})--(\ref{(QDoD) Eq: path-dependent POB E}) can be defined, but with no path dependence.

Finally, we present the two main results of this subsubsection, which give the explicit Choi matrix representation of the QCME.
\begin{thm}\label{Lemma: GKSL form}
Let $\mathcal{L}_{D}(\lambda)$ and $\mathcal{L}_{E}(\lambda)$ be the dynamical decoding and encoding generators defined in Eqs.~\eqref{(QDoD) Eq: defintion of decoding generators} and~\eqref{(QDoD) Eq: defintion of encoding generators}, respectively. For a CP-divisible quantum comb $\mathcal{T}_{\alpha}(\lambda)$ satisfying the local CP-divisibility condition~\eqref{(QDoD)Eq:DefintionForCP-divisbility1} or \eqref{(QDoD)Eq:DefintionForCP-divisbility2}, the action of these generators can be represented by the linear operators $\widetilde{\mathcal{L}}_{D}(\lambda)$ and $\widetilde{\mathcal{L}}_{E}(\lambda)$ acting on its Choi matrix $\mathsf{T}_{\alpha}(\lambda)$:
\begin{align}\label{Eq: Choi representation of the action of generators}
    \operatorname{Choi}\left(\mathcal{L}_{D}(\lambda)\circ\mathcal{T}_{\alpha}(\lambda)\right) &= \widetilde{\mathcal{L}}_{D}(\lambda)\left[\mathsf{T}_{\alpha}(\lambda)\right]\\
    \operatorname{Choi}\left(\mathcal{L}_{E}(\lambda)\circ\mathcal{T}_{\alpha}(\lambda)\right) &= \widetilde{\mathcal{L}}_{E}(\lambda)\left[\mathsf{T}_{\alpha}(\lambda)\right].
\end{align}
The decoding and encoding generators $\widetilde{\mathcal{L}}_{D}(\lambda)$ and $\widetilde{\mathcal{L}}_{E}(\lambda)$ take the following GKSL-like forms:
\begin{align}
    &\widetilde{\mathcal{L}}_{D}\left[\mathsf{T}_{\alpha}\right]:=i\left[\mathsf{T}_{\alpha},H_D\right]+\left\{\delta D_{t^\inp s^\inp}(\lambda)\otimes\mathds{1}_{t^\out s^\out},\mathsf{T}_{\alpha}\right\}\nonumber\\
    &\quad+\sum_{\mathfrak{p}\neq\mathfrak{0},\mathfrak{p}'\neq\mathfrak{0}}d_{\mathfrak{p}\mathfrak{p}'}(\lambda)\Big[K_{\mathfrak{p}}\mathsf{T}_{\alpha}K_{\mathfrak{p}'}^\dagger-\frac{1}{2}\left\{K_{\mathfrak{p}'}^\dagger K_{\mathfrak{p}},\mathsf{T}_{\alpha}\right\}\Big],\label{Eq: GKSL form of decoding generator}\\
    &\widetilde{\mathcal{L}}_{E}\left[\mathsf{T}_{\alpha}\right]:=i\left[\mathsf{T}_{\alpha},H_E\right]+\left\{\delta E_{t^\inp s^\inp}(\lambda)\otimes\mathds{1}_{t^\out s^\out},\mathsf{T}_{\alpha}\right\}\nonumber\\
    &\quad+\sum_{\mathfrak{p}\neq\mathfrak{0},\mathfrak{p}'\neq\mathfrak{0}}e_{\mathfrak{p}\mathfrak{p}'}(\lambda)\Big[K_{\mathfrak{p}}\mathsf{T}_{\alpha}K_{\mathfrak{p}'}^\dagger-\frac{1}{2}\left\{K_{\mathfrak{p}'}^\dagger K_{\mathfrak{p}},\mathsf{T}_{\alpha}\right\}\Big].\label{Eq: GKSL form of encoding generator}
\end{align}
Here, for brevity, we have used the notation $(K_{\mathfrak{p}}^{\beta_{+\epsilon}\cap\alpha}\otimes\mathds{1}^{\alpha\setminus\beta_{+\epsilon}})=K_{\mathfrak{p}}$.

The coefficients $d_{\mathfrak{p}\mathfrak{p}'}(\lambda)$ and $e_{\mathfrak{p}\mathfrak{p}'}(\lambda)$ are defined in Eqs.~\eqref{(QDoD) Eq: POB function d} and~\eqref{(QDoD) Eq: POB function e}, respectively. For the chosen evolution path, the combined generalized \emph{Kossakowski matrix}
\begin{align}
    C_{\mathfrak{p}\mathfrak{p}'}(\lambda):=\dot{\gamma}_1d_{\mathfrak{p}\mathfrak{p}'}(\lambda)+\dot{\gamma}_2e_{\mathfrak{p}\mathfrak{p}'}(\lambda)\label{(QDoD)Eq:Kossakowski}
\end{align}
is required to be positive semidefinite, $C(\lambda)\succeq0$, with respect to the matrix indices $\mathfrak{p}$ and $\mathfrak{p}'$ along the evolution path. We call this condition the \emph{causality constraint of the comb generator}. 

The path-dependent operators $H_D(\lambda)$ and $H_E(\lambda)$ are defined as
\begin{align}
    H_D(\lambda) &:= \frac{1}{2i}\left(D_{\alpha}(\lambda)-D_{\alpha}^\dagger(\lambda)\right),\label{(QDoD)Eq:Path-dependent Hamiltonian D}\\
    H_E(\lambda) &:= \frac{1}{2i}\left(E_{\alpha}(\lambda)-E_{\alpha}^\dagger(\lambda)\right),\label{(QDoD)Eq:Path-dependent Hamiltonian E}
\end{align}
where $D_{\alpha}(\lambda)$ and $E_{\alpha}(\lambda)$ are the path-dependent polarization-operator bases defined in Eqs.~\eqref{(QDoD) Eq: path-dependent POB D} and~\eqref{(QDoD) Eq: path-dependent POB E}, respectively.

We refer to the terms $\delta D_{t^\inp s^\inp}(\lambda)\otimes\mathds{1}_{t^\out s^\out}$ and $\delta E_{t^\inp s^\inp}(\lambda)\otimes\mathds{1}_{t^\out s^\out}$ as the \emph{encoder-decoder correlation terms}. The corresponding operators satisfy
\begin{align}
    \operatorname{Tr}_{t^\inp s^\inp}\left(\delta D_{t^\inp s^\inp}\right)=\operatorname{Tr}_{t^\inp s^\inp}\left(\delta E_{t^\inp s^\inp}\right)=0.
\end{align}
These traceless operators are constrained by the preservation of the generalized trace-preservation conditions under the full comb generator. Their explicit forms remain unspecified in the axiomatic treatment, while the microscopic derivation in Sec.~\ref{sec:quantumdynamicsofdynamics-microscopic} provides a concrete realization via Eq.~\eqref{(QDoD): encoder decoder mixing contribution}.
\end{thm}
\begin{proof}
    The proof is given in Appendix~\ref{(QDoD)Appendix:Proofs-2}.
\end{proof}
Note that the causality constraint on the comb generator is not required to derive the GKSL-like form in the above theorem. In principle, this form can be obtained without imposing positive semidefiniteness on the generalized Kossakowski matrix $C_{\mathfrak{p}\mathfrak{p}'}$. However, as we show in the next subsubsection, this additional condition ensures the complete positivity of the evolution generated by these generators. Together with the generalized trace-preservation conditions, this ensures that the resulting quantum comb remains compatible with the prescribed temporal ordering of its four legs. This motivates our terminology ``causality constraint''.

\begin{figure}[h]
    \centering
    \includegraphics[width=\linewidth]{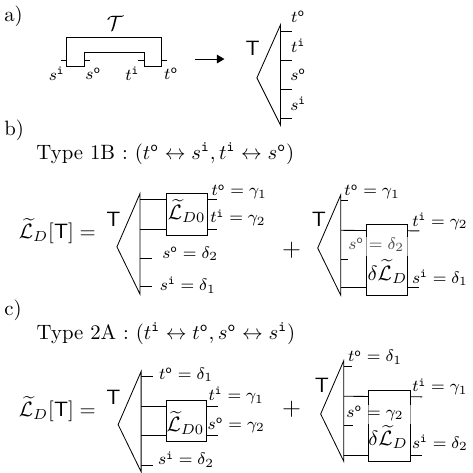}
    \caption{(a) Choi representation of the quantum comb. (b) and (c) Pictorial representations of the action of the decoding generator on quantum combs with evolution types 1B and 2A, respectively. Evolution type 1B has the contraction rule $(t^\out \leftrightarrow s^\inp,\,t^\inp \leftrightarrow s^\out)$, while type 2A has the contraction rule $(t^\inp \leftrightarrow t^\out,\,s^\out \leftrightarrow s^\inp)$. Note that type 1B is the evolution type considered in Example~\ref{(QDoD) Eg: evolution of channel}.}
    \label{fig:EncodingDecodingGenerators}
\end{figure}

Furthermore, to clarify why we call certain contributions to the generators ``correlation terms'', let us examine how they act on the quantum comb. For brevity, we introduce the following notation:
\begin{align}
    \widetilde{\mathcal{L}}_{D}[\mathsf{T}] &= \widetilde{\mathcal{L}}_{D0}[\mathsf{T}] + \delta \widetilde{\mathcal{L}}_{D}[\mathsf{T}],\\
    \widetilde{\mathcal{L}}_{E}[\mathsf{T}] &= \widetilde{\mathcal{L}}_{E0}[\mathsf{T}] + \delta \widetilde{\mathcal{L}}_{E}[\mathsf{T}],
\end{align}
where the first terms on the right-hand side represent the uncorrelated contributions, while the second terms represent the anticommutator contributions involving the encoder-decoder correlation terms. In Fig.~\ref{fig:EncodingDecodingGenerators}, we illustrate how these contributions act on quantum combs for two different evolution types. Notice that the uncorrelated contributions act only on the evolving legs, indicated by $\gamma_1$ and $\gamma_2$ in the figure, whereas the encoder--decoder correlation contributions connect the evolving legs $(\gamma_1,\gamma_2)$ with the initial-condition legs $(\delta_1,\delta_2)$. The same distinction applies to the encoding generator. This motivates our terminology.

Finally, note that if the decoder/encoder-correlation terms are zero, we recover the standard GKSL form for both decoding/encoding generators. This is explicitly stated in Appendix~\ref{(QDoD)Apx:GKSL} with proof. 

Within the present axiomatic approach, the explicit forms of the encoder--decoder correlation terms are not uniquely determined. Unlike the conventional GKSL equation, which contains no analogous terms, the QCME explicitly accounts for correlations between the encoder and decoder. We therefore introduce a canonical representation of the QCME in which these correlation terms are absorbed into Hermitian operators, without eliminating their contributions to the dynamics.
\begin{cor}\label{Alternative form of GKSL}
    The dynamical decoding/encoding generators can also be written as 
    \begin{align}
        \widetilde{\mathcal{L}}_{D}[\mathsf{T}_\alpha] &= i \left[ \mathsf{T}_{\alpha},H_{D} \right] + \left\{G_{D}, \mathsf{T}_{\alpha},\right\}\nonumber\\
        &+ \sum_{\mathfrak{p}\not=\mathfrak{0},\mathfrak{p}'\not=\mathfrak{0}} d_{\mathfrak{p}\mathfrak{p}'}K_\mathfrak{p} \mathsf{T}_{\alpha} K^{\dagger}_{\mathfrak{p}'},\label{Eq: Alternative GKSL form of decoding generator}\\
        \widetilde{\mathcal{L}}_{E}[\mathsf{T}_\alpha] &= i \left[ \mathsf{T}_{\alpha},H_{E} \right] + \left\{G_{E}, \mathsf{T}_{\alpha},\right\}\nonumber\\
        &+ \sum_{\mathfrak{p}\not=\mathfrak{0},\mathfrak{p}'\not=\mathfrak{0}} e_{\mathfrak{p}\mathfrak{p}'}K_\mathfrak{p} \mathsf{T}_{\alpha} K^{\dagger}_{\mathfrak{p}'},\label{Eq: Alternative GKSL form of encoding generator}
    \end{align}
    where path-dependent generators are defined as
    \begin{align}
        G_{D} (\lambda) &= \frac{1}{2} (D_{\alpha}(\lambda) + {D_{\alpha}(\lambda) }^\dagger) ,\label{(QDoD)Eq:Path-dependent G operator D}\\
        G_{E} (\lambda)& = \frac{1}{2} (E_{\alpha}(\lambda)  + {E_{\alpha}(\lambda) }^\dagger).\label{(QDoD)Eq:Path-dependent G operator E}
    \end{align}
\end{cor}
\begin{proof}
    The result follows from the proof of Theorem~\ref{Lemma: GKSL form}.
\end{proof}

\subsubsection{Quantum Dynamics of Dynamics}\label{(QDoD)Sec:QDoD5}
In this subsubsection, we present a formal solution to the QCME~\eqref{(QDoD)Eq:CombME} and state the main result of this paper, Theorem~\ref{Thm: main result}, which establishes that a quantum comb is CP-divisible if and only if its decoding and encoding generators take the GKSL-like forms given in Theorem~\ref{Lemma: GKSL form} (or equivalently, the alternative forms in Corollary~\ref{Alternative form of GKSL}).

\begin{lem}
    Let $\vec{\gamma}(\lambda):=(\gamma_1(\lambda),\gamma_2(\lambda))$ denote a path in the two-dimensional time plane, and let $\mathcal{T}_{\alpha}(\lambda)$ be a CP-divisible quantum comb satisfying the local CP-divisibility condition. The formal solution to the QCME~\eqref{(QDoD)Eq:CombME} is given by the path-ordered exponential
\begin{align}
    \mathcal{T}_{\alpha}(\lambda) &= \mathcal{P}\operatorname{exp}\left(\int^{\vec{\gamma}(\lambda)}_{\vec{\gamma}(0)}\mathcal{L}_{D}\,d\gamma_1+\mathcal{L}_{E}\,d\gamma_2\right)\circ\mathcal{T}_{\alpha}(0)\label{Eq: Formal solution to the covariant GKSL eq}\\
    &= \mathcal{P}\operatorname{exp}\left(\int^{\lambda}_{0}d\lambda'\,\mathcal{L}(\lambda')\right)\circ\mathcal{T}_{\alpha}(0),\label{(QDoD) Eq: solution to comb master equation}
\end{align}
where $\mathcal{L}(\lambda)$ is the quantum comb generator defined in Eq.~\eqref{(QDoD) Eq: Quantum comb generator}. The evolution path $\vec{\gamma}(\lambda)$ is chosen such that the generalized Kossakowski matrix remains positive semidefinite along the path.

    In terms of the Choi matrix representation, the above equation can be written as 
\begin{align}
    \mathsf{T}_{\alpha} &= \mathcal{P}\text{exp} \left(\int_{\vec{\gamma}(0)}^{\vec{\gamma}(\lambda)} \widetilde{\mathcal{L}}_{D} d\gamma_1  + \widetilde{\mathcal{L}}_{E} d\gamma_2 \right)\circ \mathsf{T}_{\alpha}\\
    & = \mathcal{P}\text{exp} \left(\int^{\lambda}_{0} d\lambda' \widetilde{\mathcal{L}}(\lambda') \right) \circ\mathcal{T}_{\alpha}(0)
\end{align}
where $\mathsf{T}_\alpha$ is a Choi-representation of the quantum comb and $\widetilde{\mathcal{L}}_{D}$ and $\widetilde{\mathcal{L}}_{D}$ takes the GKSL-like form shown in Eq.~(\ref{Eq: GKSL form of decoding generator})--(\ref{Eq: GKSL form of encoding generator}).
\end{lem}

\begin{proof}
    Let us begin with the QCME 
    \begin{align}
        \frac{\partial \mathcal{T}_{\alpha}(\lambda)}{\partial\lambda}  = \mathcal{L} (\lambda) \circ \mathcal{T}_{\alpha} (\lambda).
    \end{align}
    Integrating both sides with respect to $\lambda$, we obtain 
    \begin{align}
        \mathcal{T}_{\alpha}(\lambda) = \mathcal{T}_{\alpha}(0)  + \int_0^{\lambda} d\lambda'  \mathcal{L} (\lambda') \circ \mathcal{T}_{\alpha} (\lambda').\nonumber
    \end{align}
    The integral equation can be solved iteratively by recursively substituting $\mathcal{T}_{\alpha}(\lambda)$ into itself, thereby generating a sequence of nested time integrals. 
    \begin{align}
        &\mathcal{T}_\alpha (\lambda) = \mathcal{T}_\alpha (0) + \int_0^{\lambda} d\lambda' \mathcal{L}(\lambda') \circ \mathcal{T}_{\alpha} (0)\nonumber\\
        &= \int_0^{\lambda} d \lambda' \int_0^{\lambda'} d\lambda'' \mathcal{L}(\lambda')\circ \mathcal{L}(\lambda'')\circ \mathcal{T}_\alpha (0) + \dots \nonumber\\
        &= \mathcal{P}\text{exp} \left(\int^{\lambda}_{0} d\lambda' \mathcal{L}(\lambda') \right) \circ\mathcal{T}_{\alpha}(0).\nonumber
    \end{align}
    where the path-ordering exponential is defined as 
    \begin{align}
        \mathcal{P} \text{exp} \left(\int_0^{\lambda} d\lambda' \mathcal{A}(\lambda)\right)\circ \mathcal{T}:= \mathcal{T} + \int_0^{\lambda} d \lambda' \mathcal{A}(\lambda) \circ \mathcal{T} &\nonumber\\
        + \int_0^{\lambda} d\lambda' \int_0^{\lambda'}d \lambda''  \mathcal{A}(\lambda') \circ \mathcal{A} (\lambda'') \circ \mathcal{T} + \dots  &
    \end{align}
    for some linear opeator $\mathcal{A}$ acting on $\mathcal{M}$. 
\end{proof}
Notice that for the semigroup quantum comb, the path-dependence of the decoding/encoding generators drops, then the integral in the exponent of Eq.~(\ref{(QDoD) Eq: solution to comb master equation}) can be evaluated and the solution takes the following form 
\begin{align}
    \mathcal{T}_{\alpha}(\lambda) = \text{exp}\bigg[\left(\gamma_1(\lambda) - \gamma_1(0)\right)\mathcal{L}_{D}&\nonumber\\
    +\left(\gamma_2(\lambda) - \gamma_2(0)\right)\mathcal{L}_{E}\bigg]\circ \mathcal{T}(0)&
\end{align}
where $\mathcal{L}_{D}$ and $\mathcal{L}_{E}$ are path-independent versions of the decoding/encoding generators.

Finally, we state our main result:
\begin{thm}[Quantum dynamics of dynamics]\label{Thm: main result}
    Let $\mathcal{T}_\alpha (\lambda)$ be a path-dependent quantum comb. Let $C(\lambda)$ be its corresponding generalized Kossakowski matrix~\eqref{(QDoD)Eq:Kossakowski}. Then $\mathcal{T}_\alpha$ is a CP-divisible quantum comb satisfying the local CP-divisibility condition with a positive semidefinite generalized Kossakowski matrix if and only if it solves the QCME~\eqref{(QDoD)Eq:CombME}. 
\end{thm}
\begin{proof}
    The proof is given in Appendix~\ref{(QDoD)Appendix:Proofs-3}.
\end{proof}

Note that a similar theorem holds for a semigroup quantum comb with path-independent generators.

This completes our axiomatic formalism for CP-divisible quantum combs and their related master equations. In the next section, we develop a microscopic approach to the same problem of infinitesimal quantum-comb evolution. Before moving on, however, we conclude this section with a simple example illustrating how our formalism departs substantially from the conventional GKSL treatment.

\subsection{A Simple Example: Correlated Schr\"odinger and Heisenberg Picture Evolutions.}
To provide intuition for our formalism, we consider an example that illustrates how our approach differs from the conventional GKSL treatment. 
\begin{example}(Correlation of Schr\"odinger and Heisenberg picture evolutions.)\label{(QDoD)eg:SchHeiCorr}
    The aim of this example is to show that both Schr\"odinger- and Heisenberg-picture evolutions arise from our formalism. Furthermore, we show that our formalism also yields a correlation between these evolutions.

    To see this correlation, let us consider the temporal ordering $t^\out>t^\inp>s^\out>s^\inp$. An experimentalist first prepares a state $\pi^S(s^\inp)$ of the system at $s^\inp$, allows it to evolve under the system-environment dynamics of the physical sample, and then measures it at $s^\out$ with observable $A^S$. The experimentalist then prepares another state $\rho^S (t^\inp)$ at $t^\inp$, which evolves to $t^\out$. This process can be regarded as a map from $s^\inp$ to $t^\out$, with a observe-and-prepare operation at the intermediate times $s^\out$ and $t^\inp$. This is illustrated in Fig.~\ref{(QDoD)Fig:SHcorr}(a).
    \begin{figure}
        \centering
        \includegraphics[width=\linewidth]{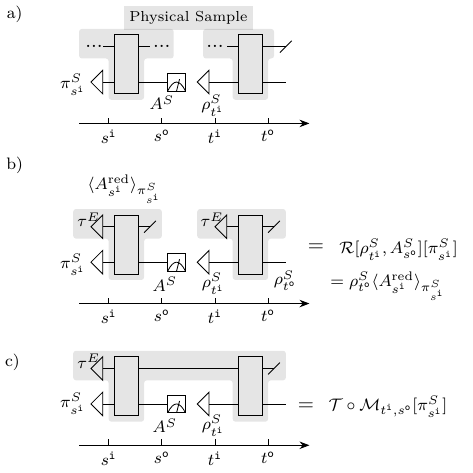}
        \caption{(a) Experimental setup of Example~\ref{(QDoD)eg:SchHeiCorr}. (b) Memoryless case of the process. (c) Memoryful case of the process.}
        \label{(QDoD)Fig:SHcorr}
    \end{figure}
    
    Next, let us suppose that the environment of the physical sample carries no memory between the two evolution intervals $(s^\out,t^\inp)$. In particular, we assume that the environment is initially in the state $\tau^E$ and remains approximately in this state throughout the dynamics. This memoryless case can then be expressed as follows:
    \begin{equation}
        \begin{split}
        &\mathcal{R}[\rho^S_{t^\inp},A^S_{s^\out}] [\pi^S_{s^\inp}]
        \\
        &=
        \operatorname{Tr}_E\Big(
        \mathcal{U}_{t^\out,t^\inp}
        [\rho^S_{t^\inp}\otimes\tau^E]
        \Big)
        \\
        &\quad\times
        \operatorname{Tr}_{SE}\Big(
        (A^S_{s^\out}\otimes\mathds{1}^E)
        \mathcal{U}_{s^\out,s^\inp}
        [\pi^S_{s^\inp}\otimes\tau^E]
        \Big)
        \\
        &=
        \mathcal{E}_{t^\out,t^\inp}[\rho^S_{t^\inp}]
        \operatorname{Tr}_S\Big(
        \mathcal{E}^*_{s^\out,s^\inp}[A^S_{s^\out}]
        \pi^S_{s^\inp}
        \Big) = \rho^S_{t^\out}\langle A^{\text{red}}_{s^\inp}\rangle _{\pi^S_{s^\inp}}.\label{(QDoD)Eq:Eg5:Step1}
        \end{split}
    \end{equation}
    The reduced CPTP map and its adjoint are denoted by $\mathcal{E}_{a,b}$ and $\mathcal{E}_{a,b}^*$, respectively, where
    \begin{align}
        \mathcal{E}_{a,b}[\bullet]
        :=
        \operatorname{Tr}_E\Big(
        \mathcal{U}_{a,b}[\bullet\otimes\tau^E]
        \Big).
    \end{align}
    As discussed in Example~\ref{(QDoD) Fundamental example 2}, the adjoint of the CPTP map induces the time evolution of the reduced Heisenberg-picture operator~\cite{breuer1997stochastic,settimo2026divisibility}:
    \begin{align}
        A^{\mathrm{red}}_{s^\inp}
        =
        \mathcal{E}^*_{s^\out,s^\inp}[A^S_{s^\out}].
    \end{align}
    From Eq.~\eqref{(QDoD)Eq:Eg5:Step1}, we see that the output of this process is the Schr\"odinger-picture evolution of the state prepared at $t^\inp$, $\mathcal{E}_{t^\out,t^\inp}[\rho^S_{t^\inp}]$, weighted by the expectation value of the reduced Heisenberg-picture operator $\mathcal{E}^*_{s^\out,s^\inp}[A^S_{s^\out}]$ in the initial state $\pi^S_{s^\inp}$. Thus, the Schr\"odinger- and Heisenberg-picture evolutions appear as independent separate factors in this expression. This process is illustrated in Fig.~\ref{(QDoD)Fig:SHcorr}(b)

    We now ask what happens if the environment retains information from before the measurement and carries it into the evolution after state preparation. This process is shown in Fig.~\ref{(QDoD)Fig:SHcorr}(c). Although the system is prepared in a fixed state at $t^\inp$, the environment can retain information about the preparation. To see how this can give rise to correlations between two previously derived evolutions, let us consider the following quantum comb with the same temporal ordering.
    \begin{equation}
        \begin{split}
             \mathcal{T}\circ \mathcal{M}_{t^\inp s^\out}[\pi^S_{s^\inp}] = \operatorname{Tr}_E \Big[\mathcal{U}_{t^\out t^\inp}\circ (\mathcal{M}_{t^\inp s^\out} \otimes \mathcal{I}^E)\\
             \circ \, \mathcal{U}_{s^\out s^\inp}[\pi^S_{s^\inp} \otimes \tau^E] \Big]
        \end{split}
    \end{equation}
    where the observe-and-prepare map is defined by
    \begin{align}
        \mathcal{M}_{t^\inp s^\out}[\bullet_{s^\out}]
        :=
        \operatorname{Tr}_S
        (A^S_{s^\out}\bullet_{s^\out})
        \rho^S_{t^\inp}.
    \end{align}
    Its Choi representation is given by 
    \begin{align}
        \mathsf{M}_{t^\inp s^\out}
        =
        \rho^S_{t^\inp}\otimes
        (A^S_{s^\out})^{\mathsf T}.
    \end{align}
    Unlike the factorized expression for $\mathcal{R}$, the two interactions in $\mathcal{T}$ share the same environment. 

    Now that we have established the physical setup and the quantum comb we wish to evolve, our next task is to solve the QCME. To this end, we identify the evolving legs as $\gamma_1=t^\out$ and $\gamma_2=s^\inp$, with the observe-and-prepare map $\mathcal{M}_{t^\inp s^\out}$ as the initial condition. We then consider the QCME:
    \begin{equation}
        \begin{split}
            \frac{d\mathsf{T}_{\alpha}(\lambda) }{d\lambda}=&
        \dot{\gamma}_1 \widetilde{\mathcal{L}}_D [\mathsf{T}_{\alpha}(\lambda)] + \dot{\gamma}_2 \widetilde{\mathcal{L}}_E [\mathsf{T}_{\alpha}(\lambda)],\\
        =& i [\mathsf{T}_\alpha,\dot{\gamma}_1H_D + \dot{\gamma}_2 H_E] + \{\dot{\gamma}_1 G_D+\dot{\gamma}_2 G_E, \mathsf{T}_\alpha\}\\
        & + \sum_{\mathfrak{p}\not=\mathfrak{0}, \mathfrak{p}'\not=\mathfrak{0}}C_{\mathfrak{p}\mathfrak{p}'}K_{\mathfrak{p}}\mathsf{T}_\alpha K_{\mathfrak{p}'}^\dagger,\label{(QDoD)Eq:SHcorr}
        \end{split}
    \end{equation}
    where $C_{\mathfrak{p}\mathfrak{p}'}=\dot{\gamma}_1d_{\mathfrak{p}\mathfrak{p}'}+\dot{\gamma}_2e_{\mathfrak{p}\mathfrak{p}'}$ is a generalized Kossakowski matrix. Here, we have adopted the canonical forms given in Corollary~\ref{Alternative form of GKSL}, which incorporate the correlation terms without requiring their explicit expressions.

    Before proceeding to the details, we first outline our approach. In the conventional axiomatic GKSL formulation, one specifies
    \begin{itemize}
        \item[(a)] the system Hamiltonian,
        \item[(b)] the Lindblad (jump) operators, and
        \item[(c)] the decay rates, or more generally, the Kossakowski matrix.
    \end{itemize}
    Analogously, in our axiomatic formalism, we specify the corresponding building blocks of the QCME:
    \begin{itemize}
        \item[(a)] the operators $H_D$, $H_E$, $G_D$, and $G_E$,
        \item[(b)] the path-dependent polarization-operator bases (POBs) defined in Eqs.~\eqref{(QDoD) Eq: path-dependent POB D} and~\eqref{(QDoD) Eq: path-dependent POB E}, and
        \item[(c)] the generalized Kossakowski matrix $C_{\mathfrak{p}\mathfrak{p}'}$.
    \end{itemize}
    Once we obtain the Choi matrix representation of the generator, we convert it back to the linear-map representation, allowing us to separate the resulting expression into the familiar GKSL contributions and the encoder-decoder correlation contributions. For clarity of presentation, we discuss these ingredients in the order (b), (c), and (a), rather than following the order listed above.

    \textbf{(b)}: Let us begin by choosing the operators used to construct the path-dependent POBs. Let $\{L_n^{\out}\}_{n}$ be a family of operators acting on $\mathcal{H}_{t^\out}$, and let $\{L_m^{\inp}\}_{m}$ be a family of operators acting on $\mathcal{H}_{s^\inp}$. These are single-leg operators, which can be extended to the two-leg Hilbert space and expanded in the POB as follows:
    \begin{align}
        L_n^{\out}\otimes\mathds{1}^{s^\inp}&=\sum_{\mathfrak p\ne\mathfrak 0}a_n^{\mathfrak p}K_{\mathfrak p},\label{(QDoD) MixingExample1-1}\\
        -\mathds{1}^{t^\out}\otimes(L_m^\inp)^{\mathsf T}&=\sum_{\mathfrak p\ne\mathfrak 0}b_m^{\mathfrak p}K_{\mathfrak p}.\label{(QDoD) MixingExample1-2}
    \end{align}
    The minus sign and transpose in the second equation are introduced by convention for later convenience. The coefficients are obtained using the orthonormality relation~\eqref{(QDoD)Proof:Kraus rep. setp 7} of the polarization-operator basis:
    \begin{align}
        a_n^{\mathfrak p} &= \operatorname{Tr}\Big[K_{\mathfrak p}^{\dagger}(L_n^\out\otimes\mathds{1})\Big],\\
        b_m^{\mathfrak p} &= -\operatorname{Tr}\Big[K_{\mathfrak p}^{\dagger}(\mathds{1}\otimes(L_m^\inp)^{\mathsf T})\Big].
    \end{align}
    Here, the traces are taken over the two evolving Hilbert-space factors.

    \textbf{(c)}: Next, we introduce the block matrix
    \begin{align}
        \Gamma:=\begin{pmatrix}\gamma^{\out D\out D}&\gamma^{\out D\inp E}\\(\gamma^{\out D\inp E})^\dagger&\gamma^{\inp E\inp E}\end{pmatrix},
    \end{align}
    where $\gamma^{\out D\out D}$, $\gamma^{\inp E\inp E}$, and $\gamma^{\out D\inp E}$ are coefficient matrices with entries $\gamma^{\out D\out D}_{nm}$, $\gamma^{\inp E\inp E}_{nm}$, and $\gamma^{\out D\inp E}_{nm}$, respectively. We also define the coefficient vectors $\vec{v}_{\mathfrak{p}}:=(\vec{a}^{\mathfrak{p}},\vec{b}^{\mathfrak{p}})^\mathsf{T}$, with $\vec{a}^{\mathfrak{p}}=(a_n^{\mathfrak{p}})_n$ and $\vec{b}^{\mathfrak{p}}=(b_m^{\mathfrak{p}})_m$.
    We choose $\Gamma$ to be positive semidefinite and construct the generalized Kossakowski matrix as
    \begin{align}
        C_{\mathfrak{p}\mathfrak{p}'}:=\vec{v}_{\mathfrak{p}'}^\dagger\Gamma\vec{v}_{\mathfrak{p}},
    \end{align}
    which is therefore also positive semidefinite. Expanding this expression yields
    \begin{equation}
        \begin{split}
            C_{\mathfrak{p}\mathfrak{p}'} &= \sum_{n,m}\gamma^{\out D\out D}_{nm}a_m^{\mathfrak p}(a_n^{\mathfrak p'})^*+\sum_{n,m}\gamma^{\inp E\inp E}_{nm}b_m^{\mathfrak p}(b_n^{\mathfrak p'})^*\\
            &\quad+\sum_{n,m}\Big[\gamma^{\out D\inp E}_{nm}b_m^{\mathfrak p}(a_n^{\mathfrak p'})^*+\bigl(\gamma^{\out D\inp E}_{nm}\bigr)^*a_n^{\mathfrak p}(b_m^{\mathfrak p'})^*\Big].
        \end{split}
        \label{Eq: Example coefficient restriction}
    \end{equation}

    Substituting the coefficient Eq.~\eqref{Eq: Example coefficient restriction} into the $K_{\mathfrak p}\mathsf{T}_{\alpha}K_{\mathfrak p'}^\dagger$ term of the QCME and using the expansions~(\ref{(QDoD) MixingExample1-1}) and (\ref{(QDoD) MixingExample1-2}), we obtain
    \begin{equation}
        \begin{split}
            &\sum_{\substack{\mathfrak p\ne\mathfrak 0\\
            \mathfrak p'\ne\mathfrak 0}}C_{\mathfrak{p}\mathfrak{p}'}K_{\mathfrak p}\mathsf{T}_{\alpha}K_{\mathfrak p'}^\dagger\\
            &=\sum_{n,m}\gamma^{\out D\out D}_{nm}(L_m^\out\otimes\mathds{1}_{s^\inp})\mathsf{T}_{\alpha}(L_n^{\out\dagger}\otimes\mathds{1}_{s^\inp})\\
            &\quad+\sum_{n,m}\gamma^{\inp E\inp E}_{nm}(\mathds{1}_{t^\out}\otimes(L_m^\inp)^{\mathsf T})\mathsf{T}_{\alpha}(\mathds{1}_{t^\out}\otimes L_n^{\inp *})\\
            &\quad-\sum_{n,m}\Big[\gamma^{\out D\inp E}_{nm}(\mathds{1}_{t^\out}\otimes(L_m^\inp)^{\mathsf T})\mathsf{T}_{\alpha}(L_n^{\out\dagger}\otimes\mathds{1}_{s^\inp})\\
            &\hspace{24mm}+\bigl(\gamma^{\out D\inp E}_{nm}\bigr)^*(L_n^\out\otimes\mathds{1}_{s^\inp})\mathsf{T}_{\alpha}(\mathds{1}_{t^\out}\otimes L_m^{\inp *})\Big].\label{Eq: Example jump contribution}
        \end{split}
    \end{equation}
    Next, let us specify the commutator and anticommutator terms of Eq.~\eqref{(QDoD)Eq:SHcorr}

    \textbf{(a)}: Next, we choose the operators $H_D$, $H_E$, $G_D$, and $G_E$, analogous to specifying the Hamiltonian in the conventional GKSL equation. Combining the commutator and anticommutator contributions, we introduce the following operator:
    \begin{align}
        A:=\dot{\gamma}_1(G_D-iH_D)+\dot{\gamma}_2(G_E-iH_E),
    \end{align}
    such that
    \begin{equation}
        \begin{split}
            &i[\mathsf{T}_\alpha,\dot{\gamma}_1H_D+\dot{\gamma}_2H_E]+\{\dot{\gamma}_1G_D+\dot{\gamma}_2G_E,\mathsf{T}_\alpha\}\\
        &=A\mathsf{T}_\alpha+\mathsf{T}_\alpha A^\dagger.
        \end{split}
    \end{equation}
    For this example, let us choose a particular form of $A$ as 
    \begin{equation}
    \begin{split}
        A&=\left(-iH^{\out}-\frac12\sum_{n,m}\gamma^{\out D\out D}_{nm}L_n^{\out\dagger}L_m^\out\right)\otimes\mathds{1}\\
        &\quad+\mathds{1}\otimes\left(iH^{\inp}-\frac12\sum_{n,m}\gamma^{\inp E\inp E}_{nm}L_n^{\inp\dagger}L_m^\inp\right)^{\mathsf T}\\
        &\quad+\sum_{n,m}\gamma^{\out D\inp E}_{nm}L_n^{\out\dagger}\otimes(L_m^\inp)^{\mathsf T}.
        \label{Eq: Example drift restriction}
    \end{split}
\end{equation}
    Here $H^{\out}$ and $H^{\inp}$ are Hermitian operators on $\mathcal{H}_{t^\out}$ and
    $\mathcal{H}_{s^\inp}$, respectively. They specify the local Hamiltonian contributions in this example. We suppress the $\lambda$ dependence for brevity.

    We next translate these expressions into the operator
    representation. First, let us recall that for an arbitrary
    linear map
    $\mathcal{E}:\mathcal{L}(\mathcal{H}_{s^\inp})
    \rightarrow\mathcal{L}(\mathcal{H}_{t^\out})$
    and operators
    $A,B\in\mathcal{L}(\mathcal{H}_{t^\out})$ and
    $C,D\in\mathcal{L}(\mathcal{H}_{s^\inp})$,
    the following relation holds:
    \begin{align}
        &(A\otimes C^{\mathsf T})
        \operatorname{Choi}\big(
        \mathcal{E}[\bullet_{s^\inp}]
        \big)
        (B\otimes D^{\mathsf T})
        \nonumber\\
        &\qquad=
        \operatorname{Choi}\Big(
        A\mathcal{E}[C\bullet_{s^\inp}D]B
        \Big).
        \label{Eq: Example Choi to channel identity}
    \end{align}
    Here the transpose is taken in the basis used to define
    the Choi representation.

    Since  $\mathcal{T}_{\alpha}\circ\mathcal{M}: \mathcal{L}(\mathcal{H}_{s^\inp})\rightarrow\mathcal{L}(\mathcal{H}_{t^\out})$, we can apply this identity to the contracted map. The operators $L_n^\out$ and $L_m^\inp$ act only on the evolving legs $t^\out$ and $s^\inp$, respectively. Therefore, their action commutes with the contraction of the initial-condition legs $t^\inp$ and $s^\out$ with $\mathcal{M}$. The same holds for the Hamiltonian operators $H^\out$ and $H^\inp$.

    We can thus first contract
    Eqs.~\eqref{Eq: Example jump contribution}
    and~\eqref{Eq: Example drift restriction} with
    $\mathcal{M}$, and then apply
    Eq.~\eqref{Eq: Example Choi to channel identity}.
    Combining the two diagonal contributions with their
    corresponding local drift terms gives
    \begin{align}
        &\mathcal{L}_D\Big[
        (\mathcal{T}_{\alpha}\circ\mathcal{M})
        [\bullet_{s^\inp}]
        \Big]
        +
        (\mathcal{T}_{\alpha}\circ\mathcal{M})
        \Big[\mathcal{L}_E[\bullet_{s^\inp}]\Big],
    \end{align}
    where the single-system generators are
    \begin{align}
        &\begin{split}
        \mathcal{L}_D[\bullet]
        =
        -i[H^\out,\bullet]
        +
        \sum_{n,m}\gamma^{\out D\out D}_{nm}
        \Big(
        L_m^\out\bullet L_n^{\out\dagger}
        \\
        -\frac12
        \{L_n^{\out\dagger}L_m^\out,\bullet\}
        \Big),
        \label{(QDoD) Eq:MixingExample2-1}
        \end{split}\\
        &\begin{split}
        \mathcal{L}_E[\bullet]
        =
        +i[H^\inp,\bullet]
        +
        \sum_{n,m}\gamma^{\inp E\inp E}_{nm}
        \Big(
        L_m^\inp\bullet L_n^{\inp\dagger}
        \\
        -\frac12
        \{L_n^{\inp\dagger}L_m^\inp,\bullet\}
        \Big).
        \label{(QDoD) Eq:MixingExample2-2}
        \end{split}
    \end{align}
    In this example, these represent the reduced actions of the decoding and encoding generators on the output and input legs, respectively. Since the block matrix $\Gamma$ is positive semidefinite, its diagonal blocks are also positive semidefinite, ensuring that the above expressions indeed take the standard GKSL form.

    To see the correlated contribution explicitly, the term
    proportional to $\gamma^{\out D\inp E}_{nm}$ in
    Eq.~\eqref{Eq: Example jump contribution} becomes
    \begin{align}
        -\gamma^{\out D\inp E}_{nm}
        (\mathcal{T}_{\alpha}\circ\mathcal{M})
        [L_m^\inp\bullet]L_n^{\out\dagger},
    \end{align}
    while the corresponding term in
    Eq.~\eqref{Eq: Example drift restriction} becomes
    \begin{align}
        \gamma^{\out D\inp E}_{nm}
        L_n^{\out\dagger}
        (\mathcal{T}_{\alpha}\circ\mathcal{M})
        [L_m^\inp\bullet].
    \end{align}
    Their sum is therefore a commutator.
    Combining all contributions, our QCME
    takes the four-term form
    \begin{equation}
        \begin{split}
        &\frac{d}{d\lambda}(\mathcal{T}_{\alpha}\circ\mathcal{M})[\bullet]\\
        &=\mathcal{L}_D\Big[(\mathcal{T}_{\alpha}\circ\mathcal{M})[\bullet]\Big]+(\mathcal{T}_{\alpha}\circ\mathcal{M})\Big[\mathcal{L}_E[\bullet]\Big]\\
        &\quad+\sum_{n,m}\gamma^{\out D\inp E}_{nm}\Big[L_n^{\out\dagger},(\mathcal{T}_{\alpha}\circ\mathcal{M})[L_m^\inp\bullet]\Big]\\
        &\quad+\sum_{n,m}\bigl(\gamma^{\out D\inp E}_{nm}\bigr)^*\Big[(\mathcal{T}_{\alpha}\circ\mathcal{M})[\bullet L_m^{\inp\dagger}],L_n^\out\Big].\label{Eq: Example mixed operator evolution}
        \end{split}
    \end{equation}
    Here $\bullet$ belongs to
    $\mathcal{L}(\mathcal{H}_{s^\inp})$.

    To see clearly how the above equation contains both Schrödinger and Heisenberg pictures, as well as a correlation of the two, set $\gamma^{\out D\inp E}_{nm}=0$. Starting from the observe-and-prepare initial condition, the solution retains the form
    \begin{align}
        (\mathcal{T}_{\alpha}\circ\mathcal{M})
        [\bullet]
        =
        \rho^S
        \operatorname{Tr}_S
        \bigl(A^{\mathrm{red}}\bullet\bigr).
        \label{Eq: Example factorized evolution}
    \end{align}
    From this first two terms of Eq.~(\ref{Eq: Example mixed operator evolution}) becomes 
    \begin{align}
        &\begin{split}
            \mathcal{L}_D\Big[(\mathcal{T}_{\alpha}\circ\mathcal{M})[\bullet]\Big]=& \mathcal{L}_D\Big[\rho^S \operatorname{Tr}_S (A^{\text{red}}\bullet)\Big]\\
            =&\operatorname{Tr}_S (A^{\text{red}}\bullet)  \mathcal{L}_D\big[\rho^S \big] 
        \end{split}\\
        &\begin{split}
        (\mathcal{T}_{\alpha}\circ\mathcal{M})\Big[\mathcal{L}_E[\bullet]\Big]=& \rho^S \operatorname{Tr}_S \Big(A^\text{red} \mathcal{L}_E [\bullet]\Big) \\
        =&  \rho^S \operatorname{Tr}_S \Big(\mathcal{L}_E^*[A^\text{red}]  \bullet \Big) 
        \end{split}
    \end{align}
    where in the last line, we used $\operatorname{Tr}_S\bigl(A^{\mathrm{red}}\mathcal{L}_E[\bullet]\bigr)=\operatorname{Tr}_S \bigl(\mathcal{L}_E^*[A^{\mathrm{red}}]\bullet\bigr)$. These are the GKSL and adjoint-GKSL equations acting on the state and reduced observable, respectively. This recovers the independent state and observable evolutions illustrated by $\mathcal{R}$.

    We now retain $\gamma^{\out D\inp E}_{nm}$. Returning to the Choi representation, the initial condition (observe--and--prepare) is 
    $\rho^S\otimes(A^{\mathrm{red}})^{\mathsf T}$. Applying this to the explicit Choi represenation Eq.~\eqref{Eq: Example jump contribution} and \eqref{Eq: Example drift restriction}, gives
    \begin{align}
        &\frac{d}{d\lambda}
        \operatorname{Choi}
        (\mathcal{T}_{\alpha}\circ\mathcal{M})
        \nonumber\\
        &=
        \mathcal{L}_D[\rho^S]
        \otimes(A^{\mathrm{red}})^{\mathsf T}
        +
        \rho^S\otimes
        \bigl(\mathcal{L}_E^*[A^{\mathrm{red}}]\bigr)^{\mathsf T}
        \nonumber\\
        &\quad+
        \sum_{n,m}\gamma^{\out D\inp E}_{nm}
        [L_n^{\out\dagger},\rho^S]
        \otimes(A^{\mathrm{red}}L_m^\inp)^{\mathsf T}
        \nonumber\\
        &\quad+
        \sum_{n,m}
        \bigl(\gamma^{\out D\inp E}_{nm}\bigr)^*
        [\rho^S,L_n^\out]
        \otimes(L_m^{\inp\dagger}A^{\mathrm{red}})^{\mathsf T}.
        \label{Eq: Example joint state observable evolution}
    \end{align}
    The first two terms are the independent GKSL and adjoint-GKSL evolutions derived above. The last two terms couple the state and observable factors. This term cannot be obtained by independently varying the two factors at the chosen initial point. It therefore demonstrates a non-trivial correlation of the Schr\"odinger and Heisenberg evolutions.
\end{example}

\section{Quantum Dynamics of Dynamics: Microscopic Approach}
\label{sec:quantumdynamicsofdynamics-microscopic}
In the preceding section, we derived the most general generator of a CP-divisible quantum comb within the axiomatic framework. In this section, we develop a complementary microscopic approach, starting from system--environment dynamics and introducing a sequence of approximations. We first formulate a superchannel analog of the von Neumann equation and then extend it to evolution along a path in the two-dimensional time plane, obtaining the path-dependent von Neumann equation in Eq.~\eqref{(QDoD): von-Neumann equation of quantum comb}. In Sec.~\ref{(QDoD)Sec:Microscopic}, we apply a series of approximations to this equation and derive the resulting microscopic master equation in Eq.~\eqref{(QDoD)Eq:FinalMicro}.

\subsection{Von Neumann Equation of Quantum Channel}\label{(QDoD)Sec:vonNeumann}
Analogous to the microscopic derivation of the GKSL equation, which starts from the von Neumann equation for quantum states, we begin by considering its generalization to quantum superchannels or, more generally, two-time quantum objects. To this end, we first construct a unitary dilation of the four-legged quantum comb such that tracing out the environmental degrees of freedom recovers the encoder--decoder representation introduced in Sec.~\ref{(QDoD)SubSubSec:QuantumCombFramework}. We then use these unitaries to obtain the path-dependent von Neumann equation. 

\subsubsection{Encoding/Decoding Unitaries}
\begin{figure}
    \centering
    \includegraphics[width=\linewidth]{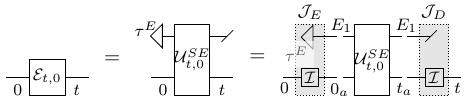}
    \caption{Diagram illustrating the decomposition of a CPTP map into a composition of two isometries and a unitary. The labels $0_a$ and $t_a$ denote auxiliary Hilbert-space labellings associated with the time variables $0$ and $t$, respectively, in the unitary $\mathcal{U}_{t,0}^{SE}$. The labels $E_1$ and $E_2$ denote environmental Hilbert spaces.}
    \label{(QDoD)Fig4-0}
\end{figure}

In the conventional GKSL formalism, one starts from the joint system--environment unitary $\mathcal{U}^{SE}_{t,0}$. To obtain the reduced system dynamics, one introduces an environment state $\tau^E$ and traces out the environmental degrees of freedom. These operations can be described using the isometries $\mathcal{J}_E[\rho^S]=\rho^S\otimes\tau^E$ and $\mathcal{J}_D[\rho^{SE}]=\operatorname{Tr}_E[\rho^{SE}]$. The resulting quantum channel is given by
\begin{align}
    \mathcal{E}_{t,0}[\rho^S]=\operatorname{Tr}_E\left(\mathcal{U}^{SE}_{t,0}[\rho^S\otimes\tau^E]\right)=\left(\mathcal{J}_D\circ\mathcal{U}^{SE}_{t,0}\circ\mathcal{J}_E\right)[\rho^S].
\end{align}
Thus, the channel $\mathcal{E}_{t,0}$ can be decomposed into a dynamical or time-dependent part given by $\mathcal{U}^{SE}_{t,0}$ and non-dynamical or time-independent parts given by $\mathcal{J}_E$ and $\mathcal{J}_D$. This is illustrated in Fig.~\ref{(QDoD)Fig4-0}.

The main idea of this subsubsection is to generalize the above decomposition to quantum combs. Specifically, we consider the general quantum comb introduced in Sec.~\ref{(QDoD)SubSubSec:QuantumCombFramework} and decompose it into a time-dependent part described by unitaries and a time-independent part described by isometries. Once this decomposition is established, we derive the dynamical equations governing these unitaries.

\begin{figure}
    \centering
    \includegraphics[width=\linewidth]{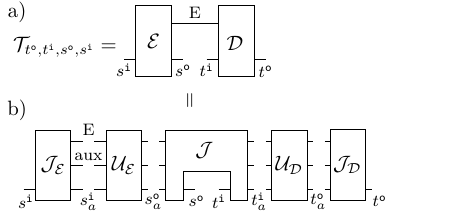}
    \caption{(a) Representation of a quantum comb $\mathcal{T}$ in terms of two isometries $\mathcal{E}$ and $\mathcal{D}$, connected through the $E$-leg, which represents an environmental or auxiliary space. (b) Decomposition of the quantum comb into unitaries and isometries, where all time dependence is contained in the two unitaries $\mathcal{U}_{\mathcal{E}}$ and $\mathcal{U}_{\mathcal{D}}$. All unlabeled legs represent environmental or auxiliary spaces, which may be finite- or infinite-dimensional. The legs labeled $t^\out_a$, $t^\inp_a$, $s^\out_a$, and $s^\inp_a$ also represent auxiliary spaces, but unlike the other auxiliary legs, they are associated with time variables.}
    \label{(QDoD) fig4-1}
\end{figure}

First, recall from Sec.~\ref{(QDoD)SubSubSec:QuantumCombFramework} that any four-legged quantum comb can be expressed in terms of two isometries as
\begin{align}
    \mathcal{T}_{t^\out t^\inp s^\out s^\inp}[\mathcal{A}_{t^\inp,s^\out}]=\mathcal{D}_{t^\out,t^\inp}\circ(\mathcal{A}_{t^\inp,s^\out}\otimes\mathcal{I}_{E})\circ\mathcal{E}_{s^\out,s^\inp},
\end{align}
where $\mathcal{A}_{t^\inp,s^\out}$ is an arbitrary two-legged quantum operation. Furthermore, as discussed in Sec.~\ref{sec:quantumdynamicsofdynamics}, any temporal ordering can be represented as a one-slot quantum comb using a delayed-input construction. Motivated by this representation, we propose to separate the time-dependent and time-independent parts of the quantum comb as follows:
\begin{equation}
    \begin{split}
        &\mathcal{D}\circ(\mathcal{A}\otimes\mathcal{I})\circ\mathcal{E}\\
        &=\mathcal{J}_{\mathcal{D}}\circ\mathcal{U}_{\mathcal{D}}(t^\out,t^\inp)\circ\mathcal{J}[\mathcal{A}]\circ\mathcal{U}_{\mathcal{E}}(s^\out,s^\inp)\circ\mathcal{J}_{\mathcal{E}},
    \end{split}
\end{equation}
where $\mathcal{U}_{\mathcal{D}}$ and $\mathcal{U}_{\mathcal{E}}$ are the time-dependent unitaries, while $\mathcal{J}_{\mathcal{E}}$ and $\mathcal{J}_{\mathcal{D}}$ represent the time-independent parts. The latter are defined on the following spaces:
\begin{align}
    \mathcal{J}_{\mathcal{E}} &: \mathcal{L}(\mathcal{H}_{s^\inp})\rightarrow\mathcal{L}(\mathcal{H}_{s^\inp}\otimes\mathcal{H}_E\otimes\mathcal{H}_{\text{aux}}),\\
    \mathcal{J}_{\mathcal{D}} &: \mathcal{L}(\mathcal{H}_{t^\out}\otimes\mathcal{H}_E\otimes\mathcal{H}_{\text{aux}})\rightarrow\mathcal{L}(\mathcal{H}_{t^\out}).
\end{align}
Here, $\mathcal{H}_{\text{aux}}$ is an auxiliary Hilbert space whose form depends on the specific setup. Finally, $\mathcal{U}_E$ and $\mathcal{U}_D$ are the time-dependent unitaries such that 
\begin{align}
    \mathcal{U}_E: & \mathcal{L}(\mathcal{H}_{s^\inp_a}\otimes \mathcal{H}_E \otimes \mathcal{H}_{\text{aux}}) \rightarrow \mathcal{L}(\mathcal{H}_{s^\out_a}\otimes \mathcal{H}_E \otimes \mathcal{H}_{\text{aux}}),\label{(QDoD)Eq:EncodingUnitary}\\
    \mathcal{U}_D: & \mathcal{L}(\mathcal{H}_{t^\inp_a}\otimes \mathcal{H}_E \otimes \mathcal{H}_{\text{aux}}) \rightarrow \mathcal{L}(\mathcal{H}_{t^\out}\otimes \mathcal{H}_E \otimes \mathcal{H}_{\text{aux}}),\label{(QDoD)Eq:DecodingUnitary}
\end{align}
where $t^\out_a$, $t^\inp_a$, $s^\out_a$, and $s^\inp_a$ are also auxiliary labels, but differ from the other auxiliary labels in that they are associated with the time variables $t^\out$, $t^\inp$, $s^\out$, and $s^\inp$, respectively. Note that for this construction, we assumed that the Hilbert spaces associated with the time-labeled legs are fixed, isomorphic copies of a common system Hilbert space $\mathcal{H}_S$, with $\mathcal{H}_{s^\inp}\cong\mathcal{H}_{s^\out}\cong\mathcal{H}_{t^\inp}\cong\mathcal{H}_{t^\out}\cong\mathcal{H}_S$. The decomposition is illustrated in more detail in Fig.~\ref{(QDoD) fig4-1}(b).

To provide some intuition for this decomposition, let us consider a simple illustrative example.
\begin{example}
In Example~\ref{(QDoD) Eg: evolution of channel}, we considered a quantum comb of the type shown in Fig.~\ref{(QDoD) fig4-2}(a), with the temporal ordering $t^\out > t^\inp > s^\inp > s^\out$. In this example, we illustrate how this quantum comb can be decomposed step by step using our proposed decomposition scheme. The full sequence is shown in Figs.~\ref{(QDoD) fig4-2}(a)--(c), which we use to explain each step.
\begin{figure}
    \centering
    \includegraphics[width=\linewidth]{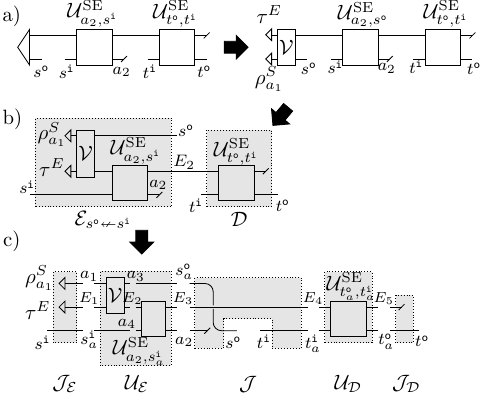}
    \caption{Pictorial representation of rewriting the strictly causally ordered quantum comb defined in the motivating example~\ref{(QDoD) Eg: evolution of channel} into the isometry form. (a) Quantum comb defined in Eq.~(\ref{(QDoD) Eq:Channel evolution comb}) and rewiring the initially correlated state into a separable state using an arbitrary unitary $\mathcal{V}$. (b) Identification of the encoder and decoder parts of the quantum comb, where we rewrite the encoder by promoting the $s^\out$--leg to an extended space. (c) Full decomposition of the isometry into the time-independent isometries and the unitaries.}
    \label{(QDoD) fig4-2}
\end{figure}

    As shown at the beginning of Sec.~\ref{sec:quantumdynamicsofdynamics}, the quantum comb can be expressed in the form illustrated in Fig.~\ref{(QDoD) fig4-2}(b). To further decompose this representation, we assume that the correlated initial state $\rho^{SE}$ can be written as $\rho^{SE}=\mathcal{V}[\rho^S\otimes\tau^E]$, where $\rho^S$ and $\tau^E$ are system and environment states, respectively, and $\mathcal{V}$ is a time-independent unitary. Under this assumption, the encoding isometry is given by
    \begin{equation}
        \begin{split}
            \mathcal{E}_{s^\out\nleftarrow s^\inp}=& \operatorname{Tr}_{a_2} \Big[(\mathcal{U}^{SE}_{a_2,s^{\inp'}_a }\otimes \mathcal{I}_{s^\out_a,s^{\out'}_a})\\
            &\circ (\mathcal{I}_{s^{\inp'}_a , s^\inp_a}\otimes \mathcal{V}_{s^{\out'}_a,a_1}) [\rho^S_{a_1} \otimes \tau^E_{E_1}]\Big].
        \end{split}
    \end{equation}
    Each operator in this expression corresponds to a part of the diagram in Fig.~\ref{(QDoD) fig4-2}(c). In the shaded region representing the encoding unitary $\mathcal{U}_{\mathcal{E}}$, the two unitary operators
    \begin{align}
        \mathcal{U}^{SE}_{a_2,s^{\inp'}_a}&: \mathcal{L}(\mathcal{H}_{s^\inp_a})\otimes \mathcal{L}(\mathcal{H}_{E_1})\rightarrow \mathcal{L}(\mathcal{H}_{a_2})\otimes \mathcal{L}(\mathcal{H}_{E_3}),\\
        \mathcal{V}_{s^{\out'}_a,a_1}&: \mathcal{L}(\mathcal{H}_{a_1})\otimes \mathcal{L}(\mathcal{H}_{E_1})\rightarrow \mathcal{L}(\mathcal{H}_{s^{\out'}_a})\otimes \mathcal{L}(\mathcal{H}_{E_2}),
    \end{align}
    are represented by rectangular boxes, while the identity maps
    \begin{align}
        \mathcal{I}_{s^\out_a,s^{\out'}_a}&:\mathcal{L}(\mathcal{H}_{s^{\out'}_a})\rightarrow \mathcal{L}(\mathcal{H}_{s^\out_a}),\\
        \mathcal{I}_{s^{\inp'}_a,s^\inp_a}&:\mathcal{L}(\mathcal{H}_{s^\inp_a})\rightarrow \mathcal{L}(\mathcal{H}_{s^{\inp'}_a})
    \end{align}
    are represented by connecting lines. For the decoding part, the unitary is simply given by the unitary operator $\mathcal{U}^{SE}_{t^\out_a, t^\inp_a}$.

The input state $\rho^S_{a_1} \otimes \tau^E$ can be observed into the encoding isometry 
\begin{align}
    \mathcal{J}_{\mathcal{E}} [\varrho^S_{s^\inp}] &= \rho^S_{a_1} \otimes \tau^E \otimes \mathcal{I}_{s^\inp_a , s^\inp }[\varrho^S_{s^\inp}]\\
    \mathcal{J}_{\mathcal{D}} [\varrho^{SE}] &= \operatorname{Tr}_{E_5}(\varrho^{SE})
\end{align}
The $a_2$ trace is absorbed into the isometry $\mathcal{J}$, as one can see from Fig.~\ref{(QDoD) fig4-2}(c). 

Following a similar procedure as above, we obtain the decomposition for the application in Sec.~\ref{sec:OQS}, shown in Fig.~\ref{(QDoD) fig4-3}. 
\end{example}

\subsubsection{Von Neumann Equation of quantum channel}
Let us first recall that, for a closed quantum system with Hamiltonian $H(t)$, the density operator satisfies the von Neumann equation
\begin{align}
    i\frac{d\rho(t)}{dt}=[H(t),\rho(t)].
\end{align}
The corresponding unitary propagator $U(t,s)$, which relates the states through $\rho(t)=U(t,s)\rho(s)U^\dagger(t,s)$, satisfies
\begin{align}
    i\frac{\partial U(t,s)}{\partial t}&=H(t)U(t,s),\\
    i\frac{\partial U(t,s)}{\partial s}&=-U(t,s)H(s).
\end{align}
These equations describe variations of the final and initial times, respectively, with the Hamiltonian at the final time, $H(t)$, acting on the propagator from the left and the Hamiltonian at the initial time, $H(s)$, acting from the right. In this subsubsection, we develop an analogous description for the encoding and decoding unitaries. We first define the Hamiltonian or generators associated with variations of their input and output times and then derive the corresponding evolution equations.

    Recall the encoding and decoding unitary channels defined in Eqs.~\eqref{(QDoD)Eq:EncodingUnitary} and~\eqref{(QDoD)Eq:DecodingUnitary}. We write their actions in terms of the corresponding unitary operators as
\begin{align}
    \mathcal{U}_E(s^\out,s^\inp)[\bullet] &= U_E(s^\out,s^\inp)\bullet U_E^\dagger(s^\out,s^\inp),\\
    \mathcal{U}_D(t^\out,t^\inp)[\bullet] &= U_D(t^\out,t^\inp)\bullet U_D^\dagger(t^\out,t^\inp),
\end{align}
where we have omitted the auxiliary subscript $a$ from the time labels for brevity. Assuming differentiability with respect to the time variables, we introduce two types of generators.

The first type describes independent variations of the input and output times. We define the two-time-dependent generators
\begin{align}
    h_E^{\out}(s^\out,s^\inp) &:= i\frac{\partial U_E}{\partial s^\out}U_E^\dagger,\qquad h_E^{\inp}(s^\out,s^\inp):=-iU_E^\dagger\frac{\partial U_E}{\partial s^\inp},\\
    h_D^{\out}(t^\out,t^\inp) &:= i\frac{\partial U_D}{\partial t^\out}U_D^\dagger,\qquad h_D^{\inp}(t^\out,t^\inp):=-iU_D^\dagger\frac{\partial U_D}{\partial t^\inp}.
\end{align}
We refer to these as the \textit{encoding and decoding generators}. By construction, they act on the corresponding output and input spaces:
\begin{align}
    h_E^{\out}\in\mathcal{L}(\mathcal{H}_{s^\out}\otimes\mathcal{H}_{EA}),\qquad h_E^{\inp}\in\mathcal{L}(\mathcal{H}_{s^\inp}\otimes\mathcal{H}_{EA}),\\
    h_D^{\out}\in\mathcal{L}(\mathcal{H}_{t^\out}\otimes\mathcal{H}_{EA}),\qquad h_D^{\inp}\in\mathcal{L}(\mathcal{H}_{t^\inp}\otimes\mathcal{H}_{EA}),
\end{align}
where $EA$ collectively labels the environment and auxiliary spaces.

The second type describes evolution along a specified path in the multi-time parameter space. Recall that such a path can be parametrized by $\lambda$ as
\begin{equation}
    \lambda\mapsto\bigl(t^\out(\lambda),t^\inp(\lambda),s^\out(\lambda),s^\inp(\lambda)\bigr).
\end{equation}
Taking the total derivative along this path and applying the chain rule, we define
\begin{align}
    H_E^\out(\lambda) &:= i\frac{dU_E}{d\lambda}U_E^\dagger=\dot{s}^\out h_E^\out-\dot{s}^\inp U_Eh_E^\inp U_E^\dagger,\label{(QDoD) Def:path-dependent output encoder generators}\\
    H_E^\inp(\lambda) &:= -iU_E^\dagger\frac{dU_E}{d\lambda}=\dot{s}^\inp h_E^\inp-\dot{s}^\out U_E^\dagger h_E^\out U_E,\label{(QDoD) Def:path-dependent input encoder generators}\\
    H_D^\out(\lambda) &:= i\frac{dU_D}{d\lambda}U_D^\dagger=\dot{t}^\out h_D^\out-\dot{t}^\inp U_Dh_D^\inp U_D^\dagger,\label{(QDoD) Def:path-dependent output decoder generators}\\
    H_D^\inp(\lambda) &:= -iU_D^\dagger\frac{dU_D}{d\lambda}=\dot{t}^\inp h_D^\inp-\dot{t}^\out U_D^\dagger h_D^\out U_D.\label{(QDoD) Def:path-dependent input decoder generators}
\end{align}
Here, a dot denotes differentiation with respect to $\lambda$, and all quantities on the right-hand sides are evaluated along the chosen path. We refer to these operators as the \textit{path-dependent encoding and decoding generators}.

From these definitions, we obtain three ways to represent the formal solution to our unitary operators 
\begin{align}
    U_E (\lambda)=& \mathcal{P}\exp\Big(-i \int_{\lambda_0}^{\lambda}d\lambda' H_E^\out (\lambda')\Big) U_E (\lambda_0)\label{(QDoD)Eq:UErep1}\\
    =& U_E (\lambda_0)\overline{\mathcal{P}}\exp\Big(i \int_{\lambda_0}^{\lambda}d\lambda' H_E^\inp (\lambda')\Big)\label{(QDoD)Eq:UErep2}\\
    =& \mathcal{P}\exp\Big(-i \int_{\lambda_0}^{\lambda}d\lambda' \dot{s}^\out (\lambda')h_E^\out (\lambda')\Big) U_E (\lambda_0)\nonumber\\
    &\times \overline{\mathcal{P}}\exp\Big(i \int_{\lambda_0}^{\lambda}d\lambda' \dot{s}^\inp(\lambda') h_E^\inp (\lambda')\Big)\label{(QDoD)Eq:UErep3}
\end{align}
where $\overline{\mathcal{P}}\exp(\dots)$ is an anti-path-ordered exponential. Similar solutions also exist for the decoder unitary operator 
\begin{align}
    U_D (\lambda)=& \mathcal{P}\exp\Big(-i \int_{\lambda_0}^{\lambda}d\lambda' H_D^\out (\lambda')\Big) U_D (\lambda_0)\label{(QDoD)Eq:UDrep1}\\
    =& U_D (\lambda_0)\overline{\mathcal{P}}\exp\Big(i \int_{\lambda_0}^{\lambda}d\lambda' H_D^\inp (\lambda')\Big)\label{(QDoD)Eq:UDrep2}\\
    =& \mathcal{P}\exp\Big(-i \int_{\lambda_0}^{\lambda}d\lambda' \dot{t}^\out (\lambda')h_D^\out (\lambda')\Big) U_D (\lambda_0)\nonumber\\
    &\times \overline{\mathcal{P}}\exp\Big(i \int_{\lambda_0}^{\lambda}d\lambda' \dot{t}^\inp(\lambda') h_D^\inp (\lambda')\Big).\label{(QDoD)Eq:UDrep3}
\end{align}

Taking the path-derivative of the encoding/decoding unitaries using the representations~\eqref{(QDoD)Eq:UErep3} and \eqref{(QDoD)Eq:UDrep3}, we find 
\begin{align}
    i\frac{\partial \mathcal{U}_E [\bullet]}{\partial \lambda} &= \dot{s}^\out \operatorname{ad}_{h_E^\out}\circ \mathcal{U}_E[\bullet]-\dot{s}^\inp \mathcal{U}_E \circ \operatorname{ad}_{h_E^\inp}[\bullet],\label{(QDoD):path-derivative of UE 1}\\
    i\frac{\partial \mathcal{U}_D [\bullet]}{\partial \lambda} &= \dot{t}^\out \operatorname{ad}_{h_D^\out}\circ \mathcal{U}_D[\bullet]-\dot{t}^\inp \mathcal{U}_D \circ \operatorname{ad}_{h_D^\inp}[\bullet],\label{(QDoD):path-derivative of UD 1}
\end{align}
where $\operatorname{ad}$ is an adjoint operation defined with the commutator $\operatorname{ad}_A [B] := [A,B]$. Alternatively, one can use the representations~\eqref{(QDoD)Eq:UErep1}, \eqref{(QDoD)Eq:UErep2}, \eqref{(QDoD)Eq:UDrep1}, or \eqref{(QDoD)Eq:UDrep2}, we obtain four more equations,
\begin{align}
    &i \frac{\partial \mathcal{U}_E}{\partial \lambda} = \operatorname{ad}_{H_E^\out}\circ \,\mathcal{U}_E [\bullet] = - \mathcal{U}_E \circ \operatorname{ad}_{H_E^\inp}[\bullet],\label{(QDoD):path-derivative of UE 2}\\
    &i \frac{\partial \mathcal{U}_D}{\partial \lambda} = \operatorname{ad}_{H_D^\out}\circ \,\mathcal{U}_D [\bullet] = - \mathcal{U}_D \circ \operatorname{ad}_{H_D^\inp}[\bullet],\label{(QDoD):path-derivative of UD 2}
\end{align}

We now combine the preceding equations to construct a von Neumann equation for the quantum comb. We first introduce the higher-order quantum operator
\begin{align}
    \mathcal{R}[\mathcal{A}]:=\mathcal{U}_D\circ\mathcal{J}[\mathcal{A}]\circ\mathcal{U}_E,\label{(QDoD)Eq:vonNeumannHOQO}
\end{align}
where $\mathcal{A}$ is a quantum operation inserted on the open legs $s^\out$ and $t^\inp$. The operator $\mathcal{R}$ corresponds to the middle three shaded regions in Fig.~\ref{(QDoD) fig4-1}(b). Applying the time-independent boundary isometries $\mathcal{J}_{\mathcal{E}}$ and $\mathcal{J}_{\mathcal{D}}$ at its input and output, respectively, recovers the original quantum comb:
\begin{align}
    \mathcal{T}[\mathcal{A}]=\mathcal{J}_{\mathcal{D}}\circ\mathcal{R}[\mathcal{A}]\circ\mathcal{J}_{\mathcal{E}}.
\end{align}

Finally, using the path derivatives of the unitary superoperators~(\ref{(QDoD):path-derivative of UE 1})--(\ref{(QDoD):path-derivative of UD 2}), we obtain the following.
\begin{lem}[Path-ordered von Neumann equation]
    Let $\mathcal{R}$ be the quantum comb defined in Eq.~\eqref{(QDoD)Eq:vonNeumannHOQO}. Then its path derivative is given by 
    \begin{align}
         i\frac{\partial \mathcal{R}}{\partial \lambda} = \mathcal{K}(\lambda) \circ  \mathcal{R},\label{(QDoD): von-Neumann equation of quantum comb}
    \end{align}
    where the operator $\mathcal{K}$ takes the following form 
    \begin{align}
        \Big(\mathcal{K}(\lambda)\circ \mathcal{R}\Big)[\mathcal{F}](\rho) :=& \dot{t}^\out \operatorname{ad}_{h_D^\out} \circ \mathcal{R}[\mathcal{F}] (\rho) \nonumber\\
        &- \dot{t}^\inp \mathcal{U}_D \circ \operatorname{ad}_{h_D^\inp}\circ \mathcal{U}_D^\dagger \circ \mathcal{R}[\mathcal{F}](\rho)\nonumber\\
        &+ \dot{s}^\out \mathcal{R}[\mathcal{F}]\circ \mathcal{U}_E^\dagger \circ \operatorname{ad}_{h_E^\out}\circ \mathcal{U}_E (\rho) \nonumber\\
        & - \dot{s}^\inp \mathcal{R}[\mathcal{F}]\circ \operatorname{ad}_{h_E^\inp}(\rho).
    \end{align}
    A more compact form is given in terms of the path-dependent generators  
    \begin{align}
        \Big(\mathcal{K}(\lambda)\circ \mathcal{R}\Big)[\mathcal{F}](\rho) :=& \operatorname{ad}_{H_D^\out} \circ \mathcal{R}[\mathcal{F}](\rho) - \mathcal{R}[\mathcal{F}]\circ \operatorname{ad}_{H_E^\inp} (\rho). \label{(QDoD): simpler form of von-Neumann equation of quantum comb}
    \end{align}
\end{lem}

\begin{example}
    As an illustration, consider the sequential decoder unitary associated with the causally ordered setup of Sec.~\ref{sec:OQS},
\begin{align}
    U_D(t_1)
    &=e^{-i(t_2-t_1)(\mathds{1}^{S_1}\otimes H^{S_2E})}
    e^{-i(t_1-t)(H^{S_1E}\otimes\mathds{1}^{S_2})},
\end{align}
where $t_2>t_1>t$, and $S_1$ and $S_2$ denote the two separate system wires. The identity operators make explicit that each interaction leaves the other system wire unchanged; the tensor factors are understood to act on the same joint Hilbert space with the appropriate ordering. These wire labels are distinct from the time labels $s^\inp$ and $s^\out$. We take the system--environment exchange Hamiltonians to be
\begin{align}
    H^{S_j E}=g\left(\sigma_+^{S_j}\sigma_-^E+\sigma_-^{S_j}\sigma_+^E\right),\qquad j=1,2.
\end{align}
Differentiating with respect to the intermediate time $t_1$ gives the effective decoder generator
\begin{align}
    G_D(t_1)&:=i\frac{dU_D}{dt_1}U_D^\dagger\nonumber\\
    &=\cos(g\tau_2)H^{S_1E}-H^{S_2E}
    +i\sin(g\tau_2)H^{S_1S_2E},
    \label{(QDoD)Eq:SequentialDecoderGenerator}
\end{align}
where $\tau_2:=t_2-t_1$ and
\begin{align}
    H^{S_1S_2E}:=g\left(\sigma_-^{S_1}\sigma_+^{S_2}-\sigma_+^{S_1}\sigma_-^{S_2}\right)\sigma_z^E.
\end{align}
Thus, the generator is not necessarily a single system--environment Hamiltonian: it can contain a combination of Hamiltonian terms and an environment-mediated term involving both system wires. We give the derivation and explain its compatibility with the causal ordering in Appendix~\ref{(QDoD)Apx:SequentialDecoderGenerator}.
\end{example}

In the next subsection, we will take the path-ordered von Neumann equation~\eqref{(QDoD): von-Neumann equation of quantum comb} as a starting point and perform a sequence of approximations to obtain the QCMS.  
 
\subsection{Microscopic Derivation of Quantum Comb Master Equation}\label{(QDoD)Sec:Microscopic}
In this section, we derive the microscopic form of the quantum comb master equation (QCME). We first establish the microscopic setup by introducing the encoding and decoding dynamics and their generators. We then extend the Born approximation to quantum combs and apply the Markov and coarse-graining approximations to obtain a time-local dynamical equation. Finally, we express the resulting equation in a GKSL-like form, identifying the encoding, decoding, and encoder--decoder correlation terms, and establish the positive semidefiniteness of the associated coefficient matrix.
\subsubsection{Microscopic setup}

In the standard microscopic derivation of the GKSL equation, one starts from the von Neumann equation for a system coupled to an environment and derives a perturbative equation for the reduced system dynamics. Here, we follow an analogous approach for quantum combs, starting from the path-dependent von Neumann equation for the higher-order quantum operator $\mathcal{R}$. We first expand this equation perturbatively in the coupling strength and apply the time-independent encoding and decoding boundary maps to obtain a dynamical equation for the quantum comb $\mathcal{T}$. We then introduce harmonic decompositions of the encoding and decoding generators and specify the boundary maps, establishing the microscopic setup for the Born--Markov coarse-graining approximation developed in the next subsection.

Consider the path-dependent von Neumann equation~(\ref{(QDoD): von-Neumann equation of quantum comb}) and its simpler generator~(\ref{(QDoD): simpler form of von-Neumann equation of quantum comb}). Acting with the time-independent boundary maps $\mathcal{J}_E$ and $\mathcal{J}_D$ on the quantum comb $\mathcal{R}$ gives the four-time dependent, four-legged quantum comb
\begin{align}
    \mathcal{T}(t^\out,t^\inp,s^\out,s^\inp)
    =\mathcal{J}_D\circ
    \mathcal{R}(t^\out,t^\inp,s^\out,s^\inp)\circ\mathcal{J}_E.
\end{align}
Our aim is to reduce the path-dependent von Neumann equation to a dynamical equation for $\mathcal{T}$ itself. Its initial condition remains an arbitrary comb compatible with the chosen boundary construction.

First, we extract the perturbative coupling constant $g$ from the encoding and decoding generators, so that
\begin{align}
    i\frac{\partial\mathcal{R}(\lambda)}{\partial\lambda}
    =g\mathcal{K}(\lambda)\circ\mathcal{R}(\lambda).
    \label{(QDoD): microscopic step1}
\end{align}
Integrating with respect to the path parameter gives
\begin{align}
    \mathcal{R}(\lambda)
    =\mathcal{R}(\lambda_0)
    -ig\int_{\lambda_0}^{\lambda}d\lambda'\,
    \mathcal{K}(\lambda')\circ\mathcal{R}(\lambda').
    \label{(QDoD): microscopic step2}
\end{align}
Substituting this into Eq.~(\ref{(QDoD): microscopic step1}) and then introducing a lag variable $\lambda'$ yields
\begin{equation}
    \begin{split}
        i\frac{\partial\mathcal{R}(\lambda)}{\partial\lambda}
    =&g\mathcal{K}(\lambda)\circ\mathcal{R}(\lambda_0)
    \nonumber\\
    &-ig^2\int_0^{\lambda-\lambda_0}d\lambda'\,
    \mathcal{K}(\lambda)\circ\mathcal{K}(\lambda-\lambda')
    \circ\mathcal{R}(\lambda-\lambda').
    \label{(QDoD): microscopic step3}
    \end{split}
\end{equation}
Next, applying the two boundary maps, we obtain
\begin{equation}
    \begin{split}
    &i\frac{\partial\mathcal{T}(\lambda)}{\partial\lambda}
    =g\mathcal{J}_D\circ\mathcal{K}(\lambda)
    \circ\mathcal{R}(\lambda_0)\circ\mathcal{J}_E
    \\
    &-ig^2\int_0^{\lambda-\lambda_0}d\lambda'\,
    \mathcal{J}_D\circ\mathcal{K}(\lambda)
    \circ\mathcal{K}(\lambda-\lambda')
    \circ\mathcal{R}(\lambda-\lambda')\circ\mathcal{J}_E.
    \label{(QDoD): microscopic step4}
    \end{split}
\end{equation}

The double composition of $\mathcal{K}$ in the integrand can be expanded as follows:
\begin{equation}
    \begin{split}
        &\big(\mathcal{K}_{\lambda}\circ\mathcal{K}_{\lambda''}
    \circ\mathcal{R}_{\lambda''}\big)
    [\bullet_{t^\inp s^\out}](\rho_{s^\inp EA})
    \\
    &\quad=
    \operatorname{ad}_{H_D^\out(\lambda)}\circ
    \operatorname{ad}_{H_D^\out(\lambda'')}\circ
    \mathcal{R}_{\lambda''}[\bullet_{t^\inp s^\out}]
    (\rho_{s^\inp EA})
    \\
    &\qquad-
    \operatorname{ad}_{H_D^\out(\lambda)}\circ
    \mathcal{R}_{\lambda''}[\bullet_{t^\inp s^\out}]\circ
    \operatorname{ad}_{H_E^\inp(\lambda'')}
    (\rho_{s^\inp EA})
    \\
    &\qquad-
    \operatorname{ad}_{H_D^\out(\lambda'')}\circ
    \mathcal{R}_{\lambda''}[\bullet_{t^\inp s^\out}]\circ
    \operatorname{ad}_{H_E^\inp(\lambda)}
    (\rho_{s^\inp EA})
    \\
    &\qquad+
    \mathcal{R}_{\lambda''}[\bullet_{t^\inp s^\out}]\circ
    \operatorname{ad}_{H_E^\inp(\lambda'')}\circ
    \operatorname{ad}_{H_E^\inp(\lambda)}
    (\rho_{s^\inp EA}).
    \label{(QDoD): microscopic step5}
    \end{split}
\end{equation}
Here, $[\bullet_{t^\inp s^\out}]$ denotes an arbitrary insertion on the $t^\inp$ and $s^\out$ legs, while $\rho_{s^\inp EA}$ denotes an input state on the system and environment--auxiliary legs. We use $\lambda'':=\lambda-\lambda'$ for brevity.

To further reduce the equation, recall the operator spaces of the encoding and decoding generators:
\begin{align}
    H_E^\out&\in
    \mathcal{L}(\mathcal{H}_{s^\out}\otimes\mathcal{H}_{EA}),
    &H_E^\inp&\in
    \mathcal{L}(\mathcal{H}_{s^\inp}\otimes\mathcal{H}_{EA}),
    \label{(QDoD): microscopic step6-1}\\
    H_D^\out&\in
    \mathcal{L}(\mathcal{H}_{t^\out}\otimes\mathcal{H}_{EA}),
    &H_D^\inp&\in
    \mathcal{L}(\mathcal{H}_{t^\inp}\otimes\mathcal{H}_{EA}).
    \label{(QDoD): microscopic step6-2}
\end{align}
We further assume that these generators admit the harmonic decompositions
\begin{align}
    H_E^\out(\lambda)
    &=\sum_{n,\omega}e^{-i\omega s^\out(\lambda)}
    e_n^\out(\omega)\otimes b_n^{\out E}(\lambda),
    \label{(QDoD): microscopic step7-1}\\
    H_E^\inp(\lambda)
    &=\sum_{n,\omega}e^{-i\omega s^\inp(\lambda)}
    e_n^\inp(\omega)\otimes b_n^{\inp E}(\lambda),
    \label{(QDoD): microscopic step7-2}\\
    H_D^\out(\lambda)
    &=\sum_{n,\omega}e^{-i\omega t^\out(\lambda)}
    d_n^\out(\omega)\otimes b_n^{\out D}(\lambda),
    \label{(QDoD): microscopic step7-3}\\
    H_D^\inp(\lambda)
    &=\sum_{n,\omega}e^{-i\omega t^\inp(\lambda)}
    d_n^\inp(\omega)\otimes b_n^{\inp D}(\lambda).
    \label{(QDoD): microscopic step7-4}
\end{align}
The operators $e_n^\out$, $e_n^\inp$, $d_n^\out$, and $d_n^\inp$ act on their respective system spaces, while the $b_n$ act on the environment--auxiliary space. The frequency components are chosen in adjoint pairs so that the full generators are Hermitian.

Next, we specify the boundary maps. In the examples considered above, the encoding and decoding boundary maps often take the following form:
\begin{align}
    \mathcal{J}_E[\bullet_{s^\inp}]
    &=\bullet_{s^\inp}\otimes\rho_{EA},
    \nonumber\\
    \mathcal{J}_D[\bullet_{t^\out EA}]
    &=\operatorname{Tr}_{EA}
       [\bullet_{t^\out EA}].
    \label{(QDoD): microscopic step9}
\end{align}
Here, $\rho_{EA}$ is a normalized state on the environment--auxiliary space, and the decoding boundary map traces out both the environment and the auxiliary system.
\subsubsection{Born-Markov coarse-graining approximation}

We introduce the \textit{quantum comb extension of the Born approximation}:
\begin{align}
    &\mathcal{R}_{\lambda''}[\bullet_{t^\inp s^\out}]
      (\bullet_{s^\inp}\otimes\bullet_{EA})
    \nonumber\\
    &\quad\simeq
    \mathcal{T}_{\lambda''}[\bullet_{t^\inp s^\out}]
      (\bullet_{s^\inp})
    \otimes\mathcal{I}_{EA}(\bullet_{EA}).
    \label{(QDoD): Comb Born approximation}
\end{align}
This factorization is consistent with the boundary maps in Eq.~\eqref{(QDoD): microscopic step9} for the normalized environment--auxiliary state $\rho_{EA}$.

Using this approximation, the integrand of Eq.~\eqref{(QDoD): microscopic step4} becomes
\begin{align}
    &\operatorname{Tr}_{EA}\Big\{
    \nonumber\\*
    &\quad\operatorname{ad}_{H_D^\out(\lambda)}\circ
    \operatorname{ad}_{H_D^\out(\lambda'')}
    \big(\mathcal{T}_{\lambda''}[\bullet_{t^\inp s^\out}]
       (\bullet_{s^\inp})\otimes\rho_{EA}\big)
    \nonumber\\*
    &\quad-\operatorname{ad}_{H_D^\out(\lambda)}\circ
    \mathcal{T}_{\lambda''}[\bullet_{t^\inp s^\out}]\circ
    \operatorname{ad}_{H_E^\inp(\lambda'')}
       (\bullet_{s^\inp}\otimes\rho_{EA})
    \nonumber\\*
    &\quad-\operatorname{ad}_{H_D^\out(\lambda'')}\circ
    \mathcal{T}_{\lambda''}[\bullet_{t^\inp s^\out}]\circ
    \operatorname{ad}_{H_E^\inp(\lambda)}
       (\bullet_{s^\inp}\otimes\rho_{EA})
    \nonumber\\*
    &\quad+\mathcal{T}_{\lambda''}[\bullet_{t^\inp s^\out}]\circ
    \operatorname{ad}_{H_E^\inp(\lambda'')}\circ
    \operatorname{ad}_{H_E^\inp(\lambda)}
       (\bullet_{s^\inp}\otimes\rho_{EA})
    \Big\}.
    \label{(QDoD): microscopic step12}
\end{align}
Here and below, we omit the identity superoperator on $EA$ when it accompanies $\mathcal{T}$.

Using the harmonic decompositions, the first term gives
\begin{align}
    &\sum_{\omega,\omega',n,m}
    e^{i\omega t^\out(\lambda)-i\omega't^\out(\lambda'')}
    \nonumber\\
    &\quad\times\Big[
    C^{\out D\out D}_{nm}(\lambda,\lambda'')
    d_n^{\out\dagger}(\omega)d_m^\out(\omega')
    \mathcal{T}_{\lambda''}[\bullet_{t^\inp s^\out}]
    (\rho_{s^\inp})
    \nonumber\\
    &\qquad-
    \widetilde C^{\out D\out D}_{nm}(\lambda,\lambda'')
    d_m^\out(\omega')
    \mathcal{T}_{\lambda''}[\bullet_{t^\inp s^\out}]
    (\rho_{s^\inp})d_n^{\out\dagger}(\omega)
    \Big]+\mathrm{h.c.},
    \label{(QDoD): microscopic step13}
\end{align}
where the two ordered correlation functions are defined as
\begin{align}
    C^{AB}_{nm}(\lambda,\lambda')
    &:=\operatorname{Tr}_{EA}
    [b_n^{A\dagger}(\lambda)b_m^B(\lambda')\rho_{EA}],
    \label{(QDoD): correlation loss}\\
    \widetilde C^{AB}_{nm}(\lambda,\lambda')
    &:=\operatorname{Tr}_{EA}
    [b_m^B(\lambda')\rho_{EA}b_n^{A\dagger}(\lambda)].
    \label{(QDoD): correlation gain}
\end{align}
The labels $A,B$ take values in $\{\out D,\inp E\}$. The definition of $\widetilde C$ follows the gain ordering $d_m\mathcal{T}(\bullet)d_n^\dagger$ in Eq.~\eqref{(QDoD): microscopic step13}; the state lies between the two bath operators. Since the decoding boundary map is now an unconditional partial trace, cyclicity gives
\begin{align}
    \widetilde C^{AB}_{nm}(\lambda,\lambda')
    =C^{AB}_{nm}(\lambda,\lambda').
\end{align}
The notation $\mathrm{h.c.}$ denotes the Hermitian conjugate. The resulting linear identities extend to arbitrary test operators and insertions.

The second and third terms are related by exchanging $\lambda$ and $\lambda''$. Before the overall minus sign in Eq.~\eqref{(QDoD): microscopic step12}, the second term is
\begin{align}
    &\sum_{\omega,\omega',n,m}
    e^{i\omega t^\out(\lambda)-i\omega's^\inp(\lambda'')}
    \nonumber\\
    &\quad\times\Big[
    C^{\out D\inp E}_{nm}(\lambda,\lambda'')
    d_n^{\out\dagger}(\omega)
    \mathcal{T}_{\lambda''}[\bullet_{t^\inp s^\out}]
       (e_m^\inp(\omega')\rho_{s^\inp})
    \nonumber\\
    &\qquad-
    \widetilde C^{\out D\inp E}_{nm}(\lambda,\lambda'')
    \mathcal{T}_{\lambda''}[\bullet_{t^\inp s^\out}]
       (e_m^\inp(\omega')\rho_{s^\inp})
    d_n^{\out\dagger}(\omega)
    \Big]+\mathrm{h.c.}
    \label{(QDoD): microscopic step14}
\end{align}
The fourth term is
\begin{align}
    &\sum_{\omega,\omega',n,m}
    e^{i\omega s^\inp(\lambda'')-i\omega's^\inp(\lambda)}
    \nonumber\\
    &\quad\times\Big[
    C^{\inp E\inp E}_{nm}(\lambda'',\lambda)
    \mathcal{T}_{\lambda''}[\bullet_{t^\inp s^\out}]
       (e_n^{\inp\dagger}(\omega)e_m^\inp(\omega')
        \rho_{s^\inp})
    \nonumber\\
    &\qquad-
    \widetilde C^{\inp E\inp E}_{nm}(\lambda'',\lambda)
    \mathcal{T}_{\lambda''}[\bullet_{t^\inp s^\out}]
       (e_m^\inp(\omega')\rho_{s^\inp}
        e_n^{\inp\dagger}(\omega))
    \Big]+\mathrm{h.c.}
    \label{(QDoD): microscopic step15}
\end{align}
In particular, both encoder operators in the fourth term remain inside $\mathcal{T}[\bullet_{t^\inp s^\out}](\cdot)$.

In the standard microscopic GKSL derivation, the first-order contribution is eliminated by assuming vanishing expectation values of the environmental interaction operators~\cite{BreuerPetruccione}. Here, we impose the corresponding condition
\begin{align}
    \operatorname{Tr}_{EA}
    [b_n^X(\lambda)\rho_{EA}]=0,
\end{align}
for every relevant $n,\lambda$ and $X\in\{\out D,\inp E\}$, including the adjoint partners in the decomposition. By cyclicity, this also implies $\operatorname{Tr}_{EA}[\rho_{EA}b_n^X(\lambda)]=0$. We also assume the Born approximation at $\lambda_0$. Under these assumptions, the first-order contribution in Eq.~\eqref{(QDoD): microscopic step4} vanishes.

Finally, we evaluate the integral in Eq.~\eqref{(QDoD): microscopic step4}. We assume that the correlations have short memory along the chosen oriented path, compared with the scale on which the reduced comb changes.

To average out rapidly oscillating contributions, the conventional GKSL derivation employs the secular approximation, which neglects terms oscillating rapidly due to differences in Bohr frequencies. In our case, however, the oscillatory phases depend on multiple time variables along a generally nonlinear path. Consequently, their effective frequencies with respect to the path parameter $\lambda$ need not be constant, and the standard secular approximation cannot be directly applied without additional assumptions on the path, such as affine time dependence with sufficiently separated effective frequencies. We therefore employ a \textit{local coarse-graining method}~\cite{SchallerBrandes2008}, which allows us to average the second-order contributions over a finite path interval without imposing such restrictions or discarding cross-frequency terms a priori.

Following the local coarse-graining method~\cite{SchallerBrandes2008}, we first consider the coarse-graining window $[\lambda,\lambda+\Delta\lambda]$ along some chosen evolution path on the two-dimensional time plane. We choose the width $\Delta\lambda$ to be longer than the environment--auxiliary correlation scale and shorter than the reduced-comb evolution scale, both measured with respect to the path parameter $\lambda$. We impose the quantum comb extension of the Born approximation~\eqref{(QDoD): Comb Born approximation} locally and neglect memory between successive windows. The comb can then be held fixed at $\mathcal{T}_\lambda$ in the second-order integrand. Integrating Eq.~\eqref{(QDoD): microscopic step4} over the coarse-graining window gives
\begin{equation}
    \begin{split}
        \frac{\mathcal T_{\lambda+\Delta\lambda}
                    -\mathcal T_\lambda}{\Delta\lambda}
    \simeq&
    \frac{-g^2}{\Delta\lambda}
    \int_\lambda^{\lambda+\Delta\lambda}du\int_\lambda^u dv\\
    &\times \mathcal J_D\circ\mathcal K_u\circ\mathcal K_v
    \circ(\mathcal T_\lambda\otimes\mathcal I_{EA})
    \circ\mathcal J_E.
    \label{(QDoD): path coarse grained kernel}
    \end{split}
\end{equation}
The quotient is identified with the coarse-grained path derivative. This is a generalized \textit{Born-Markov coarse-graining approximation}.
\subsubsection{GKSL-like generator and positivity}
Next, define the path-dependent ordered correlation integrals
\begin{align}
    &\Gamma^{\out D\out D}_{nm}(\lambda;\omega,\omega')
    \nonumber\\*
    &\quad:=\frac1{\Delta\lambda}
    \int_\lambda^{\lambda+\Delta\lambda}du\int_\lambda^u dv\,
    e^{i\omega t^\out(u)-i\omega't^\out(v)}
    C^{\out D\out D}_{nm}(u,v),\\
    &\Gamma^{\inp E\inp E}_{nm}(\lambda;\omega,\omega')
    \nonumber\\
    &\quad:=\frac1{\Delta\lambda}
    \int_\lambda^{\lambda+\Delta\lambda}du\int_\lambda^u dv\,
    e^{i\omega s^\inp(u)-i\omega's^\inp(v)}
    C^{\inp E\inp E}_{nm}(u,v),\\
    &\Gamma^{\out D\inp E}_{nm}(\lambda;\omega,\omega')
    \nonumber\\
    &\quad:=\frac1{\Delta\lambda}
    \int_\lambda^{\lambda+\Delta\lambda}du\int_\lambda^u dv\,
    e^{i\omega t^\out(u)-i\omega's^\inp(v)}
    C^{\out D\inp E}_{nm}(u,v),\\
    &\Gamma^{\inp E\out D}_{nm}(\lambda;\omega,\omega')
    \nonumber\\
    &\quad:=\frac1{\Delta\lambda}
    \int_\lambda^{\lambda+\Delta\lambda}du\int_\lambda^u dv\,
    e^{i\omega s^\inp(u)-i\omega't^\out(v)}
    C^{\inp E\out D}_{nm}(u,v).
    \label{(QDoD): one sided transforms}
\end{align}
The coefficients $\widetilde\Gamma$ are defined similarly by replacing $C$ with $\widetilde C$.

Because the decoding boundary map is the unconditional partial trace over $EA$, cyclicity of the trace gives
\begin{align}
    \widetilde C^{AB}_{nm}(u,v)&=C^{AB}_{nm}(u,v),\\
    \widetilde\Gamma^{AB}_{nm}(\lambda;\omega,\omega')
    &=\Gamma^{AB}_{nm}(\lambda;\omega,\omega').
\end{align}
The correlations can be expressed as expectation values in the environment--auxiliary state $\rho_{EA}$:
\begin{align}
    C^{AB}_{nm}(u,v)
    =\braket{b_n^{A\dagger}(u)b_m^B(v)}_{\rho_{EA}}.
\end{align}
Consequently,
\begin{align}
    C^{AB}_{nm}(u,v)^*=C^{BA}_{mn}(v,u).
    \label{(QDoD): correlation adjoint symmetry}
\end{align}

Using these coefficients, we rewrite Eq.~\eqref{(QDoD): path coarse grained kernel} in terms of the decoder, encoder, and mixed contributions. The decoder contribution takes the form
\begin{align}
    &g^2\sum_{n,m,\omega,\omega'}
      \Gamma^{\out D\out D}_{nm}(\lambda;\omega,\omega')
    \nonumber\\
    &\quad\times\Big[
    d_m^\out(\omega')
    \mathcal T_\lambda[\bullet_{t^\inp s^\out}](\rho_{s^\inp})
    d_n^{\out\dagger}(\omega)
    \nonumber\\
    &\qquad-
    d_n^{\out\dagger}(\omega)d_m^\out(\omega')
    \mathcal T_\lambda[\bullet_{t^\inp s^\out}](\rho_{s^\inp})
    \Big]+\mathrm{h.c.},
    \label{(QDoD): integrated decoder term}
\end{align}
for some state $\rho_{s^\inp}$. For $A,B\in\{\out D,\inp E\}$, define
\begin{align}
    \gamma^{AB}_{nm}(\lambda;\omega,\omega')
    &:=\Gamma^{AB}_{nm}(\lambda;\omega,\omega')
      +\Gamma^{BA}_{mn}(\lambda;\omega',\omega)^*,\\
    S^{AA}_{nm}(\lambda;\omega,\omega')
    &:=\frac{
        \Gamma^{AA}_{nm}(\lambda;\omega,\omega')
       -\Gamma^{AA}_{mn}(\lambda;\omega',\omega)^*
       }{2i}.
\end{align}
The adjoint exchanges both operator and frequency indices. Define the Hermitian operators
\begin{align}
    H_{\mathcal D}(\lambda)
    &:=\sum_{n,m,\omega,\omega'}
      S^{\out D\out D}_{nm}(\lambda;\omega,\omega')
      d_n^{\out\dagger}(\omega)d_m^\out(\omega'),\\
    H_{\mathcal E}(\lambda)
    &:=\sum_{n,m,\omega,\omega'}
      S^{\inp E\inp E}_{nm}(\lambda;\omega,\omega')
      e_n^{\inp\dagger}(\omega)e_m^\inp(\omega').
\end{align}
The decoder contribution is $g^2\mathcal L_{D,\lambda}\circ\mathcal T_\lambda$, where
\begin{equation}
    \begin{split}
        \mathcal L_{D,\lambda}[\bullet]
    :=&-i[H_{\mathcal D}(\lambda),\bullet]
    \nonumber\\
    &+\sum_{n,m,\omega,\omega'}
      \gamma^{\out D\out D}_{nm}(\lambda;\omega,\omega')
    \Big[
      d_m^\out(\omega')\bullet d_n^{\out\dagger}(\omega)
    \nonumber\\
    &\hspace{28mm}
      -\frac12\{d_n^{\out\dagger}(\omega)d_m^\out(\omega'),
                         \bullet\}
    \Big].
    \label{(QDoD): decoder GKSL contribution}
    \end{split}
\end{equation}

For the fourth contribution, the reversed encoder ordering uses the opposite time-ordering. In particular,
\begin{equation}
    \begin{split}
        &\frac1{\Delta\lambda}
    \int_\lambda^{\lambda+\Delta\lambda}du\int_\lambda^u dv\,
    e^{i\omega s^\inp(v)-i\omega's^\inp(u)}
    C^{\inp E\inp E}_{nm}(v,u)
    \nonumber\\
    &\qquad=\Gamma^{\inp E\inp E}_{mn}
                         (\lambda;\omega',\omega)^*.
    \end{split}
\end{equation}
Its integrated contribution is therefore
\begin{equation}
    \begin{split}
        &g^2\sum_{n,m,\omega,\omega'}
    \Gamma^{\inp E\inp E}_{mn}(\lambda;\omega',\omega)^*
    \nonumber\\
    &\quad\times\mathcal T_\lambda[\bullet_{t^\inp s^\out}]
    \Big(e_m^\inp(\omega')\rho_{s^\inp}
                    e_n^{\inp\dagger}(\omega)
    \nonumber\\
    &\hspace{33mm}
      -e_n^{\inp\dagger}(\omega)e_m^\inp(\omega')
                    \rho_{s^\inp}\Big)
    +\mathrm{h.c.}
    \end{split}
\end{equation}
Consequently, the encoder contribution is $g^2\mathcal T_\lambda\circ\mathcal L_{E,\lambda}$, with
\begin{equation}
    \begin{split}
        \mathcal L_{E,\lambda}[\bullet]
    :=&+i[H_{\mathcal E}(\lambda),\bullet]
    \nonumber\\
    &+\sum_{n,m,\omega,\omega'}
      \gamma^{\inp E\inp E}_{nm}(\lambda;\omega,\omega')
    \Big[
      e_m^\inp(\omega')\bullet e_n^{\inp\dagger}(\omega)
    \nonumber\\
    &\hspace{28mm}
      -\frac12\{e_n^{\inp\dagger}(\omega)e_m^\inp(\omega'),
                         \bullet\}
    \Big].
    \label{(QDoD): encoder GKSL contribution}
    \end{split}
\end{equation}
The opposite Lamb-shift sign follows from the reversed encoder commutators. Both encoder operators remain inside $\mathcal T_\lambda[\bullet_{t^\inp s^\out}](\cdot)$.

For the second and third contributions, the two time orderings combine to give
\begin{equation}
    \begin{split}
        \gamma^{\out D\inp E}_{nm}(\lambda;\omega,\omega')
    &:=\Gamma^{\out D\inp E}_{nm}(\lambda;\omega,\omega')
    \\
    &\quad+
      \Gamma^{\inp E\out D}_{mn}(\lambda;\omega',\omega)^*.
    \label{(QDoD): mixed rate coefficient}
    \end{split}
\end{equation}
Equivalently:
\begin{align}
    &\gamma^{\out D\inp E}_{nm}(\lambda;\omega,\omega')
    \nonumber\\*
    &\quad=\frac1{\Delta\lambda}
    \int_\lambda^{\lambda+\Delta\lambda}du
    \int_\lambda^{\lambda+\Delta\lambda}dv\,
    e^{i\omega t^\out(u)-i\omega's^\inp(v)}
    C^{\out D\inp E}_{nm}(u,v).
\end{align}
The mixed contribution is $g^2\mathcal L_{DE,\lambda}[\mathcal T_\lambda]$, where
\begin{equation}
    \begin{split}
        &\mathcal L_{DE,\lambda}[\mathcal T_\lambda]
       [\bullet_{t^\inp s^\out}](\rho_{s^\inp})
    \\
    &\quad:=\sum_{n,m,\omega,\omega'}\Big\{
      \gamma^{\out D\inp E}_{nm}(\lambda;\omega,\omega')
      \Big[d_n^{\out\dagger}(\omega),
    \\
    &\hspace{40mm}
         \mathcal T_\lambda[\bullet_{t^\inp s^\out}]
          (e_m^\inp(\omega')\rho_{s^\inp})\Big]
    \\
    &\qquad+
      \gamma^{\out D\inp E}_{nm}(\lambda;\omega,\omega')^*
      \Big[\mathcal T_\lambda[\bullet_{t^\inp s^\out}]
          (\rho_{s^\inp}e_m^{\inp\dagger}(\omega')),
          d_n^\out(\omega)\Big]\Big\}.
    \label{(QDoD): encoder decoder mixing contribution}
    \end{split}
\end{equation}
The phases are already contained in $\gamma$. 

The diagonal rate blocks likewise have integral representations.
For example,
\begin{align}
    &\gamma^{\out D\out D}_{nm}(\lambda;\omega,\omega')
    \nonumber\\
    &\quad=\frac1{\Delta\lambda}
    \int_\lambda^{\lambda+\Delta\lambda}du
    \int_\lambda^{\lambda+\Delta\lambda}dv\,
    e^{i\omega t^\out(u)-i\omega't^\out(v)}
    C^{\out D\out D}_{nm}(u,v).
\end{align}
To make the Gram structure explicit, define the integrated
operators
\begin{align}
    F_n^{\out D}(\lambda;\omega)
    &:=
    \frac1{\sqrt{\Delta\lambda}}
    \int_\lambda^{\lambda+\Delta\lambda}du\,
    e^{-i\omega t^\out(u)}b_n^{\out D}(u),
    \nonumber\\
    F_n^{\inp E}(\lambda;\omega)
    &:=
    \frac1{\sqrt{\Delta\lambda}}
    \int_\lambda^{\lambda+\Delta\lambda}du\,
    e^{-i\omega s^\inp(u)}b_n^{\inp E}(u).
\end{align}
Since the correlations are expectation values in the normalized state $\rho_{EA}$, every entry of the four rate blocks satisfies
\begin{align}
    &\gamma^{XY}_{nm}(\lambda;\omega,\omega')
    \nonumber\\
    &\quad=
    \operatorname{Tr}_{EA}\left[
        \rho_{EA}
        F_n^X(\lambda;\omega)^\dagger
        F_m^Y(\lambda;\omega')
    \right]
    \nonumber\\
    &\quad=
    \left\langle
        F_n^X(\lambda;\omega)\rho_{EA}^{1/2},
        F_m^Y(\lambda;\omega')\rho_{EA}^{1/2}
    \right\rangle_{\mathrm{HS}},
\end{align}
where $X,Y\in\{\out D,\inp E\}$, and $\langle V,W\rangle_{\mathrm{HS}}
=\operatorname{Tr}_{EA}(V^\dagger W)$.
Thus, including the composite labels $(n,\omega)$ within
each block, the full rate matrix
\begin{align}
    \boldsymbol{\gamma}(\lambda)
    :=
    \begin{pmatrix}
       \gamma^{\out D\out D}(\lambda)
       &\gamma^{\out D\inp E}(\lambda)\\
       \gamma^{\inp E\out D}(\lambda)
       &\gamma^{\inp E\inp E}(\lambda)
    \end{pmatrix}
\end{align}
is the Gram matrix of the combined family
$\{F_n^X(\lambda;\omega)\rho_{EA}^{1/2}\}_{X,n,\omega}$.
In particular, the mixed blocks are the inner products
between the decoder and encoder families.

Explicitly, for arbitrary complex coefficients
$z^X_{n\omega}$, let
$Z=\sum_{X,n,\omega}
z^X_{n\omega}F_n^X(\lambda;\omega)$.
Then
\begin{align}
    z^\dagger\boldsymbol{\gamma}(\lambda)z
    &=
    \sum_{\substack{X,Y\\n,m,\omega,\omega'}}
    (z^X_{n\omega})^*
    \gamma^{XY}_{nm}(\lambda;\omega,\omega')
    z^Y_{m\omega'}
    \nonumber\\*
    &=
    \operatorname{Tr}_{EA}(\rho_{EA}Z^\dagger Z)
    =
    \left\|Z\rho_{EA}^{1/2}\right\|_{\mathrm{HS}}^2
    \geq0.
\end{align}
Consequently, the full rate matrix is positive semidefinite, $\boldsymbol{\gamma}(\lambda)\succeq0$, not merely its diagonal blocks. The diagonal blocks $\gamma^{\out D\out D}$ and $\gamma^{\inp E\inp E}$ determine the dissipative parts of the decoder and encoder GKSL generators, $\mathcal L_{D,\lambda}$ and $\mathcal L_{E,\lambda}$, respectively. The off-diagonal blocks determine the encoder--decoder correlation term $\mathcal L_{DE,\lambda}$, which is incorporated into the same positive semidefinite rate matrix.

Combining all four contributions, the final equation is
\begin{align}
    &\frac{d\mathcal T_\lambda
       [\bullet_{t^\inp s^\out}](\rho_{s^\inp})}{d\lambda}
    \nonumber\\
    &\quad=g^2\Big\{
       \mathcal L_{D,\lambda}\Big[
          \mathcal T_\lambda[\bullet_{t^\inp s^\out}]
            (\rho_{s^\inp})\Big]
    \nonumber\\
    &\qquad+\mathcal T_\lambda[\bullet_{t^\inp s^\out}]
                 (\mathcal L_{E,\lambda}[\rho_{s^\inp}])
       +\mathcal L_{DE,\lambda}[\mathcal T_\lambda]
          [\bullet_{t^\inp s^\out}](\rho_{s^\inp})\Big\}.
    \label{(QDoD): microscopic GKSL like comb equation}
\end{align}
Equivalently, the dynamical equation can be expressed directly in terms of the quantum comb $\mathcal T_\lambda$, without specifying an insertion or input state:
\begin{align}
    \frac{d\mathcal T_\lambda}{d\lambda}
    =g^2\Big(
       \mathcal L_{D,\lambda}\circ\mathcal T_\lambda
       +\mathcal T_\lambda\circ\mathcal L_{E,\lambda}
       +\mathcal L_{DE,\lambda}[\mathcal T_\lambda]\label{(QDoD)Eq:FinalMicro}
       \Big).
\end{align}
Note that the above equality holds up to the second-order coarse-grained equation. All rates are defined per unit path parameter and contain the velocities of the two selected moving legs through the full path-dependent generators. 

For example, when only $t^\out$ and $t^\inp$ vary, $H_E^\inp=0$. The encoder and mixed coefficients then vanish, while the decoder coefficients still contain both decoder velocities. When one decoder time and one encoder time vary, all three contributions may be present. The same coefficient definitions cover the other choices, including those with $\dot t^\out=0$. 

When the encoder--decoder correlation term $\mathcal L_{DE,\lambda}$ vanishes, Eq.~\eqref{(QDoD)Eq:FinalMicro} reduces to two uncoupled GKSL contributions, acting at the decoder output and encoder input, respectively. As discussed in Sec.~\ref{(QDoD)Sec:QDoD4}, these contributions take the standard GKSL form. In the general case, the additional mixing term is determined by encoder--decoder correlations in the environment--auxiliary system and has the same paired-commutator structure as the interference term in Hasen\"ohrl and Caro~\cite{hasenohrl2022quantum}. This comparison concerns the mixing structure; the dynamical object here remains the full four-legged quantum comb.

The result relies on vanishing environment--auxiliary first moments, the \textit{local Born factorization}~(\ref{(QDoD): Comb Born approximation}), and the \textit{short-memory coarse-graining assumptions}~(\ref{(QDoD): path coarse grained kernel}) for the selected oriented path. 

Note that, in the coarse-graining approach, no separate secular approximation is imposed. Instead, the path-dependent oscillatory phases are moved inside the correlation integrals defining the coefficients $\Gamma^{AB}_{nm}(\lambda;\omega,\omega')$ and $\gamma^{AB}_{nm}(\lambda;\omega,\omega')$.
The frequency averaging that motivates the standard secular approximation is therefore addressed when these coefficients are evaluated, rather than by imposing a universal resonance condition beforehand. 

For a specified path, rapidly oscillating contributions may be suppressed by the integration over the coarse-graining window. Any further secular simplification must be justified from the correlation integrals rather than from the explicit phases alone. Such a simplification is optional: under the assumptions stated above, the coarse-grained rate matrix is positive without discarding cross-frequency terms.

In this section, we have derived the quantum comb master equation from the path-dependent von Neumann equation for the system--environment quantum comb. By extending the Born and Markov approximations to quantum combs and employing local coarse-graining, we obtained the microscopic master equation~\eqref{(QDoD)Eq:FinalMicro}. This equation reproduces the GKSL-like structure of the axiomatic comb master equation~\eqref{(QDoD)Eq:CombME}, established in Theorem~\ref{Thm: main result}, with decoder and encoder GKSL contributions and an additional encoder--decoder correlation term. The microscopic derivation therefore provides a physical interpretation of these contributions in terms of environment--auxiliary interactions and correlations, while identifying the assumptions under which the axiomatic master equation emerges as an effective description of quantum comb dynamics.
\section{Applications}

We now present two applications that highlight the usefulness and versatility of our approach. The first regards out-of-time-ordered correlators in the study of quantum chaos; the second pertains to open quantum dynamics with initial correlations. In both scenarios, tracking the density operator in time is insufficient to fully capture the scenario at hand, and the fundamental object of interest is the time evolution of a two-point object (e.g., a two-time correlation between observables in the former case; a preparation map in the latter case). This makes them both particularly amenable to our present framework, which allows one to propagate any such initial object in time via the QCME.

\subsection{Quantum Chaos \& OTOCs}
\label{sec:otoc}
    The \textbf{out-of-time-ordered correlator (OTOC)} is widely used to characterize quantum chaos and information scrambling~\cite{larkin1969quasiclassical,maldacena2016bound,Swingle2018}. It is closely related to the commutator of two observables, $[A(t),B]$, where $A(t)=\mathcal{U}_{t,0}^{*}[A]=U_{t,0}^{\dagger}AU_{t,0}$ denotes the observable $A$ in the Heisenberg picture. Physically, this commutator characterizes the extent to which a perturbation generated by $B$ at the initial time influences a subsequent measurement of $A$ at time $t$. This influence can be quantified by the squared commutator,
    \begin{align}
        C(t):=-\braket{[A(t),B]^2}_{\rho},
    \end{align}
    where $\braket{X}_{\rho}:=\operatorname{Tr}(X\rho)$. For Hermitian observables, $C(t)$ is non-negative and quantifies their degree of noncommutativity. Assuming $B^2=\mathds{1}$ and $A^2(t)=\mathds{1}$ for all $t$, the squared commutator can be expressed as
    \begin{align}
        C(t)=2\left(1-\operatorname{Re}F(t)\right),
    \end{align}
    where
    \begin{align}
        F(t):=\braket{A(t)BA(t)B}_{\rho}
    \end{align}
    is the four-point out-of-time-ordered correlator. In what follows, we refer to $F(t)$ as the OTOC.

    The connection between OTOCs and the quantum-comb formalism has been explored in Refs.~\cite{zonnios2022signatures,emori2023error}. Motivated by the quantum-comb formulation of measurement disturbance and OTOCs developed in Ref.~\cite{emori2023error}, we construct an ancilla-assisted quantum comb that represents the four-point correlation function via derivatives of controlled-unitary channels.
    
    To this end, we first consider the OTOC on the joint system-environment Hilbert space,
    \begin{align}
        F(t)
        =\braket{
        A^{SE}(t)(B^S\otimes\mathds{1}_E)
        A^{SE}(t)(B^S\otimes\mathds{1}_E)
        }_{\rho^{SE}},
    \end{align}
    where
    \begin{align}
        A^{SE}(t)
        :=\mathcal{U}_{t,0}^{*}
        [A^S\otimes\mathds{1}_E]
    \end{align}
    denotes the Heisenberg-evolved observable under the joint system-environment dynamics, and $\rho^{SE}$ is the initial joint state. For brevity, we omit the identity operator $\mathds{1}_E$ whenever no ambiguity arises.

    Let us define the following maps 
    \begin{align}
        \mathcal{A}^S [\bullet] :=A^S \bullet A^S, \quad \mathcal{A}^{SE}_{t}[\bullet] := A^{SE}(t) \bullet A^{SE}(t). 
    \end{align}
    One can show that the two maps above are related via a unitary operation
    \begin{align}
        \mathcal{U}^{SE *}_{t,0} \circ \mathcal{A}^S \otimes \mathcal{I}_E \circ \mathcal{U}^{SE}_{t,0} [\bullet] = \mathcal{A}^{SE}_t[\bullet]
    \end{align}
    Then the OTOC $F(t)$ can be rewritten as 
    \begin{equation}
        \begin{split}
            F(t) &= \operatorname{Tr}\Big(B \mathcal{A}_t^{SE}[B\rho_{SE}]\Big)\\
        &= \operatorname{Tr}\Big(B \mathcal{U}^{SE *}_{t,0} \circ \mathcal{A}^S \otimes \mathcal{I}_E \circ \mathcal{U}^{SE}_{t,0}[B\rho_{SE}]\Big) .\label{(QDoD) OTOC step 1}
        \end{split}
    \end{equation}
    
\begin{figure}
        \centering
        \includegraphics[width=\linewidth]{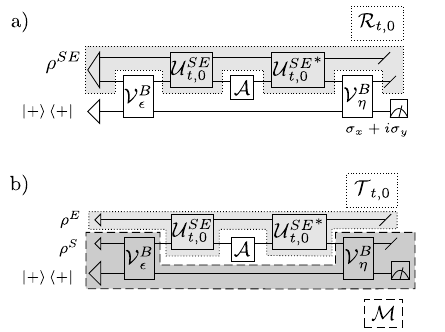}
        \caption{(a) Two diagrams showing the quantum comb with $t$, $\epsilon$, and $\eta$ dependence. Taking the $\epsilon, \eta$ derivativing and setting $\epsilon=0$ and $\eta=0$ gives the OTOC $F(t)$. The shaded region corresponds to the 6-legged quantum comb $\mathcal{R}_{t,0}$. (b) Showing the same diagram as panel (a) but with an initially uncorrelated state $\rho^{SE} = \rho^S \otimes \rho^E$. The shaded region corresponds to the four-legged quantum comb $\mathcal{T}_{t,0}$. The darker region corresponds to the higher-order quantum operator $\mathcal{M}_{\epsilon , \eta}$, where $\mathcal{M}:=-\partial_\epsilon \partial_\eta \mathcal{M}_{0,0}$ with $\mathcal{M}[\mathcal{T}_{t,0}] = F(t)$.}
        \label{(QDoD) Fig7-1}
    \end{figure}

    To express the operation of the observables $B$ in the above expression, we will extend the Hilbert space from $\mathcal{H}_S \otimes \mathcal{H}_E $ to $\mathcal{H}_a \otimes \mathcal{H}_S \otimes \mathcal{H}_E   $, where $\mathcal{H}_a$ is a qubit auxiliary space. In this extended space, let us define the following unitary 
    \begin{align}
        V_\epsilon^{O^{SE}} &:= \ketbra{0}{0}\otimes \mathds{1}_{SE} + \ketbra{1}{1}\otimes e^{-i \epsilon O^{SE}}\\
        \mathcal{V}_\epsilon^{O^{SE}} [\bullet]&:= V_\epsilon^{O^{SE}} \bullet V_\epsilon^{O^{SE}\dagger}
    \end{align}
    where $O^{SE}$ is some generic system-environment operator. Using this map, one can show the following relations 
    \begin{equation}
        \begin{split}
            i \partial_\epsilon \left.\mathcal{V}_\epsilon^{O^{SE}}\big[\ketbra{+}{+}\otimes \rho^{SE}\big]\right|_{\epsilon=0} &= - \frac{1}{2} \ketbra{0}{1}\otimes \rho^{SE} O^{SE} \\
        &\quad+ \frac{1}{2} \ketbra{1}{0}\otimes O^{SE}\rho^{SE} \\
        &+ \frac{1}{2} \ketbra{1}{1}\otimes \operatorname{ad}_{O^{SE}} [\rho^{SE}],
        \end{split}
    \end{equation}
    where $\operatorname{ad}_O[\bullet] = [O,\bullet]$ nad $\rho^{SE}$ is some arbitrary system-environment state. Furthermore, let us consider some system-environment map $\mathcal{E}$, then one can show that 
    \begin{equation}
        \begin{split}
            i \partial_\eta \left.\mathcal{V}_\eta^{O^{SE}_2}\right|_{\eta=0}\!\!\!\!\!\!\!\circ (\mathcal{I}_a \otimes \mathcal{E}_{SE})\circ i \partial_\epsilon \left.\mathcal{V}_\epsilon^{O^{SE}_1}\!\!\big[\ketbra{+}{+}\otimes \rho^{SE}\big]\right|_{\epsilon=0}\!\!\!\!\!\!&\\
            = \frac12\ketbra{0}{1}\otimes
    \mathcal E[\rho O_1]O_2+\frac12\ketbra{1}{0}\otimes
    O_2\mathcal E[O_1\rho]&\\
        +\frac12\ketbra{1}{1}\otimes
    \operatorname{ad}_{O_2}\circ
    \mathcal E\circ
    \operatorname{ad}_{O_1}[\rho]&
        \end{split}
    \end{equation}
    Notice that the $\ketbra{1}{0}$ term of the above expression has the same algebraic structure as the algebraic structure in Eq.~(\ref{(QDoD) OTOC step 1}) if we make the following identification 
    \begin{align}
        O^{SE}_1, O^{SE}_2 \rightarrow B^S \otimes \mathds{1}_E, \quad\mathcal{E}\rightarrow \mathcal{A}^{SE}_t
    \end{align}
    then we obtain 
    \begin{equation}
        \begin{split}
            &-\partial_{\epsilon } \partial_\eta 2 \bra{1} \mathcal{V}_\eta^{B^S} \circ (\mathcal{I}_a\otimes \mathcal{A}^{SE}_t )\\
        &\circ \mathcal{V}_\epsilon^{B^S}\big[\ketbra{+}{+}\otimes \rho^{SE}\big] \ket{0}\Big|_{\substack{\epsilon=0\\ \eta=0}} = B^S \mathcal{A}^{SE}_{t}[B^S \rho^{SE}]
        \end{split}
    \end{equation}
    Recalling that $2\ketbra{0}{1} =\sigma_x^{a} + i \sigma_y^a $, we obtain the following expression for the OTOC
    \begin{equation}
        \begin{split}
        F(t) =&-\partial_{\epsilon } \partial_\eta \operatorname{Tr}_{a SE}\Big[ (\sigma_x^a + i \sigma^a_y)\mathcal{V}_\eta^{B^S} \circ (\mathcal{I}_a\otimes \mathcal{A}^{SE}_t )\\
        &\circ \mathcal{V}_\epsilon^{B^S}\big[\ketbra{+}{+}\otimes \rho^{SE}\big] \Big]\Big|_{\substack{\epsilon=0\\ \eta=0}}
        \end{split}
    \end{equation}
    This is pictorially represented in Fig.~\ref{(QDoD) Fig7-1}(a), from which one can observe that the following 6-legged quantum comb 
    \begin{equation}
        \begin{split}
            \mathcal{R}[\mathcal{C}_1,\mathcal{C}_2,\mathcal{C}_3] = &\operatorname{Tr}_{SE}\Big[(\mathcal{C}_3\otimes\mathcal{I}_E)\circ \mathcal{U}^{SE*}_{t,0}\circ (\mathcal{C}_2 \otimes \mathcal{I}_{2})\\
            &\circ \mathcal{U}^{SE}_{t,0}\circ (\mathcal{C}_1\otimes \mathcal{I}_E)[\rho_SE]\Big]
        \end{split}
    \end{equation}
    where $\mathcal{C}_i$, $i=1,2,3$ are some genric maps acting on the sysrtem. 
    
\begin{figure}[h]
        \centering
        \includegraphics[width=\linewidth]{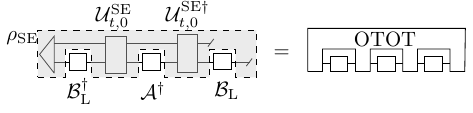}
        \caption{(a) Diagram showing the action of the quantum comb defined in Fig.~\ref{(QDoD) Fig7-1} on a quantum instrument $\mathcal{A}^S[\bullet] = A\bullet A$, which gives the reduced Heisenberg picture instrument. (b) Diagram showing how a quantum comb shown in panel (a) evolves in time as a CP-divisible quantum comb.}
        \label{(QDoD) fig:eg3-2}
    \end{figure}

    One could consider the time evolution of the above 6-legged quantum comb, but to connect to our formalism for the four-legged case, let us consider an initially uncorrelated state $\rho^{SE} = \rho^S \otimes \rho^E$; then one can reduce the above 6-legged quantum comb to the four-legged quantum comb 
    \begin{align}
        \mathcal{T}_{t,s} \circ\mathcal{C} [\bullet] := \operatorname{Tr}_E \Big[\mathcal{U}^{SE*}_{t,s}\circ (\mathcal{C}\otimes \mathcal{I}_E)\circ \mathcal{U}^{SE}_{t,0}[\bullet \otimes \rho^E]\Big] \label{(QDoD) OTOC step2}.
    \end{align}
    See Fig.~\ref{(QDoD) Fig7-1}(b) for its pictorial representation. 
    To understand the time evolution of this comb, let us define a new quantum instrument inspired by the reduced Heisenberg picture observable defined in Ref.~\cite{breuer1997stochastic}, which we denote \textit{reduced Heisenberg picture instrument}
    \begin{align}
        \mathcal{A}^{\text{red}}_t [\bullet]:= \operatorname{Tr}_{E} \Big(\mathcal{A}^{SE}_t [ \bullet \otimes \rho^E]\Big).
    \end{align}
    Then one can show that, acting with the quantum comb~(\ref{(QDoD) OTOC step2}) on the system quantum instrument $\mathcal{A}^S$, one obtains 
    \begin{align}
        \mathcal{T}_{t,0}\circ \mathcal{A}^S = \mathcal{A}^{\text{red}}_t.
    \end{align}
    See Fig.~\ref{(QDoD) fig:eg3-2}(a).
    Then, assuming that the quantum comb is CP-divisible, one obtains the following time evolution
    \begin{align}
        \mathcal{T}_{t_n , t_{n-1} } \circ \mathcal{T}_{t_{n-1},t_{n-2}}\circ \dots \mathcal{T}_{t_2 , t_1}\circ \mathcal{A}^{\text{red}}_{t_1} = \mathcal{A}^{\text{red}}_{t_n}
    \end{align}
    This evolution is pictorially represented in Fig.~\ref{(QDoD) fig:eg3-2}(b). 
    
    The OTOC in terms of this quantum comb is represented as 
    \begin{equation}
        \begin{split}
            F(t) = &-\partial_\epsilon \partial_\eta \operatorname{Tr}_{aS} \Big[2\ketbra{0}{1} \mathcal{V}_\eta^{B^S} \big(\mathcal{I}_a \otimes \mathcal{T}_{t,s}\circ \mathcal{A}^{\text{red}}_s \big)\\
        &\circ \mathcal{V}_\eta^{B^S}[\ketbra{+}{+}\otimes \rho^E] \Big]\Big|_{\substack{\epsilon=0 \\ \eta=0}}
        =: \mathcal{M}[\mathcal{T}_{t,s}\circ \mathcal{A}^{\text{red}}_s]
        \end{split}
    \end{equation}
    where we have introduced a higher-order quantum operator (HOQO) $M$ which includes all the operations on the quantity $\mathcal{T}_{t,s}\circ \mathcal{A}^{\text{red}}_s$ to give the OTOC above. 
    Notice that the above expression is linear in the quantum comb $\mathcal{T}_{t,s}$; therefore, taking the time derivative of the OTOC would directly translate to taking the time derivative of the quantum comb inside the HOQO $\mathcal{M}$, which gives the following master equation 
    \begin{equation}
        \begin{split}
             \frac{d F(t)}{dt} = &\mathcal{M}[\frac{\partial \mathcal{T}_{t,0}}{dt} \circ \mathcal{A}^{S}] = \mathcal{M}[\mathcal{L}\circ\mathcal{T}_{t,0} \circ \mathcal{A}^{S}]\\
        =&\mathcal{M}[\mathcal{L}\circ\mathcal{A}^{\text{red}}_{t}].
        \end{split}
    \end{equation}
    Therefore, the OTOC satisfies the dynamical equation whose generator is the comb generator defined in Theorem~\ref{(QDoD) Def: Comb master equation}. 

    In conclusion, we have shown that the QCME provides a dynamical equation for the OTOC. A similar treatment can be extended to other temporal correlation functions whenever the corresponding four-legged quantum comb can be identified. The experimental relevance of such a description is supported by measurements of OTOCs in nuclear magnetic resonance quantum simulators, trapped-ion systems, and digital quantum processors~\cite{li2017otoc,garttner2017otoc,green2022otoc}. Of particular relevance, Emori and Tajima developed a quantum-comb-based approach to evaluating OTOCs through quantum-process irreversibility~\cite{emori2023error} and subsequently investigated its experimental implementation on a trapped-ion quantum computer~\cite{emori2025measuring}. These developments suggest a possible route toward connecting experimentally accessible temporal correlations with the generator-based description of their evolution provided by our QCME. Establishing this connection quantitatively remains an interesting direction for future research.



\subsection{Open Quantum Dynamics in the Presence of Initial Correlations}
\label{sec:OQS}

In Example~\ref{(QDoD) Eg: evolution of channel} and Application~\ref{sec:otoc} we have considered the evolution of a channel and an observable. Our formalism also covers the evolution of a more non-trivial object: the time evolution of the intervention applied to an open quantum system. This extends the simple example of Example~\ref{(QDoD) Eg: evolution of state}, but instead of an initially uncorrelated state, we here consider an initially correlated one.

\begin{figure}[h]
        \centering
        \includegraphics[width=\linewidth]{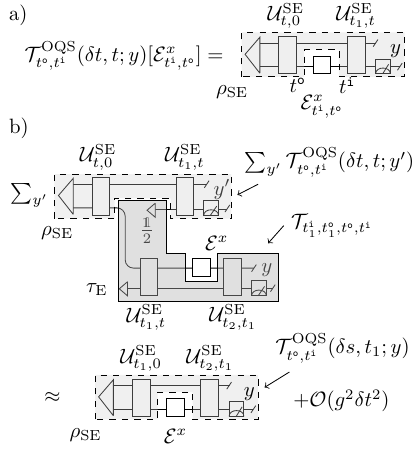}
        \caption{Two figures showing the dilation of the open quantum system (OQS) with two temporal interventions. (a) dilation of the OQS with two interventions via a instrument $\mathcal{E}^x [\bullet]= \ketbra{x}{x}\bullet \ketbra{x}{x}$ and a POVM $\xi^y = \ketbra{y}{y}$ where $x,y \in \{0,1\}$ in the computtional basis. (b) Time evolution of the OQS where the intervention to the system is replaced with an alternative intervention $\mathcal{T}_{t_1^\inp,t_1^\out,t^\out t^\inp}$, represented in a darker shade of grey. This is approximately equal to the exact time evolution of the OQS, $\mathcal{T}^{\text{OQS}}_{t^\out, t^\inp}$, up to order $\mathcal{O}(g^2 \delta t^2)$, where $\delta t$ is a shift in the intervention time representing the time evolution of the intervention time. $g$ is the coupling strength between the system and the environment.}
        \label{(QDoD) fig:eg4-1}
    \end{figure}

Consider an open quantum system $\mathcal{H}_S\otimes \mathcal{H}_E$, where $\mathcal{H}_S$ is an accessible quantum system and $\mathcal{H}_E$ is an inaccessible environment. For simplicity, we assume that the system and environment time-evolve with the time-homogeneous unitary $\mathcal{U}^{\text{SE}}_{t,s} = \mathcal{U}^{\text{SE}}_{t-s}$. Let us assume that an experimentalist performs some intervention on the system $\mathcal{H}_S$ at time $t$ via some measurement represented by a \textbf{completely positive trace-non-increasing (CPTNI)} map $ \mathcal{E}^{x}_{t^\out ,t^\inp} $. Then the system is measured immediately after at $t+\delta t$ using a \textbf{positive operator-valued measure (POVM)} $\xi_y$. This can be represented by the following 
    \begin{align}
        p(t+\delta t;y, t;x)&:= \text{Tr}_{SE} \bigg[(\xi_y\otimes \mathds{1}_E) \mathcal{U}^{\text{SE}}_{\delta t}\nonumber\\
        &\circ \mathcal{E}^{x}_{t^\out ,t^\inp}\otimes \mathcal{I}_E \circ \mathcal{U}^{\text{SE}}_{t}\big[\rho_{SE}\big]\bigg]\nonumber\\
        & =:\mathcal{T}^{\text{OQS}}_{t^\out ,t^\inp}(\delta t, t ; y) [\mathcal{E}^{x}_{t^\inp , t^\out}]
    \end{align}
    where we have defined a two-legged quantum comb $\mathcal{T}^{\text{OQS}}$ shown in Fig.~\ref{(QDoD) fig:eg4-1}(a).
    
    Then let us consider an infinitesimal time evolution of the experiment where the initial intervention is shifted by $\delta t$ and the second measurement is shifted by $\delta s$, which gives $p(t+\delta t +\delta s; y, t+\delta t ; x)$. This is implemented by the time evolution of the comb 
    \begin{align}
        \mathcal{T}^{\text{OQS}}_{t^\out ,t^\inp}(\delta t, t ; y) \rightarrow \mathcal{T}^{\text{OQS}}_{t^\out ,t^\inp}(\delta s, t+\delta t ; y)\label{(QDoD) Eq: Evolution of OQS experiment}
    \end{align}
    We first propose an evolution map that connects the two experiments above; then we give a microscopic justification for such an evolution map.

    Let us introduce a notation $t_n :=t+n \delta t $ for $n\in \mathbb{Z}$. We will use this notation explicitly in this example. Let us consider a generic four-legged quantum comb, $\mathcal{T}_{t^\inp_1 , t^\out_1 , t^\out , t^\inp}$ which acts on the open quantum system experiment as 
    \begin{align}
        \mathcal{T}_{t^\inp_1 , t^\out_1 , t^\out , t^\inp} \circ \mathcal{T}^{\text{OQS}}_{t^\out,t^\inp} (\delta t, t ; y) = \mathcal{T}^{\text{OQS}'}_{t^\out_1,t^\inp_1} (\delta s, t_1 ; y) \label{(QDoD) Eq: Proposed evolution of OQS experiment}
    \end{align}
    It remains unknown whether the resulting new experiment~(\ref{(QDoD) Eq: Proposed evolution of OQS experiment}) coincides with the actual evolved experiment~(\ref{(QDoD) Eq: Evolution of OQS experiment}). To justify such evolution, we need to give an explicit dilation to the comb. For this, we consider a simple toy model. First, let us assume that the system and environment time evolve according to the Hamiltonian $H= H_{\text{S}} + H_{\text{E}} + H_{\text{Int}}$, where 
    \begin{align}
        H_{\text{S}}=H_{\text{E}}=\frac{\omega}{2}\sigma_z, H_{\text{Int}} = g(\sigma^{\text{S}}_+ \otimes \sigma^{\text{E}}_- +\sigma^{\text{S}}_- \otimes \sigma^{\text{E}}_+ )\label{(QDoD): system-env. Hamiltonian for OQS example}
    \end{align}
    with unitary given by $\mathcal{U}^{\text{SE}}_t [\bullet] = U_t \bullet U_t^\dagger$, $U_t =e^{-i H t} $. We perform the measurement with a simple projective measurement $\xi^x = \ketbra{x}{x}$ and an instrument $\mathcal{E}^x [\bullet]= \ketbra{x}{x}\bullet\ketbra{x}{x}$, where $x\in \{0,1\}$. We will also assume that the infinitesimal evolution time before and after the intermediate measurement is equal, meaning $\delta t= \delta s$. Let $\tau_{\text{E}} $ be a reference environment state that we can choose arbitrarily. Then we define the four-legged quantum comb as 
    \begin{align}
        \mathcal{T}_{t^\inp_1 , t^\out_1 , t^\out , t^\inp} (\delta s , \delta t ; y) = \text{Tr}_{\text{SE}} \bigg[(\xi^y \otimes \mathds{1}_{\text{E}}) \mathcal{U}_{\delta s} \nonumber\\
        \circ (\bullet \otimes \mathcal{I}_{\text{E}})\circ \mathcal{U}_{\delta t} [\bullet \otimes \tau_E]\bigg] \frac{\mathds{1}_{\text{E}}}{2}.
    \end{align}
    Since this comb depends on the outcome $y$, when we act on the OQS experiment, we sum over the previous experimental result. More specifically, we consider the following evolution, 
    \begin{align}
         \mathcal{T}_{t^\inp_1 , t^\out_1 , t^\out , t^\inp} (\delta s , \delta t ; y)  \circ \sum_{y'} \mathcal{T}^{\text{OQS}}_{t^\out,t^\inp} (\delta t, t ; y') = \mathcal{T}^{\text{OQS}'}_{t^\out,t^\inp} (\delta s, t_1 ; y).
    \end{align}
    This has an operational interpretation. Instead of the measurement intervention at time $t$, the comb $\mathcal{T}_{t^\inp_1 , t^\out_1 , t^\out , t^\inp} (\delta s , \delta t ; y)  $ would reroute the system's state and instead perform a different experiment with some reference environmental state $\tau_E$ that we prepared. The previous experiment is discarded by summing over the result. Intuitively, this would yield a different result from the actual evolved experiment~(\ref{(QDoD) Eq: Evolution of OQS experiment}) because we change the experimental setup. However, we show below that the experimental result agrees up to a small correction of order $g^2 \delta t^2$. This follows from the microscopic justification provided by the comb extension of the Born approximation, which we explored in Section~\ref{sec:quantumdynamicsofdynamics-microscopic}. A pictorial representation of the above statement is shown in Fig.~\ref{(QDoD) fig:eg4-1}(b).

    To explicitly show the above claim, let us consider the following quantity 
    \begin{align}
        D(x,y):=& \mathcal{T}_{t^\inp_1 , t^\out_1 , t^\out , t^\inp} (\delta s , \delta t ; y)[\mathcal{E}^x]   \circ \sum_{y'} \mathcal{T}^{\text{OQS}}_{t^\out,t^\inp} (\delta t, t ; y')\nonumber\\
        &- \mathcal{T}^{\text{OQS}}_{t^\out ,t^\inp}(\delta s, t_1 ; y)[\mathcal{E}^x] .
    \end{align}
    Choosing a reference state $\tau_E = \mathds{1}/2$, we find the results in the computational basis is 
    \begin{align}
        D(x,y) = \mathcal{O}(g^2 \delta t^2),
    \end{align}
    for all $x,y\in \{0,1\}$. The experimental result $p(t_1;y, t;x)$ is given by 
    \begin{align}
        p(t_1;0, t;0) = \frac{1}{2}+\mathcal{O}(g^2 \delta t^2)&,~ p(t_1;1, t;1) = \frac{1}{2}+\mathcal{O}(g^2 \delta t^2)\\
        p(t_1;0, t;1) = \mathcal{O}(g^2 \delta t^2)&,~ p(t_1;1, t;0) = \mathcal{O}(g^2 \delta t^2)
    \end{align}
    We note that the two experiments coincide up to second order in the coupling-time-duration combination.
    
    If we instead consider a different reference state, namely $\tau_E = \text{diag}(1/3, 2/3)$, then we obtain
    \begin{align}
        D(x,y) = \left\{\begin{array}{cc}
            \mathcal{O}(g^4 \delta t^4) & \text{ for }x=0,y=0 \\
            \mathcal{O}(g^2 \delta t^2) & \text{ else}
        \end{array}\right.
    \end{align}
    Here, the results of the two experiments coincide up to fourth order in the weak expansion for $x=0,y=0$.

\section{Conclusions and Outlook}
\label{sec:conclusions}

We have developed a framework for the continuous-time evolution of quantum combs, in which the dynamical object is not a quantum state but a process with multiple temporal legs. The resulting \textit{quantum comb master equation} (QCME) plays the role for two-time processes that the GKSL equation plays for states. This answers the question posed in the Introduction: how to retain temporal correlations that the quantum regression theorem misses, even when the reduced dynamics is CP-divisible.

We derived the QCME through two complementary approaches. Axiomatically (Sec.~\ref{sec:quantumdynamicsofdynamics}), we introduced CP-divisibility for quantum combs and showed in Theorem~\ref{Thm: main result} that it yields the GKSL-like master equation~\eqref{(QDoD)Eq:CombME}, subject to positivity and generalized trace-preservation conditions. A distinctive feature of this equation is the presence of \textit{encoder--decoder correlation terms}, which couple different temporal legs of the comb beyond the conventional GKSL contributions. Microscopically (Sec.~\ref{sec:quantumdynamicsofdynamics-microscopic}), we started from a quantum-comb analog of the von Neumann equation and applied generalized Born--Markov and coarse-graining approximations to derive Eq.~\eqref{(QDoD)Eq:FinalMicro}. The resulting equation reproduces the GKSL-like structure obtained axiomatically, provides a physical interpretation of its terms, and identifies the assumptions under which it emerges from the underlying system--environment dynamics.

The QCME treats a range of problems within one formalism: linear response functions (Example~\ref{(QDoD) Eg: evolution of state}), channel evolution motivated by process tomography (Example~\ref{(QDoD) Eg: evolution of channel}), correlations between Heisenberg-picture evolution of an observable and Schr\"odinger-picture evolution of a state arising from observe-and-prepare operations (Example~\ref{(QDoD)eg:SchHeiCorr}), the evolution of out-of-time-order correlators (OTOCs) (Sec.~\ref{sec:otoc}), and open-system dynamics with initial system--environment correlations (Sec.~\ref{sec:OQS}). Once the initial comb is specified, the QCME determines any two-time correlator at later times, just as the state master equation determines the expectation value of any observable. Specifically, the OTOC application yields a time-local dynamical equation governed by the quantum comb generator, while the initially correlated qubit model demonstrates that an alternative comb-based evolution reproduces shifted-intervention statistics up to corrections of order $g^2\delta t^2$.

Several of the quantities unified here have previously been studied as static, fixed-time objects built from combs or process tensors: error--disturbance tradeoffs have been formulated via irreversibility of a comb-valued measurement process~\cite{emori2023error,nakajima_2024}, and chaos diagnostics have been encoded in the out-of-time-order tensor, a higher-order map capturing OTOCs for all operator choices at once~\cite{Zonnios_2022}. The QCME complements these characterizations with the dynamical equation that propagates such objects in time, in direct analogy to how the GKSL equation propagates a state once its instantaneous structure is known. In this sense, our OTOC example (Sec.~\ref{sec:otoc}) can be read as furnishing the time evolution of the out-of-time-order tensor, and the same construction extends to the comb-valued error and disturbance measures above, which we leave for future work.

Our results have limitations that suggest directions for future work. First, our treatment is presently restricted to two-time combs; the construction extends straightforwardly to combs with more temporal legs, which would give direct access to higher-order, multi-time correlators such as third- and fourth-order response functions, at the cost of a more involved generator. Treating non-CP-divisible comb dynamics would extend the framework to strongly non-Markovian regimes. Second, the cost of propagating a comb grows with its number of legs, so efficient numerical methods, for example tensor-network or trajectory unravellings of the QCME, are an important next step. Third, the framework opens a route to the quantum thermodynamics of multi-time processes and nonequilibrium open systems, and to characterizing non-Markovian processes in experimental platforms where interventions at intermediate times are routine, such as superconducting circuits and trapped ions.


\section*{Acknowledgements}
We acknowledge support from the Royal Society International Science Partnership Funds through the International Collaboration Award ICA$\backslash$R2$\backslash$252240.


\appendix
\section{Generalized Trace--Preservation of Strictly Temporal-Ordered Quantum Combs}\label{(QDoD) Apx: generalized TP}
\begin{figure}
    \centering
    \includegraphics[width=\linewidth]{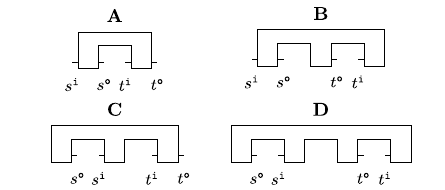}
    \caption{The four temporal-ordering representatives.
Horizontal spacing is schematic and does not specify time intervals.}
    \label{(QDoD) Fig apx-2}
\end{figure}

    Assuming without loss of generality that $t^\out, t^\inp$ pairs are always later than the $s^\out, s^\inp$ pairs, we have four possible temporal orderings of the four time variables 
    \begin{align}
        &\text{(A): } t^\out > t^\inp >s^\out >s^\inp ,\quad\text{(B): } t^\inp > t^\out >s^\out >s^\inp\\
        &\text{(C): } t^\out > t^\inp >s^\inp >s^\out , \quad\text{(D): } t^\inp > t^\out >s^\inp >s^\out.
    \end{align}
    Fig.~\ref{(QDoD) Fig apx-2} illustrates this. 

\begin{figure}
    \centering
    \includegraphics[width=\linewidth]{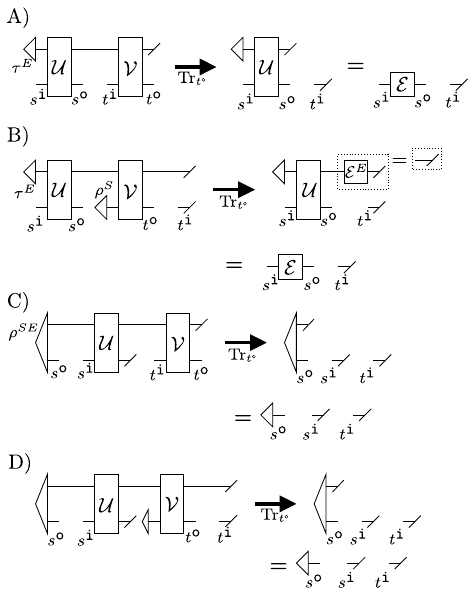}
    \caption{Four diagrams showing the generalized TP condition on the dilation of the 4 four types of temporally ordered quantum combs. Each rectangle represents some arbitrary unitary denoted by $\mathcal{U}$ and $\mathcal{V}$. Square boxes represent CPTP maps acting on either the system or the environment.}
    \label{(QDoD) Fig:apxA-1}
\end{figure}

In Fig.~\ref{(QDoD) Fig:apxA-1}, we present the general dilation of the quantum comb we consider in this paper. rectangles in Fig.~\ref{(QDoD) Fig:apxA-1} represtnes seom untiary and denotee by $\mathcal{U}$ or $\mathcal{V}$. They can be any unitary operation; for example, a standard unitary map $\mathcal{U}[\bullet] = U \bullet U^\dagger$ or its adjoint $\mathcal{U}^*[\bullet] = U \bullet U^\dagger$. Using the property of unitary and CPTP maps, each quantum comb satisfies the generalized TP conditions defined in Def.~\ref{(QDoD) Def: quantum comb}. 

For clarity, let us comment on how the generalized TP condition arises from type--B. Taking a trace over the $t^\out$--leg, the unitary $\mathcal{V} $ together with some arbitrary quantum state $\rho^S$ forms a CPTP map that acts on the environment 
\begin{align}
    \mathcal{E}^E [\bullet^E] = \operatorname{Tr}_S (\mathcal{V}[\rho^S\otimes \bullet^E]).
\end{align}
Due to the TP condition of the CPTP map, the environment trace on the map $\mathcal{E}^E$ is equivalent to the trace on the environment 
\begin{align}
    \operatorname{Tr}_E \mathcal{E}^E [\bullet] = \operatorname{Tr}_E \bullet .
\end{align}
Therefore, the $s^\out$ and $s^\inp$ legs form another CPTP map on the system with respect to an arbitrary environment state $\tau^E$. 
\begin{align}
    \mathcal{E}[\bullet^S] = \operatorname{Tr}_E (\mathcal{U}[\bullet\otimes \tau^E]). 
\end{align}
This gives rise to the generalized TP condition defined in Def.~\ref{(QDoD) Def: quantum comb}. The generalized TP condition for other types arises by following a similar argument as above.

\section{Proofs}\label{(QDoD)Appendix:Proofs}
This appendix collects the proofs of the results presented in the main text.
\subsection{Proof of Corollary~\ref{(QDoD)Coro:Kraus Representation}}\label{(QDoD)Appendix:Proofs-1}

\begin{proof}
    Given quantum combs $\mathsf{T}_\alpha$ and $\mathsf{M}_\beta$, it can be written formally as 
    \begin{align}
        \mathsf{T}_\alpha = \sum_{\mathbf{i}_\alpha ,\mathbf{j}_\alpha} \mathsf{T}_{\mathbf{i}_\alpha \mathbf{j}_\alpha} \ketbra{\mathbf{i}_\alpha}{\mathbf{j}_\alpha}, \quad \mathsf{M}_{\beta} = \sum_{\mathbf{i}_\beta ,\mathbf{j}_\beta} \mathsf{T}_{\mathbf{i}_\beta \mathbf{j}_\beta} \ketbra{\mathbf{i}_\beta}{\mathbf{j}_\beta}.\label{(QDoD)Proof:Kraus rep. setp 1}
    \end{align}
    If $\alpha = (t^\out, t^\inp, s^\out,s^\inp)$, then $\mathbf{i}_\alpha := (i_{t^\out},i_{t^\inp}, i_{s^\out},i_{s^\inp}) $ where $\ket{\mathbf{i}_\alpha} := \ket{i_{t^\out},i_{t^\inp}, i_{s^\out},i_{s^\inp}} $ is a basis of the Hilbert space $\mathcal{H}_\alpha := \mathcal{H}_{t^\out}\otimes\mathcal{H}_{t^\inp}\otimes\mathcal{H}_{s^\out}\otimes\mathcal{H}_{s^\inp}$. Let us define the following new sets 
    \begin{align}
        \gamma:= \alpha \cap \beta, \quad\gamma_-:=\beta\setminus \alpha ,\quad \gamma_+:=\alpha\setminus \beta.
    \end{align}
    By constrcution, we have $\alpha = \gamma_+ \cap \gamma$ and $\beta = \gamma_- \cap \gamma$. Using the new set, one can rewrite the Eq.~(\ref{(QDoD)Proof:Kraus rep. setp 1}) as 
    \begin{align}
        &\mathsf{T}_\alpha = \mathsf{T}_{\gamma_+ \cup \gamma} = \sum_{\substack{\mathbf{i}_{\gamma_+},\mathbf{j}_{\gamma_+}\\\mathbf{i}_{\gamma},\mathbf{j}_{\gamma}}} \mathsf{T}_{\mathbf{i}_{\gamma_+},\mathbf{j}_{\gamma_+},\mathbf{i}_{\gamma},\mathbf{j}_{\gamma}} \ketbra{\mathbf{i}_{\gamma_+}}{\mathbf{j}_{\gamma_+}}\otimes \ketbra{\mathbf{i}_{\gamma}}{\mathbf{j}_{\gamma}}\label{(QDoD)Proof:Kraus rep. step 2-1}\\
        &\mathsf{M}_\beta = \mathsf{M}_{\gamma_- \cup \gamma} = \sum_{\substack{\mathbf{i}_{\gamma_-},\mathbf{j}_{\gamma_-}\\\mathbf{i}_{\gamma},\mathbf{j}_{\gamma}}} \mathsf{M}_{\mathbf{i}_{\gamma_-},\mathbf{j}_{\gamma_-},\mathbf{i}_{\gamma},\mathbf{j}_{\gamma}} \ketbra{\mathbf{i}_{\gamma_-}}{\mathbf{j}_{\gamma_-}}\otimes \ketbra{\mathbf{i}_{\gamma}}{\mathbf{j}_{\gamma}}.\label{(QDoD)Proof:Kraus rep. setp 2}
    \end{align}
    Using the definition of the link product, we have 
    \begin{align}
        &\mathsf{T}_\alpha \star \mathsf{M}_{\beta} = \text{Tr}_{\gamma} \Big[\big(\mathds{1}_{\gamma_-}\otimes \mathsf{T}_{\gamma_+ \cup \gamma }^{T_{\gamma}}\big)\big(\mathsf{M}_{\gamma_- \cup \gamma} \otimes \mathds{1}_{\gamma_+}\big) \Big]\nonumber\\
        & = \sum \mathsf{T}_{\mathbf{i}_{\gamma_+},\mathbf{j}_{\gamma_+},\mathbf{i}_{\gamma},\mathbf{j}_{\gamma}} \mathsf{M}_{\mathbf{i}_{\gamma_-}',\mathbf{j}_{\gamma_-}',\mathbf{i}_{\gamma},\mathbf{j}_{\gamma}} \ketbra{\mathbf{i}_{\gamma_-}'}{\mathbf{j}_{\gamma_-}'}\otimes \ketbra{\mathbf{i}_{\gamma_+}}{\mathbf{j}_{\gamma_+}},\label{(QDoD)Proof:Kraus rep. setp 3}
    \end{align}
    where the sum is assumed for all indices. Note that $T_{\gamma}$ is a partial transpose defined as 
    \begin{align}
        \mathsf{T}_{\gamma_+ \cup \gamma }^{T_{\gamma}} = \sum \mathsf{T}_{\mathbf{i}_{\gamma_+},\mathbf{j}_{\gamma_+},\mathbf{i}_{\gamma},\mathbf{j}_{\gamma}} \ketbra{\mathbf{i}_{\gamma_+}}{\mathbf{j}_{\gamma_+}}\otimes \ketbra{\mathbf{j}_{\gamma}}{\mathbf{i}_{\gamma}}.
    \end{align}
    Since $\gamma_+ = \alpha \setminus \beta$, the corresponding basis $\ket{\mathbf{i}_{\gamma_+}}$ is a basis of the Hilbert spcace $\mathcal{H}^2=\mathcal{H}\otimes \mathcal{H}$ where $\mathcal{H}$ is a Hilbert space assoctaeted with each open legs. Let $\text{dim}(\mathcal{H})=d$, then one can consider a polarization operator basis of the linear operator space $\mathcal{L}(\mathcal{H}^2)$ to be
    \begin{align}
        K_{pq} &:= \sum^{d^2}_{\mathbf{i}_{\gamma_+},\mathbf{j}_{\gamma_+}} \sqrt{\frac{2p+1}{2s+1}}C^{s,m_{\mathbf{i}_{\gamma_+}}}_{s,m_{\mathbf{j}_{\gamma_+}},pq} \ketbra{\mathbf{i}_{\gamma_+}}{\mathbf{j}_{\gamma_+}}\label{(QDoD) Eq:Definition of POB 1}\\
        K_{pq}^\dagger &:= \sum^{d^2}_{\mathbf{i}_{\gamma_+},\mathbf{j}_{\gamma_+}} \sqrt{\frac{2p+1}{2s+1}}C^{s,m_{\mathbf{i}_{\gamma_+}}}_{s,m_{\mathbf{j}_{\gamma_+}},pq} \ketbra{\mathbf{j}_{\gamma_+}}{\mathbf{i}_{\gamma_+}},\label{(QDoD) Eq:Definition of POB 2}
    \end{align}
    where $s=(d^2-1)/2$, $p=0,1,2,\dots , 2s$, $q=-p , \dots , p$ and $m_1 =s , m_2 = s-1, \dots m_{d^2}=-s$. The polarization operator basis satisfies the following completeness relation 
    \begin{align}
        \sum_{pq} (K_{pq})_{\mathbf{i}\mathbf{j}}(K_{pq}^\dagger)_{\mathbf{j}'\mathbf{i}'}   = \delta_{\mathbf{i}\mathbf{i}'}\delta_{\mathbf{j}\mathbf{j}'}, \quad\text{Tr}\big(K^\dagger_{p'q'}K_{pq}\big) = \delta_{pp'}\delta_{qq'}.\label{(QDoD)Proof:Kraus rep. setp 7}
    \end{align}
    For the rest of the proof, we use the compressed notation $\mathfrak{p}:= (pq)$,.

    Next, let us define the following coefficient 
    \begin{align}
        \chi_{\mathfrak{p}\mathfrak{p}'} := \sum_{\substack{\mathbf{i}_{\gamma_+},\mathbf{j}_{\gamma_+}\\\mathbf{i}_{\gamma},\mathbf{j}_{\gamma}}} \mathsf{T}_{\mathbf{i}_{\gamma_+},\mathbf{j}_{\gamma_+},\mathbf{i}_{\gamma},\mathbf{j}_{\gamma}} (K_\mathfrak{p})_{\mathbf{i}_{\gamma_+}\mathbf{i}_{\gamma}}(K^\dagger_{\mathfrak{p}'})_{\mathbf{j}_{\gamma}\mathbf{j}_{\gamma_+}},\label{(QDoD)Proof:Kraus rep. setp 6}
    \end{align}
    where $(K_\mathfrak{p})_{\mathbf{i}_{\gamma_+}\mathbf{i}_{\gamma}}:= \bra{\mathbf{i}_{\gamma_+}}K_{\mathfrak{p}}\ket{\mathbf{i}_{\gamma}}$, and $(K_{\mathfrak{p}'}^\dagger)_{\mathbf{j}_{\gamma}\mathbf{j}_{\gamma_+}}:= \bra{\mathbf{j}_{\gamma}}K_{\mathfrak{p}'}^\dagger\ket{\mathbf{j}_{\gamma_+}}$.
    Using the completeness relation, the inverse is given by 
    \begin{align}
        \mathsf{T}_{\mathbf{i}_{\gamma_+},\mathbf{j}_{\gamma_+},\mathbf{i}_{\gamma},\mathbf{j}_{\gamma}} = \sum_{\mathfrak{p}\mathfrak{p}'} \chi_{\mathfrak{p}\mathfrak{p}'} (K^\dagger_{\mathfrak{p}})_{\mathbf{i}_{\gamma} \mathbf{i}_{\gamma_+}}(K_{\mathfrak{p}'})_{ \mathbf{j}_{\gamma_+}\mathbf{j}_{\gamma}}.\label{(QDoD)Proof:Kraus rep. step 4}
    \end{align}
    Furthermore, one can show that 
    \begin{align}
        K_{\mathfrak{p}}\ketbra{\mathbf{i}}{\mathbf{j}}K^\dagger_{\mathfrak{p}'} = \sum_{\mathbf{i}',\mathbf{i}''}(K_{\mathfrak{p}}^\dagger)_{\mathbf{i}\mathbf{i}'} \ketbra{\mathbf{i}'}{\mathbf{i}''}(K_{\mathfrak{p}'})_{\mathbf{i}'' \mathbf{j}}.\label{(QDoD)Proof:Kraus rep. step 5}
    \end{align}
    Then inserting Eq.~(\ref{(QDoD)Proof:Kraus rep. step 4}) into Eq.~(\ref{(QDoD)Proof:Kraus rep. setp 3}) and rewriting using (\ref{(QDoD)Proof:Kraus rep. step 5}) we find
    \begin{align}
        \mathsf{T}_\alpha \star \mathsf{M}_{\beta}& = \sum_{\substack{\mathbf{i}_{\gamma_-},\mathbf{i}_{\gamma}\\\mathbf{j}_{\gamma_-},\mathbf{j}_{\gamma}}}\sum_{\mathfrak{p},\mathfrak{p}'} \chi_{\mathfrak{p}\mathfrak{p}'} \mathsf{M}_{\mathbf{i}_{\gamma_-}' \mathbf{j}_{\gamma_-}' \mathbf{i}_{\gamma}\mathbf{j}_{\gamma}}\nonumber\\
        &\times K_\mathfrak{p} \ketbra{\mathbf{i}_{\gamma}}{\mathbf{j}_{\gamma}}K^\dagger_{\mathfrak{p}'} \otimes \ketbra{\mathbf{i}_{\gamma_-}'}{\mathbf{j}_{\gamma_-}'}\nonumber\\
        & = \sum_{\mathfrak{p}\mathfrak{p}'}\chi_{\mathfrak{p}\mathfrak{p}'} \big(K_{\mathfrak{p}}\otimes \mathds{1}_{\gamma_-}\big)\mathsf{M}_{\beta}\big(K^\dagger_{\mathfrak{p}'}\otimes \mathds{1}_{\gamma_-}\big)
    \end{align}

    Finally let us show that the coefficient $\chi_{\mathfrak{p}\mathfrak{p}'}$ defined in Eq.~(\ref{(QDoD)Proof:Kraus rep. setp 6}) is a positive semidefinite matrix for indices $(\mathfrak{p},\mathfrak{p}')$. First, recall that the quantum comb $\mathsf{T}_\alpha$ can be written as Eq.~(\ref{(QDoD)Proof:Kraus rep. step 2-1}). Let us define a new vector 
    \begin{align}
        \ket{v_\mathfrak{p}}&:= \sum_{i_{\gamma}i_{\gamma_{+\epsilon}}} (K^\dagger_{\mathfrak{p}})_{i_{\gamma}i_{+\epsilon}}\ket{i_{\gamma_{+\epsilon}}i_{\gamma}},\\
        \bra{v_\mathfrak{p}}&:= \sum_{i_{\gamma}i_{\gamma_{+\epsilon}}} (K_\mathfrak{p})_{i_{\gamma_{+\epsilon}}i_{\gamma}}\bra{i_{\gamma_{+\epsilon}}i_{\gamma}}.
    \end{align}
    Using the completeness relation~(\ref{(QDoD)Proof:Kraus rep. setp 7}), one can show that $\braket{v_{\mathfrak{p}}|v_\mathfrak{p}'} = \delta_{\mathfrak{p}\mathfrak{p}'}$. Furthermore, since the polarization operator basis forms a basis for the space $\mathcal{L}(\mathcal{H}^2)$, the above vectors form a basis for the extended space $\mathcal{H}^4 = \mathcal{H}_\alpha$. Then acting this on Eq.~(\ref{(QDoD)Proof:Kraus rep. step 2-1}), we find 
    \begin{align}
        \bra{v_{\mathfrak{p}}}\mathsf{T}_{\alpha} \ket{v_{\mathfrak{p}'}} = \chi_{\mathfrak{p}\mathfrak{p}'}.
    \end{align}
    where we have used the deintion~(\ref{(QDoD)Proof:Kraus rep. setp 6}). The above quantity is positive due to the positivity of the quantum comb $\mathsf{T}_{\alpha}$. Next, let us consider an arbitrary vector $\ket{w}=\in \mathcal{H}_{\alpha}$. Then the coefficient $\chi$ can be expanded as 
    \begin{align}
        \bra{w}\chi\ket{w} &= \sum_{\mathfrak{p}\mathfrak{p}'} w^{*}_{\mathfrak{p}}w_{\mathfrak{p}'}\chi_{\mathfrak{p}\mathfrak{p}'}=\sum_{\mathfrak{p}\mathfrak{p}'} w_{\mathfrak{p}}w^{*}_{\mathfrak{p}'}\bra{v_{\mathfrak{p}}}\mathsf{T}_{\alpha} \ket{v_{\mathfrak{p}'}} \nonumber\\
        &= \sum_{\mathfrak{p}\mathfrak{p}'}\bra{v^{(w)}_{\mathfrak{p}}}\mathsf{T}_{\alpha} \ket{v^{(w)}_{\mathfrak{p}'}}
    \end{align}
    where we have defined $\ket{v^{(w)}_{\mathfrak{p}}} := w_{\mathfrak{p}}\ket{v_{\mathfrak{p}}}$. Due to the positivity of the quantum comb $\mathsf{T}_{\alpha}$, we observe that $\bra{w}\chi\ket{w}\geq 0$ for all $\ket{w}\in\mathcal{H}_{\alpha}$, which confirms the positive semidefiniteness of the coefficient matrix in the index $\mathfrak{p}$.
    \end{proof}

\subsection{Proof of Theorem~\ref{Lemma: GKSL form} and Corollary~\ref{Alternative form of GKSL}}\label{(QDoD)Appendix:Proofs-2}

\begin{proof}[Proof. Theorem~\ref{Lemma: GKSL form} and Corollary~\ref{Alternative form of GKSL}.]
    We will present the proof for the decoding generator~(\ref{Eq: GKSL form of decoding generator}). A similar proof trivially extends to the encoding generator. 

    First, recall that from definition~\ref{Def: dynamical decoding/encoding generators}, we have 
    \begin{align}
        \mathcal{L}_{D} \circ \mathcal{T}_{\alpha}  =\lim_{\epsilon\rightarrow 0^+} \frac{\mathcal{T}_{\beta_{+\epsilon}} (\gamma_1 + \epsilon , \gamma_2+0) -\mathcal{I} }{\epsilon} \circ \mathcal{T}_{\alpha}. \label{(QDoD)Proof:GKSL form step 1}
    \end{align}
    Using the corollary~\ref{(QDoD)Coro:Kraus Representation}, we have 
    \begin{align}
        &\text{Choi} \big(\mathcal{T}_{\beta_{+\epsilon}} \circ \mathcal{T}_{\alpha}  \big)= \sum_{\mathfrak{p},\mathfrak{p}'} \chi_{\mathfrak{p}\mathfrak{p}'} (\gamma_1+ \epsilon , \gamma_2 + 0)\nonumber \\
        &\times (K_{\mathfrak{p}}^{\beta_{+\epsilon}\cap \alpha}\otimes \mathds{1}^{\alpha\setminus\beta_{+\epsilon}}) \mathsf{T}_\alpha  ({K_{\mathfrak{p}'}^{\beta_{+\epsilon}\cap \beta}}^\dagger \otimes \mathds{1}^{\alpha\setminus\beta_{+\epsilon}}) .\label{(QDoD)Proof:GKSL form step 2}
    \end{align}
    where the coefficient $\chi_{\mathfrak{p}\mathfrak{p}'}$ is defined in Eq.~(\ref{(QDoD) Eq: Definition of chi}) with respect to the Choi representation of the quantum comb $\mathcal{T}_{\beta_{+\epsilon}}$: 
    \begin{align}
        \chi_{\mathfrak{p}\mathfrak{p}'} (\gamma_1+ \epsilon , \gamma_2 + 0) &= \sum_{\substack{i_{\gamma_+}i_{\gamma}\\j_{\gamma_+}j_{\gamma}}}\mathsf{T}_{i_{\gamma_+}j_{\gamma_+}i_{\gamma}j_{\gamma}} (\gamma_1+ \epsilon , \gamma_2 + 0)\nonumber\\
        &\times \left(K_{pq}\right)_{i_{\gamma_+}i_{\gamma}}\left(K_{p'q'}^\dagger\right)_{j_{\gamma}j_{\gamma_+}}
    \end{align}
    where we have $\gamma = \beta_{+\epsilon} \cap \alpha$ and $\gamma_+ = \alpha \setminus \beta_{+\epsilon}$. For example~\ref{(QDoD) Eg: evolution of channel}, we have $\gamma = (t^\out,t^\inp)$ and $\gamma_{+}= (s^\out,s^\inp)$. 

    Using Eq.~(\ref{(QDoD)Proof:GKSL form step 2}), we can express the Choi representation of the generator~(\ref{(QDoD)Proof:GKSL form step 1}) in the following form 
    \begin{align}
        &\text{Choi} \big(\mathcal{L}_{D} \circ \mathcal{T}_{\alpha}\big) = \lim_{\epsilon \rightarrow 0^+} \frac{\sum_{\mathfrak{p}\mathfrak{p}'}\chi_{\mathfrak{p}\mathfrak{p}'} K_\mathfrak{p} \mathsf{T}_{\alpha} K^{\dagger}_{\mathfrak{p}'}-\mathsf{T}_{\alpha}}{\epsilon}\nonumber\\
        & =\lim_{\epsilon \rightarrow 0^+} \frac{\frac{1}{d^2_S}\chi_{\mathfrak{0}\mathfrak{0}} \mathsf{T}_{\alpha}-\mathsf{T}_{\alpha}}{ \epsilon}+ \lim_{\epsilon \rightarrow 0^+} \frac{\sum_{\mathfrak{p}\mathfrak{0}}\frac{1}{d_S}\chi_{\mathfrak{p}\mathfrak{0}} K_\mathfrak{p} \mathsf{T}_{\alpha} }{\epsilon}\nonumber\\
        & +\lim_{\epsilon \rightarrow 0^+} \frac{\sum_{\mathfrak{0}\mathfrak{p}'}\frac{1}{d_S}\chi_{\mathfrak{0}\mathfrak{p}'}  \mathsf{T}_{\alpha} K^{\dagger}_{\mathfrak{p}'}}{\epsilon} \nonumber\\
        &+ \lim_{\epsilon \rightarrow 0^+} \frac{\sum_{\mathfrak{p}\not=\mathfrak{0},\mathfrak{p}'\not=\mathfrak{0}}\chi_{\mathfrak{p}\mathfrak{p}'} K_\mathfrak{p} \mathsf{T}_{\alpha} K^{\dagger}_{\mathfrak{p}'}}{\epsilon}.
    \end{align}
    where we have wrote $(K_{\mathfrak{p}}^{\beta_{+\epsilon}\cap \alpha}\otimes \mathds{1}^{\alpha\setminus\beta_{+\epsilon}})  = K_\mathfrak{p}$ for brevity. We have also used the property of the polarization operator basis $K_{\mathfrak{0}} = K_{00} = \mathds{1} / d_S$. Let us define the following functions
    \begin{align}
        a_{\mathfrak{0}\mathfrak{0}}(\lambda) := \lim_{\epsilon \rightarrow 0^+} \frac{\chi_{\mathfrak{0}\mathfrak{0}} -d^2_S}{ \epsilon} , \quad d_{\mathfrak{p}\mathfrak{p}'} :=\lim_{\epsilon\rightarrow 0^+} \frac{\chi_{\mathfrak{p}\mathfrak{p}'}}{\epsilon} 
    \end{align}
    where $\mathfrak{p}\not= \mathfrak{0}$ and $\mathfrak{p}'\not= \mathfrak{0}$. Note that this is defined in Eq.~\eqref{(QDoD) Eq: POB function d}. Then, using definition~\ref{(QDoD)Def: path-dependent POB} the Choi representation of the decoding generator is expressed as follows, 
    \begin{align}
         \text{Choi} \big(\mathcal{L}_{D} \circ \mathcal{T}_{\alpha}\big) &= \frac{a_{\mathfrak{0}\mathfrak{0}}}{d^2_S} \mathsf{T}_{\alpha} + D \mathsf{T}_{\alpha} + \mathsf{T}_{\alpha} D^{\dagger} \nonumber\\
         &+ \sum_{\mathfrak{p}\not=\mathfrak{0},\mathfrak{p}'\not=\mathfrak{0}} d_{\mathfrak{p}\mathfrak{p}'}K_\mathfrak{p} \mathsf{T}_{\alpha} K^{\dagger}_{\mathfrak{p}'}
    \end{align}
    Recall the standard commutator-anticommutator relation $2 FT + 2 T F^\dagger = \left[F-F^\dagger , T\right]+ \left\{F+F^\dagger , T\right\}$. Also note that one can trivially write $a_{\mathfrak{0}\mathfrak{0}} \mathsf{T}_{\alpha} / d^2_S = \{a_{\mathfrak{0}\mathfrak{0}}  / 2d^2_S , \mathsf{T}_{\alpha}\} $. Then the above expression becomes 
    \begin{align}
        \text{Choi} \big(\mathcal{L}_{D} \circ \mathcal{T}_{\alpha}\big) &= i \left[ \mathsf{T}_{\alpha},H_{D} \right] + \left\{G_{D}, \mathsf{T}_{\alpha},\right\}\nonumber\\
        &+ \sum_{\mathfrak{p}\not=\mathfrak{0},\mathfrak{p}'\not=\mathfrak{0}} d_{\mathfrak{p}\mathfrak{p}'}K_\mathfrak{p} \mathsf{T}_{\alpha} K^{\dagger}_{\mathfrak{p}'},\label{(QDoD)Proof:GKSL form step 3}
    \end{align}
    where $H_{D}$ and $G_D$ are defined in Eq.~(\ref{(QDoD)Eq:Path-dependent Hamiltonian D})--(\ref{(QDoD)Eq:Path-dependent G operator D}).

    Next we note that the left-hand side of the above equation is equal to
    \begin{align}
         \text{Choi} \big(\mathcal{L}_{D} \circ \mathcal{T}_{\alpha}\big) = \text{Choi}\left(\frac{\partial \mathcal{T}_\alpha}{\partial \gamma_1} \right) = \frac{\partial \mathsf{T}_{\alpha}}{\partial \gamma_1}.
    \end{align}
    Recall the generalized trace-preservation condition~\ref{(QDoD) Def: quantum comb} on the quantum comb. Although different types of evolution give slightly different generalized TP conditions, for all evolution types, we have the following condition, 
    \begin{align}
        \text{Tr}_{t^\out s^\out} (\mathsf{T}_{\alpha}) = \mathds{1}_{t^\inp}\otimes \mathds{1}_{s^\inp} ,\quad \text{Tr}_\alpha (\mathsf{T}_{\alpha}) = d^2_S .
    \end{align}
    Therefore, the derivative of $\text{Tr}_{\alpha}(\mathsf{T}_\alpha)$ with respect to $\gamma_1$ is equal to zero. Furthermore, let us consider some arbitrary operator defined on the input space $\delta D_{t^\inp s^\inp} \in \mathcal{L}(\mathcal{H}_{t^{\inp}}\otimes \mathcal{H}_{s^\inp})$. Then, from the generalized trace-preservation condition, one can show that 
    \begin{align}
        \text{Tr}_{\alpha} \left[\left(\delta D_{t^\inp s^\inp}\otimes \mathds{1}_{t^\out s^\out}\right)\mathsf{T}_{\alpha}\right] = \text{Tr}_{t^\inp s^\inp} \left(\delta D_{t^\inp s^\inp}\right).
    \end{align}
    Therefore, if the above arbitrary operator is traceless, then taking the trace over Eq.~(\ref{(QDoD)Proof:GKSL form step 3}), leads to 
    \begin{align}
        0 &= \text{Tr}\left(\frac{\partial \mathsf{T}_{\alpha}}{\partial\gamma_1}\right) = 2\text{Tr}\left(G \mathsf{T}_{\alpha}\right)\nonumber\\
        &+  \sum_{\mathfrak{p}\not=\mathfrak{0},\mathfrak{p}'\not=\mathfrak{0}} d_{\mathfrak{p}\mathfrak{p}'} \text{Tr}\left(K^{\dagger}_{\mathfrak{p}'}K_\mathfrak{p} \mathsf{T}_{\alpha} \right).
    \end{align}
    Therefore, up to a freedom of $\delta D$, we have the following operator equality 
    \begin{align}
        G = - \frac{1}{2}\sum_{\mathfrak{p}\not=\mathfrak{0},\mathfrak{p}'\not=\mathfrak{0}} a_{\mathfrak{p}\mathfrak{p}'} K^{\dagger}_{\mathfrak{p}'}K_\mathfrak{p} + \delta D_{t^\out s^\out} \otimes \mathds{1}_{t^\inp s^\inp}.
    \end{align}
    Then we find our final expression 
    \begin{align}
        &\text{Choi} \big(\mathcal{L}_{D} \circ \mathcal{T}_{\alpha}\big) =  i \left[ \mathsf{T}_{\alpha},H_{D} \right]\nonumber\\
        &+\left\{\delta D_{t^\out s^\out}(\lambda)\otimes \mathds{1}_{t^\inp s^\inp},\mathsf{T}_{\alpha}\right\}\nonumber\\
         &+ \sum_{\mathfrak{p}\not=\mathfrak{0},\mathfrak{p}'\not=\mathfrak{0}}  d_{\mathfrak{p}\mathfrak{p}'}\Big(K_\mathfrak{p} \mathsf{T}_{\alpha} K^{\dagger}_{\mathfrak{p}'}- \frac{1}{2} \left\{ K^{\dagger}_{\mathfrak{p}'}K_\mathfrak{p} ,  \mathsf{T}_{\alpha}\right\}\Big)
    \end{align}
    The proof for the encoding generator trivially follows.
\end{proof}

\subsection{Proof of Theorem~\ref{Thm: main result}}\label{(QDoD)Appendix:Proofs-3}
Before presenting the proof, we state the following lemma, which we use in the derivation.
\begin{lem}[Path-splitting formula]
Let $\{\sigma_k\}_{k=0}^{N}$ be a partition of the interval $[0,1]$, with $\sigma_0=0$, $\sigma_N=1$, and $\Delta\sigma_k:=\sigma_{k+1}-\sigma_k$. For a continuous operator-valued function $\mathsf{A}_\sigma$ on a finite-dimensional space, the path-ordered exponential can be expressed as the following product integral:
\begin{align}
    \mathcal{P}\operatorname{exp}\left(\int_0^1 d\sigma\,\mathsf{A}_{\sigma}\right) = \lim_{N\rightarrow\infty}\prod_{k=0}^{N-1}e^{\mathsf{A}_{\sigma_k}\Delta\sigma_k}.
\end{align}
Here, the product is path ordered, with larger values of $k$ appearing to the left, and the limit is taken as $\max_k\Delta\sigma_k\rightarrow0$.
\end{lem}
This lemma follows from the well-known equivalence between the product integral and the Peano--Dyson series expansion~\cite{slavik2007product}, a result rooted in Volterra's theory of product integration.

We now turn to the proof of Theorem~\ref{Thm: main result}.
\begin{proof}
    The necessity part (CP-divisibility implies the GKSL-like form) follows from Theorem~\ref{Lemma: GKSL form}. To establish sufficiency, we need to show that the following path-ordered quantum comb satisfies the generalized CPTP conditions and is CP-divisible:
    \begin{align}
        \mathcal{T}_\alpha (\lambda) &= \mathcal{P}\text{exp} \left(\int^{\lambda}_{0} d\lambda' \mathcal{L}(\lambda') \right) \circ\mathcal{T}_{\alpha}(0)\\
        &=: \mathcal{P}\text{exp} \left(\int^{\vec{\gamma}(\lambda)}_{\vec{\gamma}(0)} d\mathcal{L} \right) \circ\mathcal{T}_{\alpha}(0),
    \end{align}
    where, for brevity, we have defined the differential form of the quantum comb generator $\mathcal{L}$. 

    To show trace preservation, we first recall the decoding/encoding generators~(\ref{Eq: GKSL form of decoding generator})--(\ref{Eq: GKSL form of encoding generator}) acting on the Choi space. By construction, they are traceless
    \begin{align}
        0=\text{Tr}\left(\widetilde{\mathcal{L}}_{D}[ \mathsf{T}_{\alpha}]\right)=\text{Tr}\left(\widetilde{\mathcal{L}}_{E}[ \mathsf{T}_{\alpha}]\right),
    \end{align}
    where $\mathsf{T}_\alpha$ is the Choi representaiton of the quantum comb.
    This implies that the quantum comb generator is also traceless
    \begin{align}
        \text{Tr}\left(d\widetilde{\mathcal{L}}[\mathsf{T}_\alpha]\right) = 0,\label{(QDoD)Eq: Main theorem step1}
    \end{align} 
    where $d\widetilde{\mathcal{L}}$ is the differential form of the quantum comb generator acting on the Choi space. 
    Then it follows that 
    \begin{align}
        \text{Tr}\left(\mathcal{P}e^{\int^{\vec{\gamma}(\lambda)}_{\vec{\gamma}(0)} d\widetilde{\mathcal{L}}}\circ \mathsf{T}_{\alpha}(0)\right) = \text{Tr}\left(\mathsf{T}_{\alpha}(0)\right)
    \end{align}
    This follows from the traceless property~(\ref{(QDoD)Eq: Main theorem step1}) applied to each order of the path-ordered exponential expansion. Since $\mathsf{T}_{\alpha} (0)$ is a quantum comb, it satisfies the generalized trace-preservation properly~\cite{chiribella2008quantum,Taranto2025} (see also Def.~\ref{(QDoD) Def: quantum comb}). Therefore, the path-ordered exponential above, acting on any quantum comb, also satisfies generalized trace preservation.

   Next, we establish CP-divisibility. We first show divisibility, which follows directly from the composition property of the path-ordered exponential:
\begin{align}
    &\mathcal{P}e^{\int_{\vec{\gamma}(\lambda_1)}^{\vec{\gamma}(\lambda)}d\mathcal{L}} \circ \mathcal{P}e^{\int_{\vec{\gamma}(0)}^{\vec{\gamma}(\lambda_1)}d\mathcal{L}} = \mathcal{P}e^{\int_{\vec{\gamma}(0)}^{\vec{\gamma}(\lambda)}d\mathcal{L}},\\
    &\mathcal{P}e^{\int_{\vec{\gamma}(\lambda)}^{\vec{\gamma}(\lambda)}d\mathcal{L}} = \mathcal{I}.
\end{align}
However, these properties alone do not establish complete positivity of the intermediate propagators. To show this, we express the path-ordered exponential in Kraus form. Using the path-splitting formula, we have
\begin{align}
    \mathcal{P}e^{\int_0^{\lambda} d\lambda' \widetilde{\mathcal{L}}(\lambda')} = \lim_{N\rightarrow\infty}\prod_{k=0}^{N-1}e^{\widetilde{\mathcal{L}}(\lambda_k)\Delta\lambda_k}.
\end{align}

To construct the Kraus representation, we first recall that the coefficient matrices $d_{\mathfrak p\mathfrak p'}$ and $e_{\mathfrak p\mathfrak p'}$ are positive semidefinite. They therefore admit the factorizations
\begin{align}
    d_{\mathfrak p\mathfrak p'}(\lambda) &= \sum_\xi v_{\mathfrak p\xi}(\lambda)v_{\mathfrak p'\xi}^*(\lambda),\\
    e_{\mathfrak p\mathfrak p'}(\lambda) &= \sum_\xi w_{\mathfrak p\xi}(\lambda)w_{\mathfrak p'\xi}^*(\lambda),
\end{align}
where $\xi$ labels the components of the factorization. We then define the following operators:
\begin{align}
    \widetilde{V}_{\xi}(\lambda) &:= \sum_{\mathfrak p}v_{\mathfrak p\xi}(\lambda)K_{\mathfrak p},\\
    \widetilde{W}_{\xi}(\lambda) &:= \sum_{\mathfrak p}w_{\mathfrak p\xi}(\lambda)K_{\mathfrak p}.
\end{align}
Using these operators, we introduce the following linear maps:
\begin{align}
    \widetilde{\mathcal{L}}_{D0}[\bullet] &:= i[\bullet,H_D]+\left\{\delta D_{t^\out s^\out}(\lambda)\otimes\mathds{1}_{t^\inp s^\inp},\bullet\right\}\nonumber\\
    &\quad-\frac{1}{2}\sum_\xi\left\{\widetilde{V}_\xi^\dagger\widetilde{V}_\xi,\bullet\right\},\\
    \widetilde{\mathcal{L}}_{E0}[\bullet] &:= i[\bullet,H_E]+\left\{\delta E_{t^\out s^\out}(\lambda)\otimes\mathds{1}_{t^\inp s^\inp},\bullet\right\}\nonumber\\
    &\quad-\frac{1}{2}\sum_\xi\left\{\widetilde{W}_\xi^\dagger\widetilde{W}_\xi,\bullet\right\}.
\end{align}
These maps can be rewritten by introducing the following non-Hermitian effective operators:
\begin{align}
    H_D^{\mathrm{eff}} &:= H_D-\frac{i}{2}\sum_\xi\widetilde{V}_\xi^\dagger\widetilde{V}_\xi+i\delta D_{t^\out s^\out}(\lambda)\otimes\mathds{1}_{t^\inp s^\inp},\\
    H_E^{\mathrm{eff}} &:= H_E-\frac{i}{2}\sum_\xi\widetilde{W}_\xi^\dagger\widetilde{W}_\xi+i\delta E_{t^\out s^\out}(\lambda)\otimes\mathds{1}_{t^\inp s^\inp}.
\end{align}
We then obtain
\begin{align}
    \widetilde{\mathcal{L}}_{D0}[\bullet] &= -iH_D^{\mathrm{eff}}\bullet+i\bullet H_D^{\mathrm{eff}\dagger},\\
    \widetilde{\mathcal{L}}_{E0}[\bullet] &= -iH_E^{\mathrm{eff}}\bullet+i\bullet H_E^{\mathrm{eff}\dagger}.
\end{align}

Let us combine the above generators into the following path-dependent map:
\begin{align}
    \widetilde{\mathcal{L}}_0[\bullet] := \frac{d\gamma_1}{d\lambda}\widetilde{\mathcal{L}}_{D0}[\bullet]+\frac{d\gamma_2}{d\lambda}\widetilde{\mathcal{L}}_{E0}[\bullet].
\end{align}
The full generator can therefore be written as
\begin{align}
    \widetilde{\mathcal{L}}[\bullet] = \widetilde{\mathcal{L}}_0[\bullet]+\sum_{\mathfrak p,\mathfrak p'}C_{\mathfrak p\mathfrak p'}(\lambda)K_{\mathfrak p}\bullet K_{\mathfrak p'}^\dagger,
\end{align}
where
\begin{align}
    C_{\mathfrak p\mathfrak p'}(\lambda) := \dot{\gamma}_1(\lambda)d_{\mathfrak p\mathfrak p'}(\lambda)+\dot{\gamma}_2(\lambda)e_{\mathfrak p\mathfrak p'}(\lambda).
\end{align}
Although $d$ and $e$ are individually positive semidefinite, their path-dependent combination need not be, since the rates $\dot{\gamma}_1$ and $\dot{\gamma}_2$ can take either sign. We therefore invoke the positive semidefiniteness condition on the combined generalized Kossakowski matrix, $C(\lambda)\succeq0$, as required by the theorem~\ref{Lemma: GKSL form}. Under this condition, $C(\lambda)$ admits a factorization $C(\lambda)=R(\lambda)R^\dagger(\lambda)$, allowing us to define the operators
\begin{align}
    J_\mu(\lambda) := \sum_{\mathfrak p}R_{\mathfrak p\mu}(\lambda)K_{\mathfrak p}.
\end{align}
Consequently, the combined contribution takes the Kraus form
\begin{align}
    \widetilde{\mathcal{J}}(\lambda)[\bullet] := \sum_\mu J_\mu(\lambda)\bullet J_\mu^\dagger(\lambda),
\end{align}
so that $\widetilde{\mathcal{L}}=\widetilde{\mathcal{L}}_0+\widetilde{\mathcal{J}}$.

We now show that $e^{\widetilde{\mathcal{L}}_0(\lambda_k)\Delta\lambda_k}$ and $e^{\widetilde{\mathcal{J}}(\lambda_k)\Delta\lambda_k}$ are individually completely positive by expressing them in Kraus form. First, using the effective operators defined above, we obtain
\begin{equation}
    \begin{split}
        &e^{\widetilde{\mathcal{L}}_0(\lambda_k)\Delta\lambda_k}[\bullet]\\
        &=\exp\Big\{-i\big[\dot{\gamma}_1(\lambda_k)H_D^{\mathrm{eff}}(\lambda_k)+\dot{\gamma}_2(\lambda_k)H_E^{\mathrm{eff}}(\lambda_k)\big]\Delta\lambda_k\Big\}\bullet\\
        &\quad\times\exp\Big\{i\big[\dot{\gamma}_1(\lambda_k)H_D^{\mathrm{eff}\dagger}(\lambda_k)+\dot{\gamma}_2(\lambda_k)H_E^{\mathrm{eff}\dagger}(\lambda_k)\big]\Delta\lambda_k\Big\}.
    \end{split}
\end{equation}
This expression takes the Kraus form with a single Kraus operator and is therefore completely positive.

Second, expanding the exponential of the combined map, we obtain
\begin{align}
    e^{\widetilde{\mathcal{J}}(\lambda_k)\Delta\lambda_k}[\bullet] &= \sum_{n=0}^{\infty}\frac{(\Delta\lambda_k)^n}{n!}\widetilde{\mathcal{J}}^n(\lambda_k)[\bullet]\\
    &=\sum_{n=0}^{\infty}\sum_{\mu_1,\ldots,\mu_n}J_{n,\boldsymbol{\mu}}(\lambda_k)\bullet J_{n,\boldsymbol{\mu}}^\dagger(\lambda_k),
\end{align}
where
\begin{align}
    J_{n,\boldsymbol{\mu}}(\lambda_k) := \sqrt{\frac{(\Delta\lambda_k)^n}{n!}}\,J_{\mu_1}(\lambda_k)\cdots J_{\mu_n}(\lambda_k),
\end{align}
with $J_{0,\boldsymbol{\mu}}=\mathds{1}$. Since $\Delta\lambda_k\geq0$, this expansion provides a Kraus representation, establishing that $e^{\widetilde{\mathcal{J}}(\lambda_k)\Delta\lambda_k}$ is completely positive.

Having established the complete positivity of both exponentials, we apply the Suzuki--Trotter decomposition:
\begin{align}
    e^{\widetilde{\mathcal{L}}(\lambda_k)\Delta\lambda_k} &= e^{(\widetilde{\mathcal{L}}_0(\lambda_k)+\widetilde{\mathcal{J}}(\lambda_k))\Delta\lambda_k}\\
    &=\lim_{M\rightarrow\infty}\left(e^{\widetilde{\mathcal{L}}_0(\lambda_k)\Delta\lambda_k/M}\circ e^{\widetilde{\mathcal{J}}(\lambda_k)\Delta\lambda_k/M}\right)^M.
\end{align}
Since compositions of completely positive maps are completely positive, together with the composition property established at the beginning of the proof, this establishes CP-divisibility. 
\end{proof}

\subsection{Proof of Corollary~\ref{(QDOD)thm:ZeroCurvatureCondtion}}\label{(QDoD)Appendix:Proofs-4}
\begin{proof}
    Consider the following two separate paths with the same endpoints $(\eta_1,\eta_2)$ and $(\eta_1+\delta, \eta_2 + \epsilon)$:  
    \begin{align}
        &\text{Path A: }(\eta_1 , \eta_2) \xrightarrow{\text{(A1)}}  (\eta_1+ \delta , \eta_2) \xrightarrow{\text{(A2)}} (\eta_1+\delta , \eta_2 + \epsilon)\nonumber\\
        &\text{Path B: }(\eta_1 , \eta_2) \xrightarrow{\text{(B1)}}  (\eta_1 , \eta_2+\epsilon) \xrightarrow{\text{(B2)}}  (\eta_1+\delta, \eta_2 + \epsilon)\nonumber
    \end{align}
    for some infinitesimal parameters $\epsilon$ and $\delta$. A1, A2, B1, and B2 indicate the intermediate paths. Two paths are parametrized by the following
    \begin{align}
        \vec{\gamma}^{\text{(A1)}}(\lambda) := (\eta_1 + \lambda \delta , \eta_2) ,\quad \vec{\gamma}^{\text{(A2)}}(\lambda) := (\eta_1 +  \delta , \eta_2+ \lambda \epsilon)\nonumber\\
        \vec{\gamma}^{\text{(B1)}}(\lambda) :=(\eta_1  , \eta_2+ \lambda \epsilon) ,\quad \vec{\gamma}^{\text{(B2)}}(\lambda) := (\eta_1 + \lambda \delta , \eta_2+\epsilon) \nonumber
    \end{align}
    where $\lambda\in[0,1]$. 

   Next, let us consider the path-ordered representation of the CP-divisible comb~(\ref{(QDoD) Eq: solution to comb master equation}) for the intermediate path A1. From the above parametrization, we observe that $\dot{\gamma}_1 = \delta$ and $\dot{\gamma}_2 =0$ for the A1 path; therefore, the CP-divisible comb simplifies in the following way 
    \begin{align}
        \mathcal{P}\text{exp}\left(\int_{\text{(A1)}}d\mathcal{L}\right) = \mathcal{P}\text{exp}\left(\int_0^1 d\lambda \delta \mathcal{L}_{D}(\eta_1 + \lambda \delta , \eta_2)\right)&\nonumber\\
        = \text{exp}\left(\delta \mathcal{L}_{D}(\eta_1,\eta_2) + \frac{\delta^2}{2} \frac{\partial\mathcal{L}_{D}}{\partial\gamma_1}+\mathcal{O}(\delta^3) \right)&
    \end{align}
    Similarly, other intermediate paths simplify to 
    \begin{align}
        \mathcal{P}\text{exp}\left(\int_{\text{(A2)}}\!\!\!\!\!d\mathcal{L}\right)=& \text{exp}\bigg(\epsilon\mathcal{L}_{E}(\eta_1,\eta_2) + \epsilon \delta \frac{\partial\mathcal{L}_{E}}{\partial\gamma_1} \nonumber\\
        &+ \frac{\epsilon^2}{2} \frac{\partial \mathcal{L}_{E}}{\partial\gamma_2}+\mathcal{O}(\epsilon^3) \bigg),\\
        \mathcal{P}\text{exp}\left(\int_{\text{(B1)}}\!\!\!\!\!d\mathcal{L}\right)=&\text{exp}\bigg(\epsilon \mathcal{L}_{E}(\eta_1,\eta_2)+\frac{\epsilon^2}{2}\frac{\partial\mathcal{L}_{E}}{\partial\gamma_2}\nonumber\\
        &+ \mathcal{O}(\epsilon^3) \bigg) ,\\
        \mathcal{P}\text{exp}\left(\int_{\text{(B2)}}\!\!\!\!\!d\mathcal{L}\right)=& \text{exp}\bigg(\delta\mathcal{L}_{D}(\eta_1,\eta_2) + \epsilon \delta \frac{\partial\mathcal{L}_{D}}{\partial\gamma_2} \nonumber\\
        &+ \frac{\delta^2}{2} \frac{\partial \mathcal{L}_{D}}{\partial\gamma_1}+\mathcal{O}(\delta^3) \bigg).
    \end{align}
    The evolution map is path-independent if the following equality holds 
    \begin{align}
        \mathcal{P}e^{\int_{\text{(A2)}}d \mathcal{L}} \circ \mathcal{P}e^{\int_{\text{(A1)}}d \mathcal{L}} = \mathcal{P}e^{\int_{\text{(B2)}}d \mathcal{L}} \circ \mathcal{P}e^{\int_{\text{(B1)}}d \mathcal{L}} .
    \end{align}
    Setting $\epsilon = \delta$ and expanding the exponential in $\delta$, the above equality holds up to second order in $\delta$, if the following equation is satisfied. 
    \begin{align}
        \frac{\partial \mathcal{L}_{D}}{\partial\gamma_1} - \frac{\partial \mathcal{L}_{E}}{\partial\gamma_2} + \left[\mathcal{L}_{E},\mathcal{L}_{D}\right]=0.
    \end{align}
    This is true for any infinitesimal closed loop; due to the CP-divisibility of the path-ordered exponential, it holds for any closed loop. 
\end{proof}
\section{GKSL Form of Decoding/Encoding Generators}\label{(QDoD)Apx:GKSL}
\begin{cor}\label{(QDoD) Coro: diagonalization of GKSL}
    The uncorrelated term of the decoding/encoding generators can be diagonalized with respect to the polarization operator basis index $(\mathfrak{p}, \mathfrak{p}')$ to give the following GKSL forms
    \begin{align}
         &\widetilde{\mathcal{L}}_{D0}\left[ \mathsf{T}_{\alpha}\right]= i \left[\mathsf{T}_{\alpha} , H_{D}\right]\label{Eq: diagonalized GKSL form of decoding generator}\\
        &+ \sum_{\xi} \delta_{\xi}(\lambda)\Big[V_{\xi} \mathsf{T}_{\alpha} {V_{\xi }}^\dagger\!- \!\frac{1}{2} \left\{{V_{\xi }}^\dagger V_{\xi }, \mathsf{T}_{\alpha}\right\}\Big].\nonumber\\
        &\widetilde{\mathcal{L}}_{E0}\left[ \mathsf{T}_{\alpha}\right]= i \left[\mathsf{T}_{\alpha} , H_{E}\right]\label{Eq: diagonalized GKSL form of encoding generator}\\
        &+ \sum_{\xi} \epsilon_{\xi}(\lambda) \Big[W_{\xi}  \mathsf{T}_{\alpha} {W_{\xi}}^\dagger\!- \!\frac{1}{2} \left\{{W_{\xi}}^\dagger W_{\xi}, \mathsf{T}_{\alpha}\right\}\Big].\nonumber
    \end{align}
    where path-dependent functions $\delta_{\xi} (\lambda)$ and $\epsilon_{\xi} (\lambda)$ are non-negative eigenvalues of the functions $d_{\mathfrak{p}\mathfrak{p}'}(\lambda)$ and $e_{\mathfrak{p}\mathfrak{p}'}(\lambda)$:
    \begin{align}
        &d_{\mathfrak{p}\mathfrak{p}'} (\lambda)= \sum_{\xi} \delta_{\xi}(\lambda) \left(U_{D}(\lambda)\right)_{\mathfrak{p}\xi}\left(U^{\dagger}_{D}(\lambda)\right)_{\xi\mathfrak{p}'},\label{(QDoD)Eq:Decay rate d}\\
        &e_{\mathfrak{p}\mathfrak{p}'}(\lambda) = \sum_{\xi} \epsilon_{\xi}(\lambda)  \left(U_{E}(\lambda)\right)_{\mathfrak{p}\xi}\left(U^{\dagger}_{E}(\lambda)\right)_{\xi\mathfrak{p}'}.\label{(QDoD)Eq:Decay rate e}
    \end{align}
    with some path-dependent unitaries $U_{D} (\lambda)$ and $U_{E} (\lambda)$.
    The path-dependent polarization operator basis is defined with the above unitary
    \begin{align}
        V_{\xi } (\lambda) := \sum_{\mathfrak{p}}\left(U_{D} (\lambda)\right)_{\mathfrak{p}\xi} K_\mathfrak{p},\label{(QDod)Eq:Diagonalized POB of D}\\
        W_{\xi } (\lambda) := \sum_{\mathfrak{p}}\left(U_{E} (\lambda)\right)_{\mathfrak{p}\xi} K_\mathfrak{p}.\label{(QDod)Eq:Diagonalized POB of E}
    \end{align}
\end{cor}
For the semigroup quantum comb, the same analysis follows but with the path-independent version of the functions~(\ref{(QDoD)Eq:Decay rate d})--(\ref{(QDoD)Eq:Decay rate e}) and the basis~(\ref{(QDod)Eq:Diagonalized POB of D})--(\ref{(QDod)Eq:Diagonalized POB of E}).
\begin{proof}
    Recall the definition of $d_{\mathfrak{p}\mathfrak{p}'} (\sigma)$ and $e_{\mathfrak{p}\mathfrak{p}'} (\sigma)$ in Eq.~\eqref{(QDoD) Eq: POB function d} and \eqref{(QDoD) Eq: POB function e}. Both definitions are defined with respect to the coefficient matrix $\chi_{\mathfrak{p}\mathfrak{p}'}$. It has been proven in Corollary~\ref{(QDoD)Coro:Kraus Representation} that the coefficient matrix $\chi_{\mathfrak{p}\mathfrak{p}'}$ is positive semidefinite with respect to the indices $(\mathfrak{p},\mathfrak{p}')$ due to the positivity of the quantum comb. This implies that $\chi_{\mathfrak{p}\mathfrak{p}'}$ can be diagonalized with unitary matrices and has positive eigenvalues, as shown above.  
\end{proof}
\section{Family of Dynamical Equations}
In a first-order dynamical system, a solution is obtained once the initial condition is specified. Similarly, our Quantum Dynamics of Dynamics (QDoD) stated in Theorem~\ref{Thm: main result} has well-established dynamical equations as a "solution" to QDoD once the initial condition is specified. We will consider a few well-established dynamical equations as demonstrative examples, such as the GKSL equation in Schr\"odinger and Heisenberg pictures and the Hasenöhrl--Caro equation for quantum channels.  

To organize the infinitely many quantum dynamics that arise from our QDoD, we first give a few constraining conditions below. 
\begin{cor}[Zero-curvature condition]\label{(QDOD)thm:ZeroCurvatureCondtion}
    If decoding/encoding generators $\mathcal{L}_{D}$ and $\mathcal{L}_{E}$ of the dynamical CP-divisible quantum comb satisfies the zero-curvature condition
    \begin{align}
        \partial_{\gamma_1 } \mathcal{L}_{D} - \partial_{\gamma_2} \mathcal{L}_{E} + \left[  \mathcal{L}_{D},\mathcal{L}_{E}\right] =0,
    \end{align}
    then the solution~(\ref{Eq: Formal solution to the covariant GKSL eq}) is path-independent and only depends on the end points $(\gamma_1(\lambda),\gamma_2 (\lambda))$ and $(\gamma_1(0),\gamma_2 (0))$.
\end{cor}
\begin{proof}
    The proof is given in Appendix~\ref{(QDoD)Appendix:Proofs-4}.
\end{proof}

\begin{figure}[h]
    \centering
    \includegraphics[width=\linewidth]{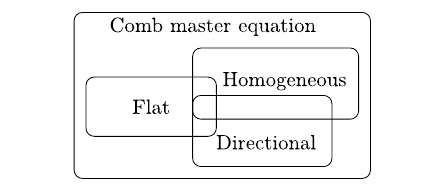}
    \caption{Venn diagram showing the different types of master equations. We note that some of the intersections may be empty.}
    \label{(QDoD) Fig: Hierarchical table}
\end{figure}

\begin{defn}[Family of Dynamical equations]\label{(QDoD) Definition: Family of QDoD}
For the QCME~\eqref{(QDoD)Eq:CombME}, a family of dynamical equations is obtained by imposing the following restrictions independently. 
\begin{enumerate}
    \item \textbf{Globally TP QCME} is obtained if the decoder/encoder--correlation terms $\delta D$ and $\delta D$ are zero.
    \item \textbf{Flat QCME} is obtained after imposing the zero-curvature condition on the generators. 
    \item \textbf{Directional QCME} is obtained after imposing the constraint on the evolution path as 
    \begin{align}
        \gamma_1 (\lambda) = \eta_1 + \lambda v_1 , ~~\gamma_2 (\lambda) = \eta_2 + \lambda v_2.
    \end{align}
    where $(\eta_2 ,\eta_2 , v_1 ,v_2)$ are fixed constants and $(v_1, v_2)\not=(0,0)$ is a non-zero vector indicating the direction of evolution on the time-plane $(\gamma_1, \gamma_2)$. If either $v_1$ or $v_2$ is zero, then we refer to the equation as uni-directional. 
    \item \textbf{Homogeneous QCME} is obtained after imposing the time-homogenety~(\ref{(QDoD)Def:TimeHomogeneity}) to the quantum comb. In this case, the QCME will be path-independent. Note that a homogeneous equation can be non-flat, but it can not be directional. 
\end{enumerate}
\end{defn}
Fig.~\ref{(QDoD) Fig: Hierarchical table} summarizes a hierarchical table of different families of master equations.

    We can construct many quantum dynamical equations by choosing an initial condition. As a demonstration below, we will derive the dynamical semigroup of the superchannel considered by Hasenöhrl and Caro in Ref.~\cite{hasenohrl2022quantum}. 

\begin{example}[Hasenöhrl--Caro superchannel semigroup equation]\label{Hasenohrl--Caro superchannel semigroup equation}
    Previous work on the master equation for quantum channels is limited. To our knowledge, only Hasenöhrl and Caro~\cite{hasenohrl2022quantum} have studied the semigroup superchannel and obtained the explicit quantum master equation for quantum channels. In this example, we show how their master equation appears as a special case of a particular evolution type out of 20 types with specific initial codniton. 
    
    First, let us consider a special case where the quantum channel and the evolution are time-homogeneous:
       \begin{equation}
           \begin{split}
               \overleftarrow{\mathcal{T}}_{t^\out,t^\inp, s^\inp,s^\out} (t^\out,t^\inp, s^\inp,s^\out)&=\overleftarrow{\mathcal{T}}_{t^\out,t^\inp, s^\inp,s^\out} (t^\out-t^\inp, s^\inp- s^\out) \\
            &=\overleftarrow{\mathcal{T}}_{t^\out,t^\inp, s^\inp,s^\out} (t,s) ,\\
            \mathcal{E}_{s^\inp , s^\out} (s^\inp , s^\out) &= \mathcal{E}_{s^\inp , s^\out} (s^\inp - s^\out) = \mathcal{E}_{s^\inp , s^\out} (s).
           \end{split}
       \end{equation}
        where $t:= t^\out - t^\inp$ and $s:= s^\inp - s^\out$. We keep the Hilbert-space indexing. This is because the number of open legs does not change even after time-homogeneity; therefore, the Hilbert-space indexing must remain unchanged. The infinitesimal evolution of the quantum channel is now mapping the quantum channel defined at time $s$, $\mathcal{E}_{s^\inp ,s^\out}(s)$ to a quantum channel defiend at time $t$, $\mathcal{E}_{t^\out,t^\inp} (t)$. This is the type of evolution considered in Ref.~\cite{hasenohrl2022quantum}.
    
    The above contraction rule is $(t^\out \leftrightarrow s^\inp, t^\inp \leftrightarrow s^\out)$, which corresponds to the type 1B of our evolution table in Appendix~\ref{(QDoD)Appendix:EvolutionFamily}. We further impose the following constraints
    \begin{enumerate}
        \item As an initial condition, we choose a single time-dependent quantum channel $\mathcal{E}$, whose Choi representation we denote by $\mathsf{E}$.
        \item we impose the time-homogeneity constraint on pairs $(t^\out,t^\inp)$ and $(s^\out, s^\inp)$.
        \item We will choose the uni-directional path 
        \begin{align}
            \gamma_1(\lambda) = \eta_1 + \lambda  , \quad \gamma_2(\lambda) = \eta_2 .
        \end{align}
        \item We will assume that the TP--freedom operators only act on $t^\inp $--leg, namely 
        \begin{align}
            \delta \widetilde{\mathcal{L}}_{D t^\inp s^\inp} = \delta \widetilde{\mathcal{L}}_{D t^\inp} \otimes \mathcal{I}_{s^\inp}.
        \end{align}
    \end{enumerate}
    Therefore, we have a \textit{uni-directional time-homogeneous QCME with a single time-dependent quantum channel as an initial condition}.  

    To see the connection with the Hansenöhrl--Caro equation, let us consider the QCME with the above constraint. Since the TP--freedom term is non-vanishing, we cannot express the equation in GKSL form; therefore, we consider the canonical form shown in Corollary~\ref{Alternative form of GKSL}.
    \begin{align}
        \frac{\partial \mathsf{T}_\alpha}{\partial \lambda} =& i \left[ \mathsf{T}_{\alpha},H_{D} \right] + \left\{G_{D}, \mathsf{T}_{\alpha},\right\}\nonumber\\
        &+ \sum_{\mathfrak{p}\not=\mathfrak{0},\mathfrak{p}'\not=\mathfrak{0}} d_{\mathfrak{p}\mathfrak{p}'}K_\mathfrak{p} \mathsf{T}_{\alpha} K^{\dagger}_{\mathfrak{p}'}
    \end{align}
    where $\alpha = (t^\out (\lambda), t^\inp(\lambda),s^\out, s^\inp) $. Note that the above generator only acts on $(t^\out , t^\inp)$--legs; therefore, applying the above equation to the initial condition $\mathsf{E}_{s^\out s^\inp}(s)$, one obtains 
    \begin{align}
        \frac{\partial \mathsf{E}_{t^\out,t^\inp }(\lambda)}{\partial \lambda} =& i \left[  \mathsf{E}_{t^\out,t^\inp }(\lambda),H_{D} \right] + \left\{G_{D},  \mathsf{E}_{t^\out,t^\inp }(\lambda),\right\}\nonumber\\
        &+ \sum_{\mathfrak{p}\not=\mathfrak{0},\mathfrak{p}'\not=\mathfrak{0}} d_{\mathfrak{p}\mathfrak{p}'}K_\mathfrak{p}  \mathsf{E}_{t^\out,t^\inp } (\lambda) K^{\dagger}_{\mathfrak{p}'}\label{(QDoD)Eq:Hasenohrl Caro Step1}
    \end{align}
    where $\mathsf{T}_\alpha\star \mathsf{E}_{s^\out,s^\inp}=\mathsf{T}_{t^\out (\lambda) t^\inp(\lambda) s^\out s^\inp}\star\mathsf{E}_{s^\out,s^\inp} = \mathsf{E}_{t^\out,t^\inp}(\lambda)$. 
    
    Next, let us consider the following operator, $V: \mathcal{H}_{t^\out}\otimes \mathcal{H}_{t^\inp} \rightarrow \mathcal{H}_{t^\out}\otimes \mathcal{H}_{t^\inp}\otimes \mathcal{H}_E$. The spectral decomposition of such an operator is 
    \begin{align}
        V = \sum_{\mu \mathfrak{p}} v_{\mu \mathfrak{p}} K_{\mathfrak{p}}\otimes \ket{e_{\mu}}
    \end{align}
    where $\{\ket{e_\mu}\}$ are orthonormal basis of the environment Hilbert space $\mathcal{H}_{E}$. Let us consider the following quantity 
    \begin{align}
        \text{Tr}_E (V\mathsf{E}V^\dagger) = \sum_{\mu \mathsf{p}\mathfrak{p}'} v_{\mu \mathfrak{p}}v^*_{\mu \mathfrak{p}'} K_\mathfrak{p} \mathsf{E}K_{\mathfrak{p}'}^\dagger\label{(QDoD)Eq:Hasenohrl Caro Step2}. 
    \end{align}
    If we identify $d_{\mathfrak{p}\mathfrak{p}'} = \sum_\mu v_{\mu \mathfrak{p}}v^*_{\mu \mathfrak{p}'}$, then the last term of the master equation~(\ref{(QDoD)Eq:Hasenohrl Caro Step1}) can be rewritten as $\text{Tr}_E (V \mathsf{E}V^\dagger)$. Next, let us recall that using the definitions of $H_D$ and $G_{D}$, the first two terms of the equation~(\ref{(QDoD)Eq:Hasenohrl Caro Step1}) can be rewritten as $\mathsf{E}D_{\alpha} + D_\alpha^\dagger \mathsf{E}$. Therefore, the master equation can be rewritten as 
    \begin{align}
        \frac{\partial \mathsf{E}_{t^\out,t^\inp }(\lambda)}{\partial \lambda} =& \mathsf{E}_{t^\out,t^\inp }(\lambda) D_{\alpha}(\lambda) + D_{\alpha}^{\dagger}(\lambda)\mathsf{E}_{t^\out,t^\inp }(\lambda)\nonumber\\
        &+ \text{Tr}_E (V \mathsf{E}V^\dagger) 
    \end{align}
    This corresponds to the equation in Corollary V.15 of Hasenöhrl--Caro~\cite{hasenohrl2022quantum}. 

    Next we note that Hasenöhrl--Caro introduces the notion of semi-causality: A linear map $L_*:\mathcal{L}(\mathcal{H}_A\otimes\mathcal{H}_B)\to\mathcal{L}(\mathcal{H}_A\otimes\mathcal{H}_B)$ is Schr\"odinger $B\nrightarrow A$ semicausal if there exists a reduced linear map $L_*^A:\mathcal{L}(\mathcal{H}_A)\to\mathcal{L}(\mathcal{H}_A)$ such that
    \[
    \operatorname{Tr}_B[L_*(\mathsf{M})] = L_*^A\!\left(\operatorname{Tr}_B[\mathsf{M}]\right)
    \]
    for all $\mathsf{M}\in\mathcal{L}(\mathcal{H}_A\otimes\mathcal{H}_B)$. This condition is equivalent to the generalized CP condition of the quantum comb. To see this, let us label each leg of the linear map $L_* $ as 
    \begin{align}
        L_*:\mathcal{L}(\mathcal{H}_{s^\inp}\otimes\mathcal{H}_{t^\inp})\to\mathcal{L}(\mathcal{H}_{s^\out}\otimes\mathcal{H}_{t^\out}). 
    \end{align}
    Then the correspondong quantum comb is $\mathcal{T}_{t^\out t^\inp s^\out s^\inp}$ with corresponding Choi representation $\mathsf{T}_{t^\out t^\inp s^\out s^\inp}$. Due to the generalized CP condition, we have 
    \begin{align}
        \operatorname{Tr}_{t^\out} (\mathsf{T}_{t^\out t^\inp s^\out s^\inp}) = \mathsf{T}_{s^\out s^\inp}\otimes \mathds{1}_{t^\inp}, \quad \operatorname{Tr}_{s^\out} (\mathsf{T}_{s^\out s^\inp}) = \mathds{1}_{s^\inp}
    \end{align}
    for some reduced two-legged quantum comb $\mathsf{T}_{s^\out s^\inp}$, which corresponds to the reduced linear map $L_*^A$ for the definition of semi-causality. 

    Therefore, the generalized CPTP property ensures that our evolution map is semi-causal in the sense of Hasenöhrl--Caro. By Theorem V.18, the above equation~(\ref{(QDoD)Eq:Hasenohrl Caro Step2}) is equivalent to equation~(1), which is one of the main results of Hasenöhrl--Caro~\cite{hasenohrl2022quantum}. 
\end{example}

\section{Generator of a sequential decoder}
\label{(QDoD)Apx:SequentialDecoderGenerator}

\begin{figure}
    \centering
    \includegraphics[width=0.9\linewidth]{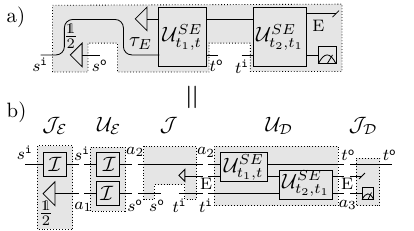}
    \caption{(a) Pictorial representation of the causally ordered quantum comb used in Sec.~\ref{sec:OQS}. (b) Identification of its encoding and decoding isometries and unitaries. Labels $a_1$, $a_2$, and $a_3$ denote auxiliary system spaces of the same dimension as the system Hilbert space.}
    \label{(QDoD) fig4-3}
\end{figure}

Here we derive the generator in Eq.~\eqref{(QDoD)Eq:SequentialDecoderGenerator} for the sequential decoder shown in Fig.~\ref{(QDoD) fig4-3}. The decoder consists of two consecutive system--environment interactions, with $t_2>t_1>t$. The first couples the system wire $S_1$ to the environment $E$, and the second couples $S_2$ to $E$. The wire labels are defined by $S_1:a_1\to t^\out$ and $S_2:t^\inp\to a_2$, as shown in Fig.~\ref{(QDoD) fig4-3}; they are distinct from the time labels $s^\inp$ and $s^\out$. We write
\begin{align}
    U_D(t_1)&=V_2(t_2,t_1)V_1(t_1,t),\\
    V_1(t_1,t)&:=e^{-i(t_1-t)H^{S_1E}},\\
    V_2(t_2,t_1)&:=e^{-i(t_2-t_1)H^{S_2E}},
\end{align}
where each operator acts as the identity on the other system wire. We have chosen the interaction picture, in which the exchange Hamiltonians are time independent and take the form
\begin{align}
    H^{S_j E}:=g\left(\sigma_+^{S_j}\sigma_-^E+\sigma_-^{S_j}\sigma_+^E\right),\qquad j=1,2.
\end{align}

Since the two time intervals depend on $t_1$ with opposite signs,
\begin{align}
    \frac{dV_1}{dt_1}&=-iH^{S_1E}V_1,\\
    \frac{dV_2}{dt_1}&=+iH^{S_2E}V_2.
\end{align}
The output-side generator of this one-parameter family is therefore
\begin{align}
    G_D(t_1)&:=i\frac{dU_D}{dt_1}U_D^\dagger\nonumber\\
    &=V_2\left(H^{S_1E}-H^{S_2E}\right)V_2^\dagger\nonumber\\
    &=V_2H^{S_1E}V_2^\dagger-H^{S_2E}.
\end{align}
The corresponding input-side generator is
\begin{align}
    G_D^\inp(t_1)&:=-iU_D^\dagger\frac{dU_D}{dt_1}\nonumber\\
    &=V_1^\dagger\left(H^{S_2E}-H^{S_1E}\right)V_1.
\end{align}
For a path $t_1=t_1(\lambda)$, these give $H_D^\out(\lambda)=\dot t_1G_D(t_1)$ and $H_D^\inp(\lambda)=\dot t_1G_D^\inp(t_1)$. Both are defined by a single derivative of $U_D$ along the chosen one-parameter family.

For the exchange interaction, conjugation by $V_2$ yields
\begin{align}
    V_2\sigma_-^E V_2^\dagger
    &=\cos(g\tau_2)\sigma_-^E
    -i\sin(g\tau_2)\sigma_-^{S_2}\sigma_z^E,\\
    V_2\sigma_+^E V_2^\dagger
    &=\cos(g\tau_2)\sigma_+^E
    +i\sin(g\tau_2)\sigma_+^{S_2}\sigma_z^E,
\end{align}
where $\tau_2=t_2-t_1$. Substituting these identities into $V_2H^{S_1E}V_2^\dagger$ gives
\begin{align}
    G_D(t_1)&=\cos(g\tau_2)H^{S_1E}-H^{S_2E}\nonumber\\
    &\quad+i\sin(g\tau_2)H^{S_1S_2E},\\
    H^{S_1S_2E}&:=g\left(\sigma_-^{S_1}\sigma_+^{S_2}-\sigma_+^{S_1}\sigma_-^{S_2}\right)\sigma_z^E.
\end{align}
The last term is environment mediated and contains operators on both system wires, $S_1$ and $S_2$. Its appearance in the generator does not represent an additional microscopic interaction or allow signalling from the later input $t^\inp$ to the earlier output $t^\out$: the unitary $V_1$ produces the earlier output before $V_2$ acts on the later input and the environment. Consequently, the sequential circuit preserves the assumed causal ordering despite the mixed term in its generator.

\section{Evolution Family of Four-Legged Quantum Combs}\label{(QDoD)Appendix:EvolutionFamily}
\begin{figure}
    \centering
    \includegraphics[width=\linewidth]{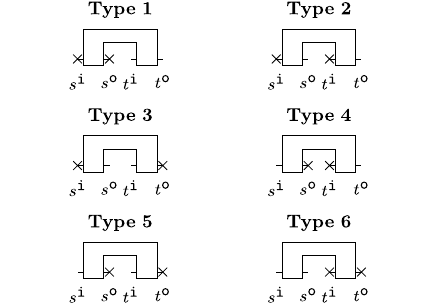}
    \caption{Showing the six possible combinations of leg pairs corresponding to the initial condition, indicated by crosses. Quantum combs are not yet temporally ordered.}
    \label{(QDoD) Fig apx-1}
\end{figure}

In this appendix, we collect 20 evolution types of the four-legged quantum combs with strict temporal orderings. For each evolution type, we refer to the organizational system introduced in Sec.~\ref{sec:quantumdynamicsofdynamics}. First, let us recall that one can choose two legs as the initial condition of the quantum comb; this would give 6 types of evolutions without temporal ordering, organized in Fig.~\ref{(QDoD) Fig apx-1}

For each of the 6 types, the quantum combs can have four different temporal arrangements, as shown in Fig.~\ref{(QDoD) Fig apx-2}. For types 1 and 6, because of the contraction causal restriction, the only possible temporal orderings are $B$ and $C$; therefore, we have $6\times 4 -4 = 20$ different CP-divisible combs.

\begin{figure}
    \centering
    \includegraphics[width=\linewidth]{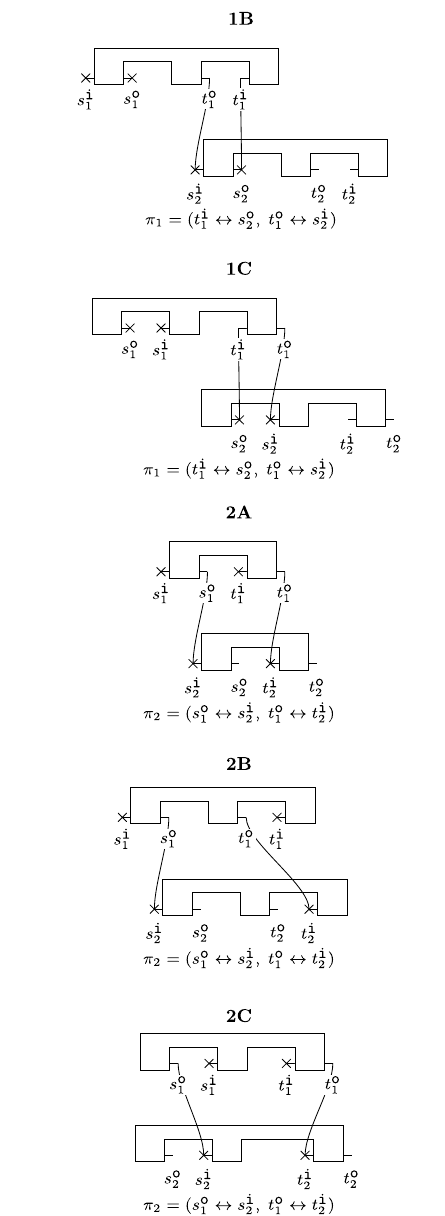}
    \caption{Type 1 admits only B and C. The first three
representatives of Type 2 are shown here; 2D continues in the next column.}
    \label{(QDoD) Fig apx-3}
\end{figure}

\begin{figure}
    \centering
    \includegraphics[width=\linewidth]{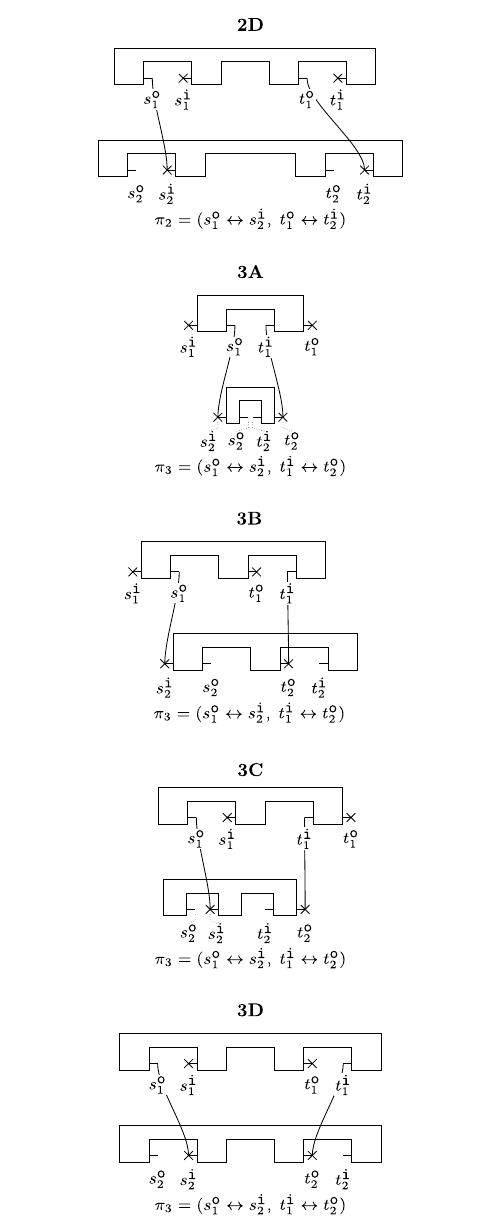}
    \caption{Completion of Type 2 and all four Type 3 diagrams.
The compressed lower shapes in 3A and 3C preserve the leg order.}
    \label{(QDoD) Fig apx-4}
\end{figure}

\begin{figure}
    \centering
    \includegraphics[width=\linewidth]{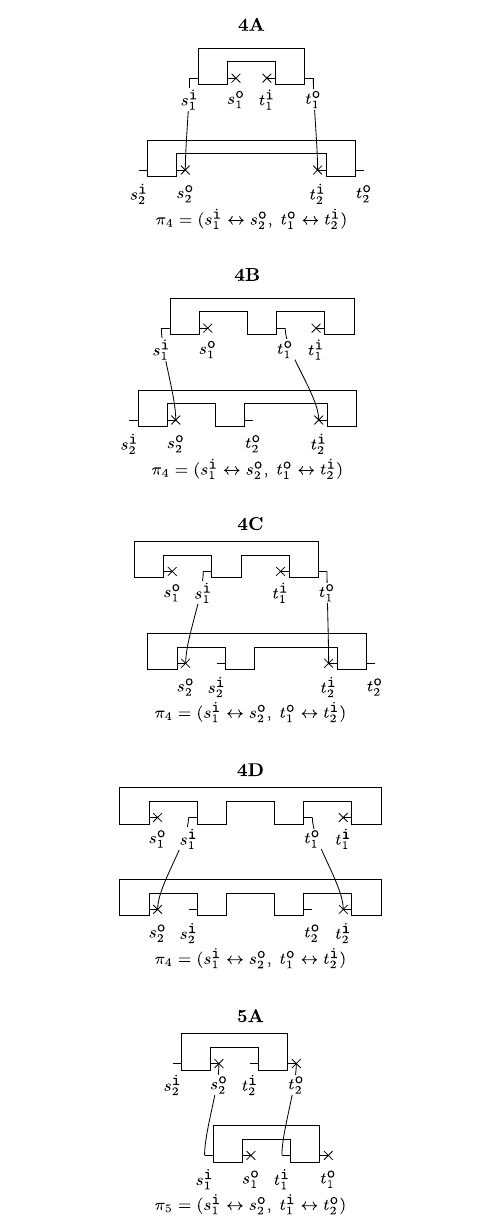}
    \caption{All four Type 4 diagrams and 5A. In 5A, the connected
marked legs are on the upper copy, as in the original drawing.}
    \label{(QDoD) Fig apx-5}
\end{figure}

\begin{figure}
    \centering
    \includegraphics[width=\linewidth]{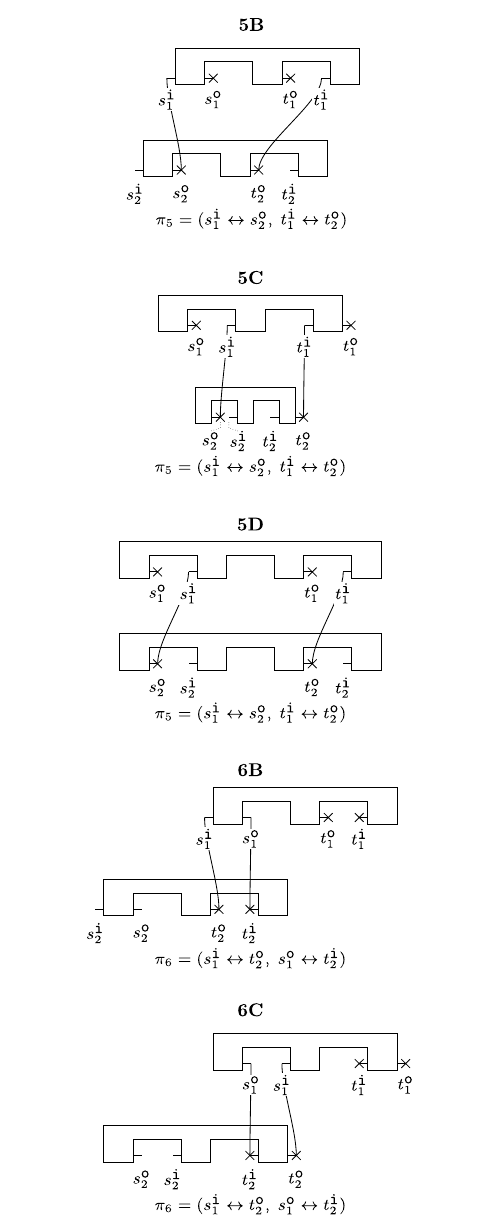}
    \caption{Completion of Type 5 and the two admissible Type 6
diagrams. The exclusion of 6A and 6D follows from temporal ordering.}
    \label{(QDoD) Fig apx-6}
\end{figure}

In Figs.~\ref{(QDoD) Fig apx-1}--\ref{(QDoD) Fig apx-6}, we give all 20 possible evolutions of the quantum combs. 

\clearpage
\bibliography{references.bib}

\end{document}